\documentclass[aps,pra,superscriptaddress,footinbib,showpacs,twocolumn]{revtex4-1}
\usepackage{comment}

\usepackage{amssymb,latexsym,mathrsfs}
\usepackage{amsfonts}
\usepackage{graphicx}
\usepackage{lmodern}
\usepackage{fix-cm}
\usepackage{xfrac}
\usepackage{color}    
\usepackage{lipsum}
\usepackage{textcomp}
\usepackage{amsmath}
\usepackage{mathtools}
\usepackage{subfigure}
\usepackage{float}
\usepackage{datetime}
\usepackage{hyperref}
 \hypersetup{
           breaklinks=true,   
           colorlinks=true,   
           pdfusetitle=true,  
    linkcolor=red,
    filecolor=magenta,      
    urlcolor=cyan,
        }
\usepackage{tabularx,ragged2e,booktabs}

\newcommand{\be}{\begin{equation}}
\newcommand{\ee}{\end{equation}}
\newcommand{\bearr}{\begin{array}}
\newcommand{\enarr}{\end{array}}

\def\bea{\begin{eqnarray}}
\def\eea{\end{eqnarray}}
\def\ba{\begin{array}}
\def\ea{\end{array}}

\definecolor{dgreen}{rgb}{0,0.7,0}

\begin{document}

\title{Multiple quantum Mpemba effect: exceptional points and oscillations}

\author{Amit Kumar Chatterjee}\email{amit.fc@vidyamandira.ac.in}
\affiliation{Yukawa Institute for Theoretical Physics, Kyoto University, Kitashirakawa Oiwakecho, Sakyo-ku, Kyoto 606-8502, Japan}
\affiliation{Department of Physics, Ramakrishna Mission Vidyamandira, Belur Math, Howrah 711 202, India}
\thanks{present address}
\author{Satoshi Takada}\email{takada@go.tuat.ac.jp}
\affiliation{Department of Mechanical Systems Engineering and Institute of Engineering, Tokyo University of Agriculture and Technology, 2–24–16 Naka-cho, Koganei, Tokyo 184–8588, Japan}
\author{Hisao Hayakawa}\email{hisao@yukawa.kyoto-u.ac.jp}
\affiliation{Yukawa Institute for Theoretical Physics, Kyoto University, Kitashirakawa Oiwakecho, Sakyo-ku, Kyoto 606-8502, Japan}

\begin{abstract} 
We explore the role of exceptional points and complex eigenvalues on the occurrence of the quantum Mpemba effect. To this end, we study a two-level driven dissipative system subjected to an oscillatory electric field and dissipative coupling with the environment. We find that both exceptional points and complex eigenvalues can lead to {\it multiple} quantum Mpemba effect. It occurs in an observable when time evolved copies corresponding to two different initial conditions, one initially having higher observable value compared to the other and both relaxing towards the same steady state, intersect each other more than once during their relaxation process. Each of the intersections denotes a quantum Mpemba effect and marks the reversal of identities between the two copies i.e. the copy with higher observable value before the intersection becomes the lower valued copy (and vice versa) after the intersection. Such multiple intersections originate from additional algebraic time dependence at the exceptional points and due to oscillatory relaxation in the case of complex eigenvalues. We provide analytical results for quantum Mpemba effect in the density matrix in presence of coherence. Depending on the control parameters (drive and dissipation), observables such as energy, von Neumann entropy, temperature etc. exhibit either single or multiple quantum Mpemba effect. However, the distance from steady state  measured in terms of the Kullback-Leibler divergence shows only single quantum Mpemba effect although the corresponding speed gives rise to either single or multiple quantum Mpemba effect. 
\end{abstract}

\maketitle

\section{Introduction}
Memory effect describes the influence of past information on present events \cite{Keim_2019}. Such memory effects often lead to unexpected events. In physical systems, several forms of interesting memory effects have been observed in different contexts e.g. pulse-sign memory in charge-density wave conductors (memory of last input signal direction) \cite{Gill_1981,Fleming_1986}; Kaiser effect in rocks and metals \cite{Kaiser_1950,Zang_2014,Heimisson_2015} and Mullins effect in rubbers \cite{Mullins_1969,Dorfmann_2004,Diani_2009} (both bearing memories of largest applied stress); Kovacs effect (memory of bath temperature that affects the thermalization process) \cite{Kovacs_1963,Kovacs_1979,Bertin_2003,Prados_2014,Militaru_2021}; aging and rejuvenation in spin glasses \cite{Jonason_1998,Rodriguez_2003,Baity-Jesi_2018} and structural glasses \cite{Zheng2003,Di2011,McKenna2017,Scalliet2019}, granular materials \cite{Josserand_2000,Kumar16,Otsuki20,Kawasaki20}, dense suspensions \cite{Pradipto20} and active matter \cite{Janssen_2017}. 

The Mpemba effect (MPE) was originally conceived as a counter-intuitive thermal phenomenon of faster cooling of hotter liquids, observed long ago \cite{Aristotle81} and lately rediscovered by Mpemba and Osborne \cite{Mpemba69}. The MPE thus bears the imprint of initial memory in a surprising way and constitutes a separate class of fascinating memory effects. Later, the MPE has been regarded as one of the typical anomalous relaxations where a system with initially higher observable (e.g. temperature, energy, entropy, distance function etc.) value can relax faster than its copy with initially lower observable value, when both of them approach towards the same steady state value. In classical systems, MPE has been  found in variety of classical systems including colloids \cite{Kumar20,Kumar22}, granular fluids  \cite{Santos17,Torrente19,Biswas20,Biswas21,Mompo20,Megias22b,Patron23}, optical resonators \cite{Keller18,Santos20,Patron21}, inertial suspensions \cite{Takada21a,Takada21b}, Markovian models \cite{Lu17, Klich19, Busiello21,Lin22}, Langevin system \cite{Biswas23a} and many others \cite{Greaney11,Baity-Jesi19,Gonzalez2021,Yang20,Yang22,Gonzalez21,Chetrite21,Holtzman22,Megias22a,Biswas23,Pemartin23}.     

In spite of considerable amount of works on MPE in the classical domain, only a few studies have been done on the quantum Mpemba effect (QMPE), such as quantum Ising model \cite{Nava19}, Dicke model \cite{Carollo21}, few-level systems \cite{Manikandan21}, V-type system \cite{Ivander23}, XXZ spin chain \cite{Ares23}, single level quantum dot \cite{Chatterjee_2023}, XY spin chain \cite{Murciano23} and free and interacting integrable systems \cite{Rylands23}. In fact, only quite recently the quantum analogue of the original thermal MPE i.e. QMPE in temperature, has been theoretically revealed in a quantum dot \cite{Chatterjee_2023} and the significant role of relaxation modes other than the slowest relaxation mode to create QMPE, has been shown \cite{Chatterjee_2023}. However, the semi-classical approach taken in Ref.~\cite{Chatterjee_2023} naturally raises a question about the generality of QMPE in temperature and other observable, in presence of {\it coherence} in purely quantum mechanical systems. This constitutes one of the motivations of the present work.   

Since the phenomenon of QMPE is a comparatively new area of research, it is important to explore the plausible connections of QMPE to other interesting features of quantum systems. In the present work, we would like to address some such connections. In particular, the study of {\it exceptional} points (EPs) and surfaces in non-Hermitian quantum systems and Liouvillian dynamics has caught a lot of attention in recent days \cite{Heiss12,Hashimoto15,Doppler16,Zhong19,McDonald20,Soleymani22,Ding22,Minganti19,Arkhipov20,Arkhipov21,Muller08,Naghiloo19,Saha23,Roccati23,Khandelwal21} along with their applications in quantum optics \cite{Miri19,Ozdemir19}. The  occurrence of EPs is directly related to eigen-spectrum of a system and hence to its relaxation process. Therefore a natural curiosity would be to understand the role of EPs on the possibility of QMPE. In fact, the occurrence of double MPE (i.e. two intersections between trajectories) at an EP has already been pointed out for MPE in classical inertial suspensions \cite{Takada21a}. Interestingly, the acceleration of relaxation process in open quantum systems by maximizing the gap between the zero mode and the slowest eigenmode at an EP, has recently been studied \cite{Zhou23}. In general, such speed up is desired for several application purposes and therefore it has been a subject of immense interest in both classical and quantum systems \cite{Lapolla20,Vu21,Kochsiek22,Ghimenti22,Bao22,Zhou23}. Since QMPE allows faster relaxation of initially higher distant trajectories, it is important to find criteria for QMPE and to understand how to utilize QMPE to achieve faster approach to steady state. Another intriguing feature in non-Hermitian quantum mechanical systems is the appearance of complex eigenvalues \cite{Bender2007,Moiseyev2011,Nakano2011,Hatano2015,Li2023}. It is noteworthy that the analysis of QMPE in the existing examples \cite{Nava19,Carollo21,Manikandan21,Ivander23,Chatterjee_2023} is  based on purely exponential relaxation processes, neither the EP nor the oscillations have been considered yet. There is only an exception in Ref.~\cite{Ares23} where more than one intersections between trajectories (two intersections to be precise, termed as {\it revivals}) has been observed in the entanglement entropy in the XXZ spin chain, although not discussed much and simply explained as a finite-size effect. 

In this paper, we explore the following two topics: (i) the effect of EPs and complex eigenvalues on QMPE by analysing the Quantum Master Equation, and (ii) the existence of thermal QMPE in a purely quantum mechanical setting. We show that in the presence of EPs and complex eigenvalues, two time-evolved trajectories of an observable (e.g., temperature, energy) can intersect each other more than once before reaching the same final steady state. 
In the context of thermal QMPE, the first intersection implies that the initially hotter quantum system relaxes faster to intersect the initially colder quantum system and thereby reverses their roles, i.e. hotter becomes colder and vice versa.
This can be followed by one or more such intersections so that the present hotter system can become colder again and so on. 
Similar observations have been obtained in average energy. 
Such multiple intersections of trajectories give rise to the remarkable phenomenon of multiple QMPE. 
To demonstrate multiple QMPE as a consequence of EP and complex eigenvalues, we study a two-level driven dissipative quantum system developed in Ref.~\cite{Hatano_2019}, which we call Hatano's model in this paper, as a typical and minimal model of open quantum systems. 
It is well known that many complicated quantum problems are mapped or reduced to two-level systems; needless to mention they are experimentally realizable (e.g., two internal electronic energy levels of an atom controlled by laser) and have applications in qubits, nuclear magnetic resonance, neutrino oscillations, etc. 
We also find thermal QMPE in presence of coherence, some cases remarkably with multiple intersections, following the same definition of temperature used in Ref.~\cite{Chatterjee_2023} . 
Importantly, we provide analytical results for the time evolution of the density matrix in the two-level driven dissipative system considered here. 
Consequently, we derive analytical criteria for QMPE in energy and in principle, one can analyze exactly (analytically and  numerically) QMPE in any other observable. Note that there have been exciting recent advances in experimental demonstration of QMPE in a single $\prescript{88}{\,}{\mathrm{Sr}^+}$ trapped ion qubit \cite{Shapira24} as well as isolated many body quantum systems using twelve calcium ions' trapped-ion quantum simulator \cite{Joshi24}. In particular, the single qubit considered in Ref.~\cite{Shapira24} in presence of coherent driving and decay, constitutes a two level driven dissipative system. It would be interesting to find connections between the QMPEs studied in the present article and those observed in Ref.~\cite{Shapira24} in future.

The paper is arranged as follows. In Sec.~\ref{sec2}, we define the details of the model and the protocol used to study QMPE. The Sec.~\ref{sec3} describes different regions in the control parameter plane depending on the types of eigenvalues of the Lindbladian and the multiple QMPE is defined. In articular, we focus on regions with exceptional points and complex eigenvalues to discuss their effects on QMPE in the density matrix and observables such as energy, entropy, Kullback-Leibler (KL) divergence \cite{Kullback51,Sagawa22,Shiraishi23,Strasberg22,Yoshii23} and temperature; by providing analytical results. In Sec.~\ref{sec4}, we discuss QMPE in a region with second-order EPs. Section~\ref{sec5} demonstrates the effect of complex eigenvalues on the time evolution of the density matrix leading to multiple QMPE in the observables. We summarize the results and discuss possible future directions in Sec.~\ref{sec6}. Explicit expressions of the eigenvectors used in the analyses in Secs.~\ref{sec4} and \ref{sec5} are provided in Appendices \ref{appen1} and \ref{appen2}, respectively. We discuss QMPE briefly in region of purely exponential relaxation in Appendix~\ref{appen3} and at a third-order EP in Appendix~\ref{appen4}. In Appendix~\ref{appen5}, we explicitly present and discuss a diagram containing different eigenvalue regions with the variations of drive and dissipation.

\section{Model and protocol}
\label{sec2}
We consider a two-level open quantum system, subjected to a periodic drive and dissipatively coupled to the environment. Following Ref.~\cite{Hatano_2019}, we hereafter refer to this model as Hatano's model in the present manuscript. In this paper, we adopt $\hbar=1$. The two levels of the system, ground state and excited state, have energies $E_{\mathrm{g}}$ and $E_{\mathrm{e}}$, respectively. The gap between the two levels is denoted by $\Delta:=E_{\mathrm{e}}-E_{\mathrm{g}}$. In Hatano's model, the transfers between the energy levels are enabled by the external oscillatory electric field  $E=E_0\mathrm{cos}(\omega t)$ and the dissipation of the system to the environment via coupling $\Gamma$. The parameter $\delta:=\Delta-\omega$ represents the {\it detuning} of the periodic drive with respect to the system and $d:=DE_0$, where $D$ is the electric dipole moment, denotes the {\it effective electric field}. The tuning parameters in Hatano's model are $d$, $\Gamma$ and $\delta$ . However, one of them fixes the unit of energy, so we essentially have two free parameters $\tilde{d}:=d/\delta$ and $\tilde{\Gamma}:=\Gamma/\delta$. 

Hatano's model, within Markovian approximation, can be described by the Gorini-Kossakowski-Sudarshan-Lindblad (GKSL) equation \cite{Lindblad76,Gorini76,Manzano20,Chruscinski17} (see \cite{Hatano_2019} for the present context):
\begin{equation}
\mathrm{i} \frac{d}{dt}|\widehat{\rho}(t)\rangle=\widehat{\mathcal{L}} |\widehat{\rho}(t)\rangle.
\label{eq:lindblad_eq} 
\end{equation}
The original $2\times2$ density matrix is recast as the column vector $|\widehat{\rho}(t)\rangle=(\rho_{\mathrm{eg}}(t),\rho_{\mathrm{ge}}(t),\rho_{\mathrm{ee}}(t),\rho_{\mathrm{gg}}(t))^{T}$ in Eq.~(\ref{eq:lindblad_eq}), where $T$ denotes the  transpose. The terms $\rho_{\mathrm{eg}}(t)$ and $\rho_{\mathrm{ge}}(t)$ are the off-diagonal density matrix elements, characterizing the purely quantum mechanical nature of the system. The elements $\rho_{\mathrm{gg}}(t)$ and $\rho_{\mathrm{ee}}(t)$ correspond to the ground state and the excited state probabilities, respectively. For convenience of notations, we denote the density matrix elements as $\rho_j$ where $j=1,2,3,4$ correspond to $\rho_{\mathrm{eg}}(t),\rho_{\mathrm{ge}}(t),\rho_{\mathrm{ee}}(t),\rho_{\mathrm{gg}}(t)$ respectively. The transition between these density matrix elements is described by the Lindbladian of the form below
\begin{equation}
\widehat{\mathcal{L}}=
\left(\begin{array}{cccc}
1-\mathrm{i}\tilde{\Gamma}/2 & 0 & -\tilde{d}/2 & \tilde{d}/2 \\
 0 & -1-\mathrm{i}\tilde{\Gamma}/2 & \tilde{d}/2 & -\tilde{d}/2 \\ 
 -\tilde{d}/2 & \tilde{d}/2 & -\mathrm{i}\tilde{\Gamma} & 0 \\
 \tilde{d}/2 & -\tilde{d}/2 & \mathrm{i}\tilde{\Gamma} & 0 \\
\end{array}
\right).
\label{eq:super-operator} 
\end{equation}
Let us clarify a few points about the matrix $\widehat{\mathcal{L}}$ at this point. Note that the dissipation $\Gamma$ makes the system non-Hermitian. More importantly, each column sum $\widehat{\mathcal{L}}$ does not equal to zero, which distinguishes Hatano's model from the classical Markovian models studied previously \cite{Lu17}. A recent work \cite{Chatterjee_2023} showing thermal quantum Mpemba effect also deals with an effective 4-state quantum system. However, the semi-classical approach in Ref.~\cite{Chatterjee_2023} considered a diagonal form of the density matrix, while the present model includes quantum coherence in the form of off-diagonal density matrix elements $\rho_{\mathrm{eg}}(t)$ and $\rho_{\mathrm{ge}}(t)$. Naturally, it is interesting to ask whether Hatano's  model, in the presence of coherence, can exhibit QMPE in observables such as temperature, energy, entropy and KL divergence. 

{\it Protocol.-} We consider two different sets I and II of initial conditions $\widehat{\rho}^{\mathrm{I}}(t=0)$ and $\widehat{\rho}^{\mathrm{II}}(t=0)$. Both of them are chosen in the form of steady state distributions corresponding to the zero  eigenvalue of $\widehat{\mathcal{L}}$. We choose steady-state initial conditions because they are easily achievable in experiments without requiring fine tuning.
Theoretically, other types of initial states could be considered that are not steady states, thereby exploring QMPE in a broader space of initial conditions.
However, achieving and verifying such initial states experimentally may necessitate more extensive fine tuning.
Therefore, we adhere to steady-state initial conditions.  
The difference between the two initial conditions I and II lies in the choice of the control parameters. To elaborate, we choose $(\tilde{d}_{\mathrm{I}},\tilde{\Gamma}_{\mathrm{I}})$ for initial condition I and $(\tilde{d}_{\mathrm{II}},\tilde{\Gamma}_{\mathrm{II}})$ for initial condition II. Both of them are instantaneously quenched to the same set of values $(\tilde{d},\tilde{\Gamma})$, so that they will eventually reach the same steady state. Thus, the entire set of control parameters consists of six parameters $(\tilde{d}_{\mathrm{I}},\tilde{\Gamma}_{\mathrm{I}},\tilde{d}_{\mathrm{II}},\tilde{\Gamma}_{\mathrm{II}}; \tilde{d},\tilde{\Gamma})$, where the first four are before-quench parameters and the last two are after-quench parameters. In most parts of the present work, we perform {\it fixed dissipation protocol} $\tilde{\Gamma}_{\mathrm{I}}=\tilde{\Gamma}_{\mathrm{II}}=\tilde{\Gamma}$ where we only tune the drive to obtain QMPE. However, in some cases we will use {\it variable dissipation protocol} $\tilde{\Gamma}_{\mathrm{I}}\neq\tilde{\Gamma}_{\mathrm{II}}\neq\tilde{\Gamma}$ where both drive and dissipation are tuned to search for QMPE in the whole parameter space. 

To mathematically formulate the initial conditions before quench, let us denote the right eigenvectors and left eigenvectors of $\widehat{\mathcal{L}}$ [Eq.~(\ref{eq:super-operator})]  by $|r_j\rangle$ and $\langle \ell_j|$ ($j=1,2,3,4$), respectively. The vector $|r_1\rangle$ corresponds to the steady probability distribution and $\langle \ell_1|$ is the corresponding left eigenvector. Below we provide their explicit forms:
\begin{eqnarray}
|r_1\rangle&=& \left(\begin{array}{cccc}
                      -\displaystyle\frac{\tilde{d}(2+\mathrm{i}\tilde{\Gamma})}{4+\tilde{d}^2+\tilde{\Gamma}^2}, & -\displaystyle\frac{\tilde{d}(2-\mathrm{i}\tilde{\Gamma})}{4+\tilde{d}^2+\tilde{\Gamma}^2}, & \displaystyle\frac{\tilde{d}^2}{4+\tilde{d}^2+\tilde{\Gamma}^2}, & 1 \\
                     \end{array}
\right)^T,\cr
\langle \ell_1|&=&\left(\begin{array}{cccc}
0, & 0, & \displaystyle\frac{4+\tilde{d}^2+\tilde{\Gamma}^2}{4+2\tilde{d}^2+\tilde{\Gamma}^2}, & \displaystyle\frac{4+\tilde{d}^2+\tilde{\Gamma}^2}{4+2\tilde{d}^2+\tilde{\Gamma}^2} \\
                      \end{array}
\right).
\label{eq:m1} 
\end{eqnarray}
The steady state $|\widehat{\rho}_{\mathrm{ss}}(\tilde{d},\tilde{\Gamma})\rangle$ is simply related to  $|r_1\rangle$ as 
\begin{eqnarray}
|\widehat{\rho}_{\mathrm{ss}}(\tilde{d},\tilde{\Gamma})\rangle=\frac{4+\tilde{d}^2+\tilde{\Gamma}^2}{4+2\tilde{d}^2+\tilde{\Gamma}^2}|r_1\rangle,
\label{eq:m2} 
\end{eqnarray}
where $\tilde{d}$ and $\tilde{\Gamma}$ are drive and dissipation after the quench, respectively. We construct the two initial conditions I and II as follows:
\begin{eqnarray}
&&|\widehat{\rho}^{\mathrm{I}}(t=0)\rangle=|\widehat{\rho}_{\mathrm{ss}}(\tilde{d}_{\mathrm{I}},\tilde{\Gamma}_{\mathrm{I}})\rangle,\cr
&&|\widehat{\rho}^{\mathrm{II}}(t=0)\rangle=|\widehat{\rho}_{\mathrm{ss}}(\tilde{d}_{\mathrm{II}},\tilde{\Gamma}_{\mathrm{II}})\rangle.
\label{eq:m3} 
\end{eqnarray}
The coherence terms $\rho_{\mathrm{eg}}$ and $\rho_{\mathrm{ge}}$ being complex conjugates of each other, we decompose them into real (re) and imaginary (im) parts as
\begin{eqnarray}
\rho_{\mathrm{eg}}=\rho_{\mathrm{re}} +\mathrm{i} \rho_{\mathrm{im}},\hspace*{0.2 cm}
\rho_{\mathrm{ge}}=\rho_{\mathrm{re}} -\mathrm{i} \rho_{\mathrm{im}}.
\label{eq:m4} 
\end{eqnarray}
Specifically, in steady state, $\rho_{\mathrm{re}}$ and $\rho_{\mathrm{im}}$ have the forms (using Eqs.~(\ref{eq:m1}) and (\ref{eq:m2})):
\begin{eqnarray}
\rho_{\mathrm{re}}=-\frac{2\tilde{d}}{4+2\tilde{d}^2+\tilde{\Gamma}^2}, \hspace*{0.2 cm} \rho_{\mathrm{im}}=-\frac{\tilde{d}\,\tilde{\Gamma}}{4+2\tilde{d}^2+\tilde{\Gamma}^2}.
\label{eq:m5} 
\end{eqnarray}
\section{Different regimes in $\tilde{d}-\tilde{\Gamma}$ plane and multiple QMPE}
\label{sec3}
As discussed in Ref.~\cite{Hatano_2019}, Hatano's model exhibits a rich spectrum of eigenvalues. In the plane of control parameters $\tilde{d}-\tilde{\Gamma}$, it has six different regions ($\mathrm{a}_{1}$), ($\mathrm{a}_{2}$), (b), (c), (d) and (e) depending on the nature of the eigenvalues of $\widehat{\mathcal{L}}$ [Eq.~(\ref{eq:super-operator})]. The regions ($\mathrm{a}_{1}$) and ($\mathrm{a}_{2}$) consist of complex eigenvalues, region ($\mathrm{b}$) with purely imaginary eigenvalues, regions ($\mathrm{c}$) and ($\mathrm{d}$) with {\it second-order} EPs and region ($\mathrm{e}$) containing a single {\it third-order} EP. At a $n$-th order EP, $n$ eigenvalues become equal and their eigenvectors coalesce. The regions ($\mathrm{a_1}$) and ($\mathrm{a_2}$) can be distinguished from each other depending on the largest nonzero eigenvalue which is purely imaginary in ($\mathrm{a_1}$) but complex in ($\mathrm{a_2}$). They can be also differentiated by their locations in the parameter space of $(\tilde{d},\tilde{\Gamma})$. The region ($\mathrm{a_1}$) is placed in relatively larger $\tilde{d}$ and smaller $\tilde{\Gamma}$, whereas the region ($\mathrm{a_2}$) is placed in comparatively smaller $\tilde{d}$ and larger $\tilde{\Gamma}$. All the nonzero eigenvalues in region (b) are purely imaginary and they are unequal from one another. The line (c) of second-order EPs separate region ($\mathrm{a_1}$) from region (b), whereas the line (d) of second-order EPs demarcate (b) from ($\mathrm{a_2}$). The two lines (c) and (d) of second-order EPs meet at a single point in the $\tilde{d}-\tilde{\Gamma}$ plane that defines the region (e) which is a third-order EP. See Appendix~\ref{appen5} for the visual descriptions of these regions (also see Figs. 3 and 4 in \cite{Hatano_2019}). We would like to analyze the full relaxation of the system that contains all three relaxation modes (apart from the eigenvalue zero that corresponds to the steady state) of the Lindbladian $\widehat{\mathcal{L}}$ [Eq.(\ref{eq:super-operator})]. The presence of multiple relaxation modes, EPs  and complex eigenvalues naturally triggers the possibility of interesting features of QMPE in Hatano's model. 

Indeed, in the space of control parameters $(\tilde{d}_{\mathrm{I}},\tilde{\Gamma}_{\mathrm{I}},\tilde{d}_{\mathrm{II}},\tilde{\Gamma}_{\mathrm{II}}; \tilde{d},\tilde{\Gamma})$, we find {\it multiple} QMPE where the two trajectories corresponding to some observable, intersect each other multiple times before reaching the same steady state. Specifically, we have obtained two or more intersections in a oscillatory manner, in regions ($\mathrm{a}_{1}$) and ($\mathrm{a}_{2}$) with complex eigenvalues. Whereas, at most two intersections or {\it double} QMPE can be found in regions (c), (d) and (e) with EP. Interestingly, even number of intersections (e.g. double QMPE or quadruple QMPE) imply that the initial identities of the two copies (higher valued copy and lower valued copy of an observable, more explicitly hotter and colder copies if the observable is temperature) are restored finally after the last intersection. Therefore, in such cases the initially hotter or higher valued copy remains hotter or higher valued after all the intersections have occurred. On the other hand, odd number of intersections (single QMPE or triple QMPE) indicates the ultimate reversal of the initial identities after the final intersection. In these cases the initially hotter or higher valued copy finally becomes colder or lower valued copy after the completion of all intersections. This classification has important consequences when one is not only interested in faster relaxations or intersections at intermediate times, but also interested in faster reach to {\it final or steady state}. It would be intriguing in future to find connections between such odd or even QMPE and faster approach to steady state. Notably, if we restrict ourselves to monotonic time evolution, then even number of QMPE is of no use. Only odd number of QMPE can help to reach faster to steady state using QMPE. However, more conditions are required like avoiding overshoot etc. to use odd QMPE to get faster reach to the steady state. In the next sections, we discuss multiple QMPE in different regimes in detail.

In this paper, we illustrate QMPEs in region (d) for quench onto the second-order EP in Sec.~\ref{sec4} (with more analytical details in Appendix~\ref{appen1}) and region ($\mathrm{a_1}$) for the quench onto the point with complex eigenvalues in Sec.~\ref{sec5} (with more analytical details in Appendix~\ref{appen2}). We do not discuss QMPE in regions (c) and ($\mathrm{a_2}$) separately, because the analyses of QMPEs in these regions are similar to those in regions (d) and ($\mathrm{a_1}$), respectively. The QMPEs in other regions are discussed in the appendices. Appendix~\ref{appen3} discusses QMPE in region (b) with purely exponential relaxations and Appendix~\ref{appen4} contains the description of the QMPE at the third-order EP in region (e).
\section{QMPE in region ($\mathrm{d}$): second-order EP}
\label{sec4}
In this section, we consider the possibility of single and multiple QMPE in region (d) of Hatano's model. This region contains second order EP where two of the nonzero eigenvalues of the Lindbladian $\widehat{\mathcal{L}}$ [Eq.~(\ref{eq:super-operator})] become same and their eigenvectors coalesce. In the $(\tilde{d},\tilde{\Gamma})$ parameter plane, the region (d) is defined by the set of points
\begin{eqnarray}
\tilde{\Gamma}=\sqrt{\frac{\tilde{d}^4}{2}+10\tilde{d}^2-4+\frac{\tilde{d}}{2}(\tilde{d}^2-8)^{3/2}}.
\label{eq:d1} 
\end{eqnarray}
The eigenvalues of $\widehat{\mathcal{L}}$ are denoted by $-\mathrm{i}\lambda_j$ ($j=1,2,3,4$) where $\lambda_1=0$. Specifically, in region (d), the eigenvalues obey $0<\lambda_2=\lambda_3<\lambda_4$. To analyze QMPE analytically, we need to diagonalize $\widehat{\mathcal{L}}$. The diagonalization is usually done by the relation $\widehat{\mathcal{L}}_d=\widehat{L}\widehat{\mathcal{L}}\widehat{R}$, where $\widehat{\mathcal{L}}_d$ is the diagonal form of the Lindbladian and the matrix $\widehat{R}$ consists of the right eigenvectors $|r_j\rangle$ of $\widehat{\mathcal{L}}$ as its columns and the matrix $\widehat{L}$ contains the left eigenvectors $\langle \ell_j|$ as its rows, $j=1,2,3,4$. More explicitly,
\begin{eqnarray}
\widehat{R}&=&\left(\begin{array}{cccc}
|r_1\rangle, & |r_2\rangle, & |r_3\rangle, & |r_4\rangle \\                  
                  \end{array}
                  \right),\cr
\widehat{L}&=&\left(\begin{array}{cccc}
\langle \ell_1|, & \langle \ell_2|, & \langle \ell_3|, & \langle \ell_4| \\                  
                  \end{array}
\right)^T.                  
\label{eq:d2} 
\end{eqnarray}
The explicit expressions for $|r_{k}\rangle$ and $\langle\ell_{k}|$ $(k=1,2,3,4)$ [Eq.~(\ref{eq:d2})] in region (d) are provided in Appendix~\ref{appen1}.

It is known that, at the second-order EPs, such {\it completely diagonal} form $\widehat{\mathcal{L}}_d$ cannot be obtained. Instead, we achieve a Jordan normal form $\widehat{\mathcal{L}}_{\mathrm{J}}$ that is not completely diagonal, through the equation below
\begin{equation}
\widehat{\mathcal{L}}_{\mathrm{J}}=\widehat{L}\widehat{\mathcal{L}}\widehat{R},
\label{eq:d3} 
\end{equation}
where $\widehat{\mathcal{L}}_{\mathrm{J}}$ has the following form
\begin{eqnarray}
\widehat{\mathcal{L}}_{\mathrm{J}}=\left(\begin{array}{cccc}
                                            0 & 0 & 0 & 0 \\
                                            0 & -\mathrm{i}\lambda_2 & 1 & 0 \\
                                            0 & 0 & -\mathrm{i}\lambda_2 & 0 \\
                                            0 & 0 & 0 & -\mathrm{i}\lambda_4 \\
                                          \end{array}
\right).
\label{eq:d4} 
\end{eqnarray}
Note the presence of the off-diagonal unit element in Eq.~(\ref{eq:d4}). The explicit expressions for $\lambda_2$ and $\lambda_4$ are obtained as
\begin{eqnarray}
\lambda_2&=& \frac{2\tilde{\Gamma}}{3}+2\mathrm{cos}\left(\frac{2\pi}{3}\right)\left(\frac{\tilde{\Gamma}}{6}\left(1-\frac{\tilde{d}^2}{2}+\frac{\tilde{\Gamma}^2}{36}\right)\right)^{1/3},\cr
\lambda_4&=& \frac{2\tilde{\Gamma}}{3}+2\left(\frac{\tilde{\Gamma}}{6}\left(1-\frac{\tilde{d}^2}{2}+\frac{\tilde{\Gamma}^2}{36}\right)\right)^{1/3}.
\label{eq:d5} 
\end{eqnarray} 

We primarily seek the exact analytical solution for the time evolution of $|\widehat{\rho}(t)\rangle$ with elements $\rho_j(t)$ ($j=1,2,3,4$ correspond to $\rho_{\mathrm{eg}},\rho_{\mathrm{ge}},\rho_{\mathrm{ee}},\rho_{\mathrm{gg}}$ respectively). In principle, once the time evolution of the density matrix is obtained, one can calculate any other observable of interest (e.g. energy, entropy, temperature, the Kullback-Leibler (KL) divergence etc.) to explore QMPE.
\subsection{Density matrix elements}
\begin{figure}[t]
  \centering \includegraphics[width=8.6 cm]{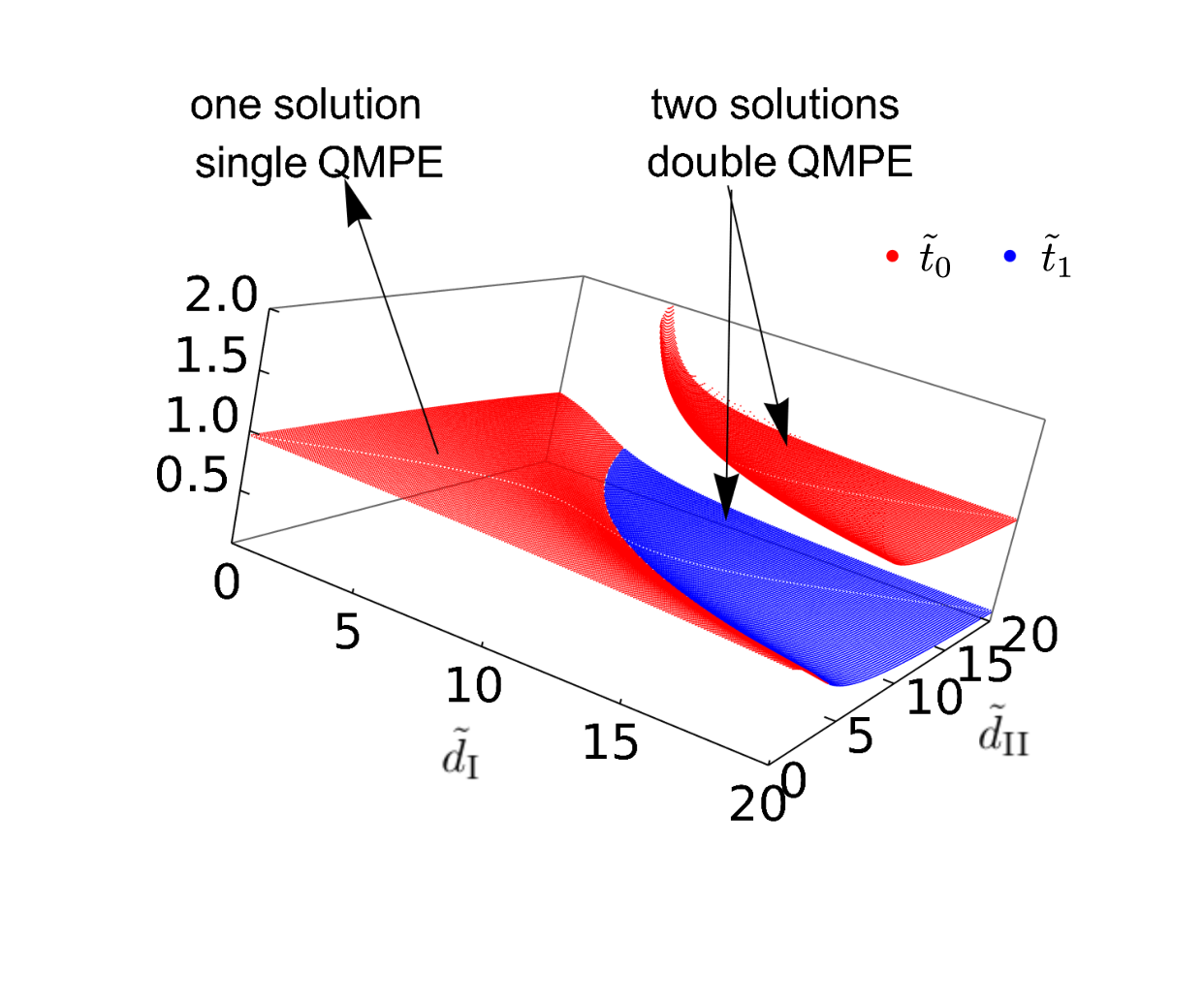}
\caption{The figure shows the intersection times $\tilde{t}_0$ and $\tilde{t}_{-1}$ from Eq.~(\ref{eq:d10}), as functions of the control parameters $\tilde{d}_{\mathrm{I}}$ and $\tilde{d}_{\mathrm{II}}$. The regions with only one solution $\tilde{t}_0$ indicate single QMPE in the ground state probability $\rho_{\mathrm{gg}}(t)$ whereas the regions with both solutions imply double QMPE. Parameters used are $\tilde{d}=4.0, \tilde{\Gamma}_{\mathrm{I}}=\tilde{\Gamma}_{\mathrm{II}}=\tilde{\Gamma}=\sqrt{(568+64\sqrt{2})/2}$ [Eq.~(\ref{eq:d1})].}
\label{fig:d1}
\end{figure}
Using Eqs.~(\ref{eq:d3}) and (\ref{eq:d4}) in Eq.~(\ref{eq:lindblad_eq}), we have obtained the following expressions of the density matrix elements $\rho_j(t)$ $(j=1,2,3,4)$:
\begin{eqnarray}
&&\rho_j(t)=\sum_{k=1}^{4} e^{-\lambda_k t} r_{k,j} a_k-\mathrm{i} t e^{-\lambda_2 t}r_{2,j}a_3, \cr && a_k=\sum_{n=1}^{4} \ell_{k,n}\rho_n(0),
\label{eq:d6} 
\end{eqnarray}
where the initial density matrix elements $\rho_n(0)$ before quench are given by Eq.~(\ref{eq:m3}) and the elements $\ell_{k,n}$ are provided in Appendix~\ref{appen1}. Interestingly, in the presence of the second-order EP in region (d), the density matrix elements in Eq.~(\ref{eq:d6}) have picked up an additional algebraic term proportional to time $t$. This is remarkably different from the time evolution in regions (b) which has only purely exponential relaxations [see Appendix~\ref{appen3}] and region ($\mathrm{a_1}$) that contains damped oscillations [see Sec.~\ref{sec5}]. The linear $t$ dependence originates from the off-diagonal defect in the Jordan normal form in Eq.~(\ref{eq:d4}). 

To investigate QMPE, we focus on the ground state density matrix element $\rho_{\mathrm{gg}}(t)$ $(\equiv\rho_4(t))$. To detect the QMPE, we would like to find the intersections of the two copies I and II during their relaxation, i.e. solution(s) to the equation $\Delta\rho_{\mathrm{gg}}(t)=\rho_{\mathrm{gg}}^{\mathrm{I}}(t)-\rho_{\mathrm{gg}}^{\mathrm{II}}(t)=0$, where $\rho_{\mathrm{gg}}^{\mathrm{I}}(t)$ and $\rho_{\mathrm{gg}}^{\mathrm{II}}(t)$ are ground state
probabilities for copies I and II, respectively. We obtain the following form of $\Delta\rho_{\mathrm{gg}}(t)$
\begin{eqnarray}
\Delta\rho_{\mathrm{gg}}(t)=-e^{-\lambda_2 t}\left[\alpha_1 e^{-(\lambda_4-\lambda_2)t}+ t\,\alpha_2+\alpha_3\right],
\label{eq:d7} 
\end{eqnarray}
where the co-efficients $\alpha_1,\alpha_2,\alpha_3$ are given by
\begin{eqnarray}
\alpha_1=a^{\mathrm{I}}_4-a^{\mathrm{II}}_4,\,\,
\alpha_2=-\mathrm{i}(a^{\mathrm{I}}_3-a^{\mathrm{II}}_3),\,\,
\alpha_3=a^{\mathrm{I}}_2-a^{\mathrm{II}}_2.
\label{eq:d8} 
\end{eqnarray}
The solutions to such transcendental equation in Eq.~(\ref{eq:d7}) are given by the  Lambert $W$ function, where the equation $ye^y=x$ has only one solution $y=W_0(x)$ if $x\geqslant0$ and two solutions $y=W_0(x)$ and  $y=W_{-1}(x)$ if $-1/e\leqslant x<0$. Indeed, such properties of Lambert $W$ function imply that there can be more than one solutions to $\Delta\rho_{\mathrm{gg}}(t)=0$, meaning multiple QMPE. More precisely, at most two intersection times i.e. double QMPE can be observed in region (d). Below we provide the criteria for double QMPE and single QMPE in this region:
\begin{figure}[t]
  \centering \includegraphics[width=8.6 cm]{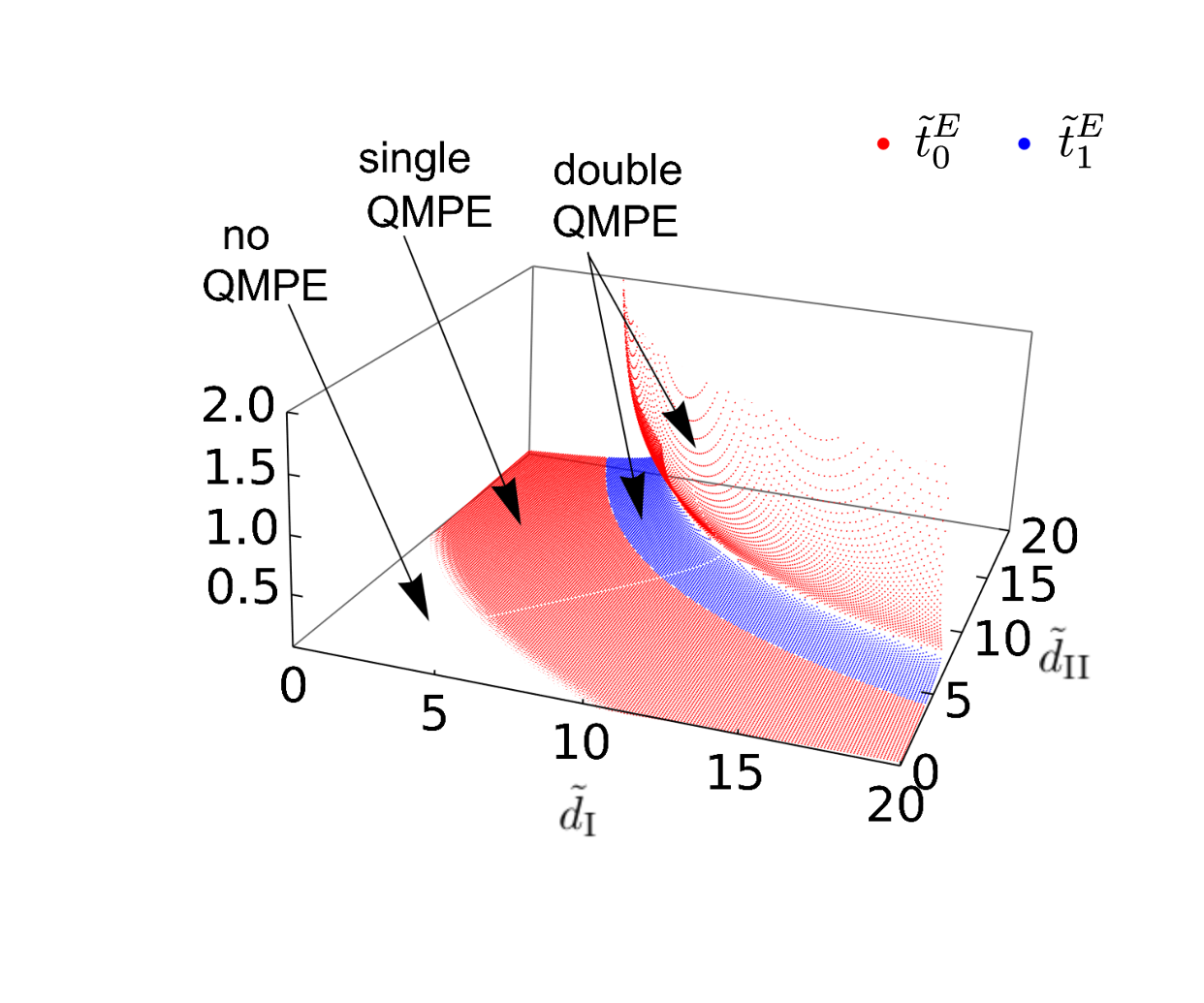}
\caption{The figure illustrates the intersection times $\tilde{t}^E_0$ and $\tilde{t}^E_{-1}$ [Eq.~(\ref{eq:d14})] in energy $E(t)$. The parameter regions with only one solution $\tilde{t}^E_0$ give  single QMPE in energy whereas regions with both solutions mark double QMPE. In contrast to intersection times in $\rho_{\mathrm{gg}}(t)$ [Fig.~\ref{fig:d1}], here there are regions where QMPE is absent in energy. Parameters used are same as Fig.~\ref{fig:d1}.}
\label{fig:d2}
\end{figure}
\begin{eqnarray}
(\lambda_4-\lambda_2)\frac{\alpha_1}{\alpha_2} e^{(\lambda_4-\lambda_2)\frac{\alpha_3}{\alpha_2}} >\frac{1}{e}:&& \hspace*{0.1 cm}\mathrm{no}\,\,\mathrm{QMPE},\cr
(\lambda_4-\lambda_2)\frac{\alpha_1}{\alpha_2} e^{(\lambda_4-\lambda_2)\frac{\alpha_3}{\alpha_2}} >\frac{1}{e}\, \&\, \frac{\alpha_1}{\alpha_2}>0:&& \hspace*{0.1 cm}\mathrm{double}\,\,\mathrm{QMPE},\cr
\frac{\alpha_1}{\alpha_2}<0:&& \hspace*{0.1 cm}\mathrm{usual}\,\,\mathrm{QMPE}.\nonumber\\
\label{eq:d9} 
\end{eqnarray}
The two possible intersection times for $\Delta \rho_{\mathrm{gg}}(t)=0$ in the form of Lambert W functions are given below:
\begin{eqnarray}
\tilde{t}_0&=&\frac{1}{(\lambda_4-\lambda_2)} W_0\left((\lambda_4-\lambda_2)\frac{\alpha_1}{\alpha_2} e^{(\lambda_4-\lambda_2)\frac{\alpha_3}{\alpha_2}}\right)-\frac{\alpha_3}{\alpha_2},\cr
\tilde{t}_{-1}&=&\frac{1}{(\lambda_4-\lambda_2)} W_{-1}\left((\lambda_4-\lambda_2)\frac{\alpha_1}{\alpha_2} e^{(\lambda_4-\lambda_2)\frac{\alpha_3}{\alpha_2}}\right)-\frac{\alpha_3}{\alpha_2}.\hspace*{ 0.6 cm}
\label{eq:d10}
\end{eqnarray}
The asymptotic expansions of the functions $W_0(x)$ and $W_{-1}(x)$ can be found in Ref.~\cite{Corless96}. Note that Lambert $W$ functions have found applications in several fields such as  viscous flows \cite{More06}, crystal growth \cite{Asadian14}, epitaxial film growth \cite{Braun03}, phase separation of polymer mixtures \cite{Bot21}, criticality in temporal networks \cite{Moran23} etc. 

In Fig.~\ref{fig:d1}, we plot the solutions $\tilde{t}_0$ and $\tilde{t}_{-1}$ [Eq.~(\ref{eq:d10})] as functions of the driving strengths before quench. Indeed, we observe that there are regions with single QMPE where only $\tilde{t}_0$ exists and there are other regions where both solutions prevail leading to double QMPE. In particular, we observe double QMPE for large values of $\tilde{d}_{\mathrm{I}}$ and $\tilde{d}_{\mathrm{II}}$. Note that QMPE appears to be unavoidable in the ground state probability $\rho_{\mathrm{gg}}(t)$ in Fig.~\ref{fig:d1} i.e. there is no such region in the $(\tilde{d}_{\mathrm{I}},\tilde{d}_{\mathrm{II}})$ where there is no QMPE.
\subsection{Energy}
Here we analyze QMPE in average energy $E(t)=\mathrm{Tr}[\widehat{\rho}(t)H]$ where $H$ is the system Hamiltonian presented as a $2\times2$ matrix with elements $H_{11}=1,H_{22}=0,H_{12}=H_{21}=\tilde{d}/2$ \cite{Hatano_2019}. Consequently, $E(t)$ can be expressed as
\begin{eqnarray}
E(t)=1-\rho_{4}(t)+\frac{\tilde{d}}{2}(\rho_1(t)+\rho_2(t)).
\label{eq:d11} 
\end{eqnarray}
\begin{figure}[t]
  \centering
  \subfigure[]{\includegraphics[width=\linewidth]{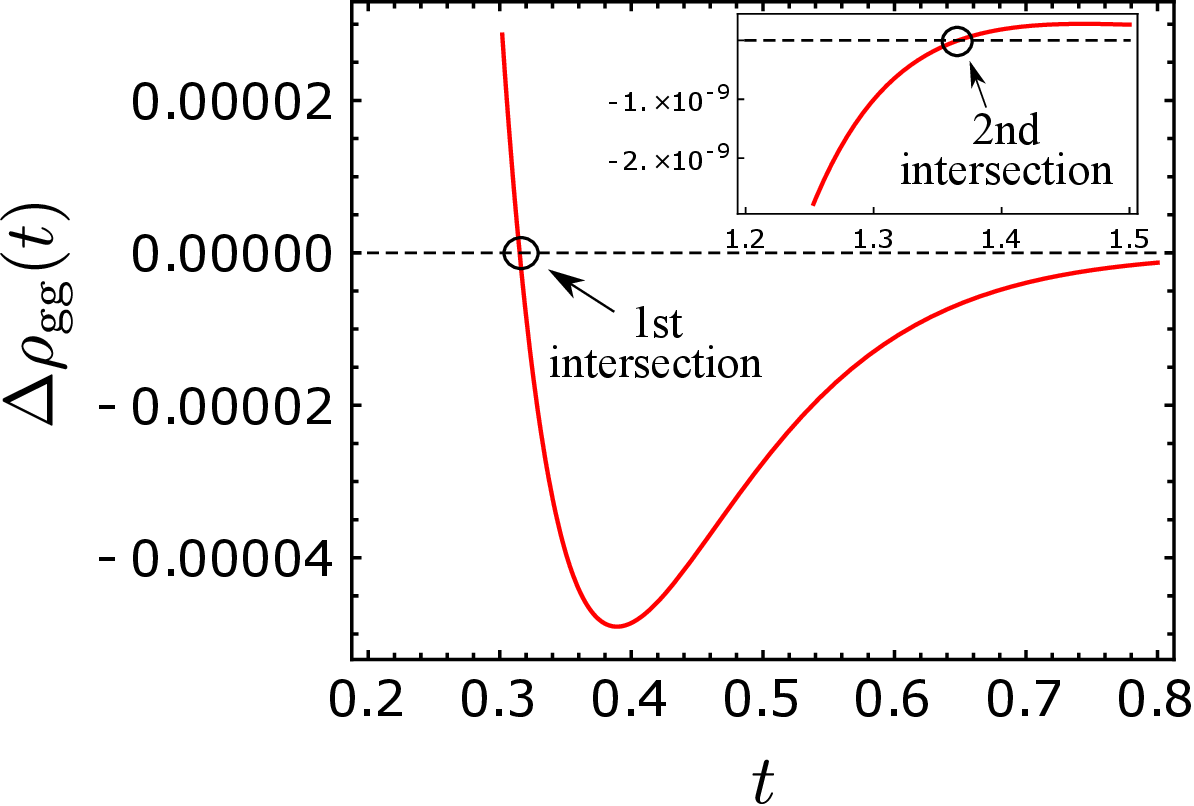}}\hfill
  \subfigure[]{\includegraphics[width=\linewidth]{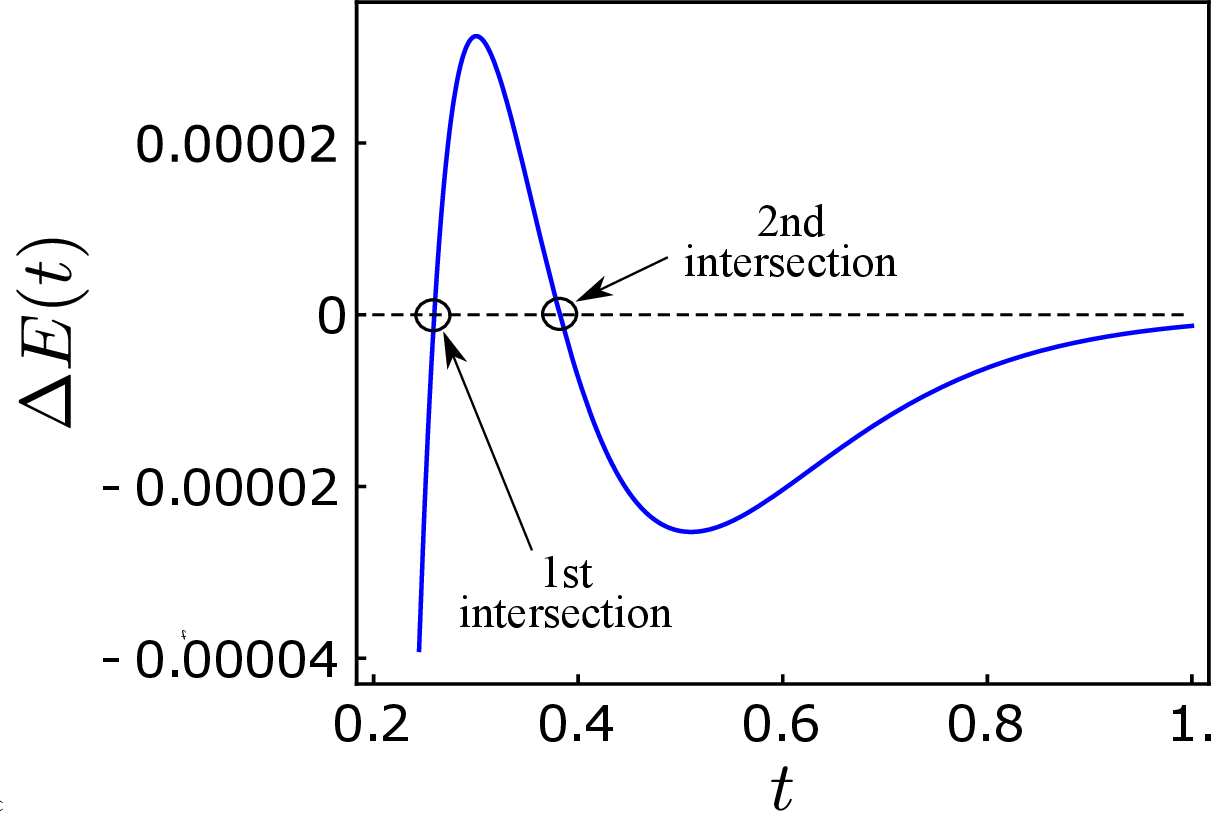}}
  \caption{The figure shows an example where both ground state probability [(a)] and energy [(b)] exhibit double QMPE. We have plotted the difference between two copies I and II in each case. For (a), the second QMPE in $\rho_{\mathrm{gg}}(t)$ is much weaker in magnitude whereas for (b) both QMPE in $E(t)$ are of comparative magnitudes Parameters used are $\tilde{d}=4.0, \tilde{\Gamma}_{\mathrm{I}}=\tilde{\Gamma}_{\mathrm{II}}=\tilde{\Gamma}=\sqrt{(568+64\sqrt{2})/2}, \tilde{d}_{\mathrm{I}}=10.0, \tilde{d}_{\mathrm{II}}=12.0$.}
\label{fig:d3}
\end{figure}
Note that the average energy directly captures the effect of coherence due to the presence of the off-diagonal density matrix elements in Eq.~(\ref{eq:d11}). To explore the QMPE in $E(t)$, we consider the time evolution of the energies starting from the two different initial conditions I and II and look for their intersections during the relaxations. More precisely, we would like to find the solutions to $\Delta E(t)=0$ where $\Delta E(t)=E_{\mathrm{I}}(t)-E_{\mathrm{II}}(t)$. We obtain the following expression
\begin{eqnarray}
\Delta E(t)=-e^{-\lambda_2 t}\left[\gamma_1 e^{-(\lambda_4-\lambda_2)t}+ t\,\gamma_2+\gamma_3\right],
\label{eq:d12} 
\end{eqnarray}
where the coefficients $\gamma_1,\gamma_2,\gamma_3$ are given by
\begin{eqnarray}
\gamma_1&=&\alpha_1\left(1+\frac{\tilde{d}^2}{1+(\Gamma/2-\lambda_4)^2}\right),\cr
\gamma_2&=&\alpha_2\left(1+\frac{\tilde{d}^2}{1+(\Gamma/2-\lambda_2)^2}\right),\cr
\gamma_3&=&\alpha_3\left(1+\frac{\tilde{d}^2}{1+(\Gamma/2-\lambda_2)^2}\right)-\alpha_2(\Gamma/2-\lambda_4),\,\,\,\,
\label{eq:d13} 
\end{eqnarray}
where the elements $\alpha_1,\alpha_2,\alpha_3$ are given by Eq.~(\ref{eq:d8}). Note that Eq.~(\ref{eq:d12}) is a similar kind of transcendental equation as Eq.~(\ref{eq:d7}), only with different coefficients. Therefore, we can analyze the possibility of QMPE in energy using Lambert $W$ functions as we did for the ground state probability. The two possible solutions to $\Delta E(t)=0$ are given by
\begin{eqnarray}
\tilde{t}^E_0&=&\frac{1}{(\lambda_4-\lambda_2)} W_0\left((\lambda_4-\lambda_2)\frac{\gamma_1}{\gamma_2} e^{(\lambda_4-\lambda_2)\frac{\gamma_3}{\gamma_2}}\right)-\frac{\gamma_3}{\gamma_2},\cr
\tilde{t}^E_{-1}&=&\frac{1}{(\lambda_4-\lambda_2)} W_{-1}\left((\lambda_4-\lambda_2)\frac{\gamma_1}{\gamma_2} e^{(\lambda_4-\lambda_2)\frac{\gamma_3}{\gamma_2}}\right)-\frac{\gamma_3}{\gamma_2},\nonumber \\
\label{eq:d14} 
\end{eqnarray}
where the superscript $E$ denotes energy and is used to differentiate the solutions from Eq.~(\ref{eq:d10}) obtained for ground state probability. 

Figure~\ref{fig:d2} exhibits the solutions in Eq.~(\ref{eq:d14}) as functions of the control parameters $(\tilde{d}_{\mathrm{I}},\tilde{d}_{\mathrm{II}})$. We observe region with single QMPE (only $\tilde{t}^E_0$ exists) as well as region for double QMPE (both $\tilde{t}^E_0$ and $\tilde{t}^E_{-1}$ exist). In addition, we observe regions where both the solutions are absent i.e. no QMPE. This crucial difference in Fig.~\ref{fig:d2} compared to Fig.~\ref{fig:d1} indicates that QMPE in ground state probability does not necessarily implies QMPE in average energy. However, Figs.~\ref{fig:d1} and \ref{fig:d2} also show that there are common parameter values that lead to QMPE both in $\rho_{\mathrm{gg}}(t)$ and $E(t)$. We present an explicit example in Fig.~\ref{fig:d3} where both of these quantities exhibit double QMPE, for the same set of parameters. However, the two intersection times for $\rho_{\mathrm{gg}}(t)$ and $E(t)$ are different from each other. The intersection times for $\rho_{\mathrm{gg}}(t)$ are comparatively far from each other with the second QMPE being much weaker in magnitude [Fig.~\ref{fig:d3}(a)], whereas the intersection times for $E(t)$ are comparatively much closer to each other, with both the first and second QMPE being comparable in magnitude [Fig.~\ref{fig:d3}(b)].       
\subsection{von Neumann entropy}
Next we consider the von Neumann entropy $S_{\mathrm{vN}}(t)$, which is a function of $\widehat{\rho}(t)$, defined as 
\begin{eqnarray}
S_{\mathrm{vN}}(t):=-\mathrm{Tr}[\widehat{\rho}(t)\,\mathrm{ln}(\widehat{\rho}(t))].
\label{eq:d15} 
\end{eqnarray}
\begin{figure}[t]
  \centering \includegraphics[width=8.6 cm]{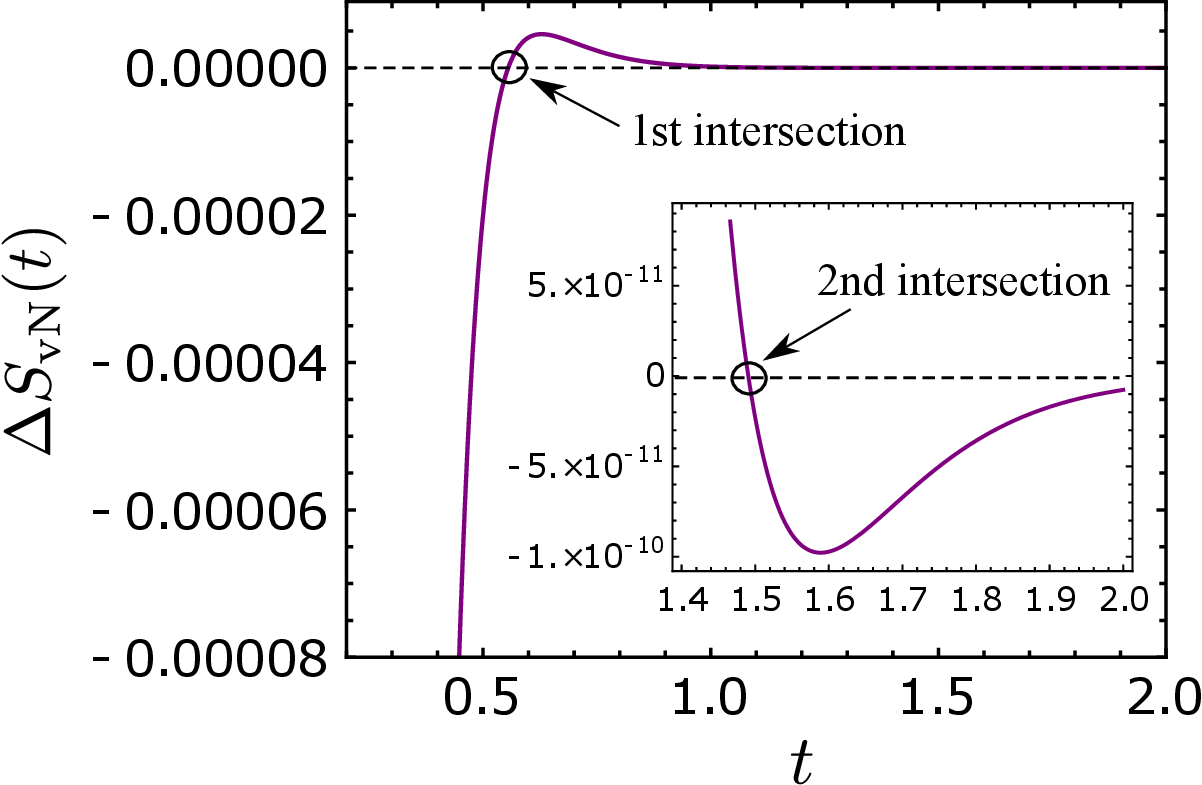}
\caption{The figure illustrates double QMPE in von Neumann entropy for the same set of parameters used in Fig.~\ref{fig:d3}. The inset shows the second intersection where the magnitude of the second QMPE is much weaker than the first QMPE shown in the main figure.}
\label{fig:d4}
\end{figure}
The nonlinearity of $S_{\mathrm{vN}}(t)$ as a function of the density matrix elements poses difficulties to perform analytical treatment of the QMPE in von Neumann entropy, in contrast to what we could achieve previously for ground state probability and energy. To observe QMPE in $S_{\mathrm{vN}}(t)$, we consider the intersection time(s) for the quantity $\Delta S_{\mathrm{vN}}(t)=S^{\mathrm{I}}_{\mathrm{vN}}-S^{\mathrm{II}}_{\mathrm{vN}}$. In Fig.~\ref{fig:d4}, we present the behavior of $\Delta S_{\mathrm{vN}}(t)$ for the same set of parameter values used in Fig.~\ref{fig:d3}. Indeed, we observe that $\Delta S_{\mathrm{vN}}(t)$ intersects zero twice, implying the existence of double QMPE in the von Neumann entropy, similar to $\rho_{\mathrm{gg}}(t)$ and $E(t)$. 
\subsection{Temperature}
\begin{figure}[t]
  \centering \includegraphics[width=8.6 cm]{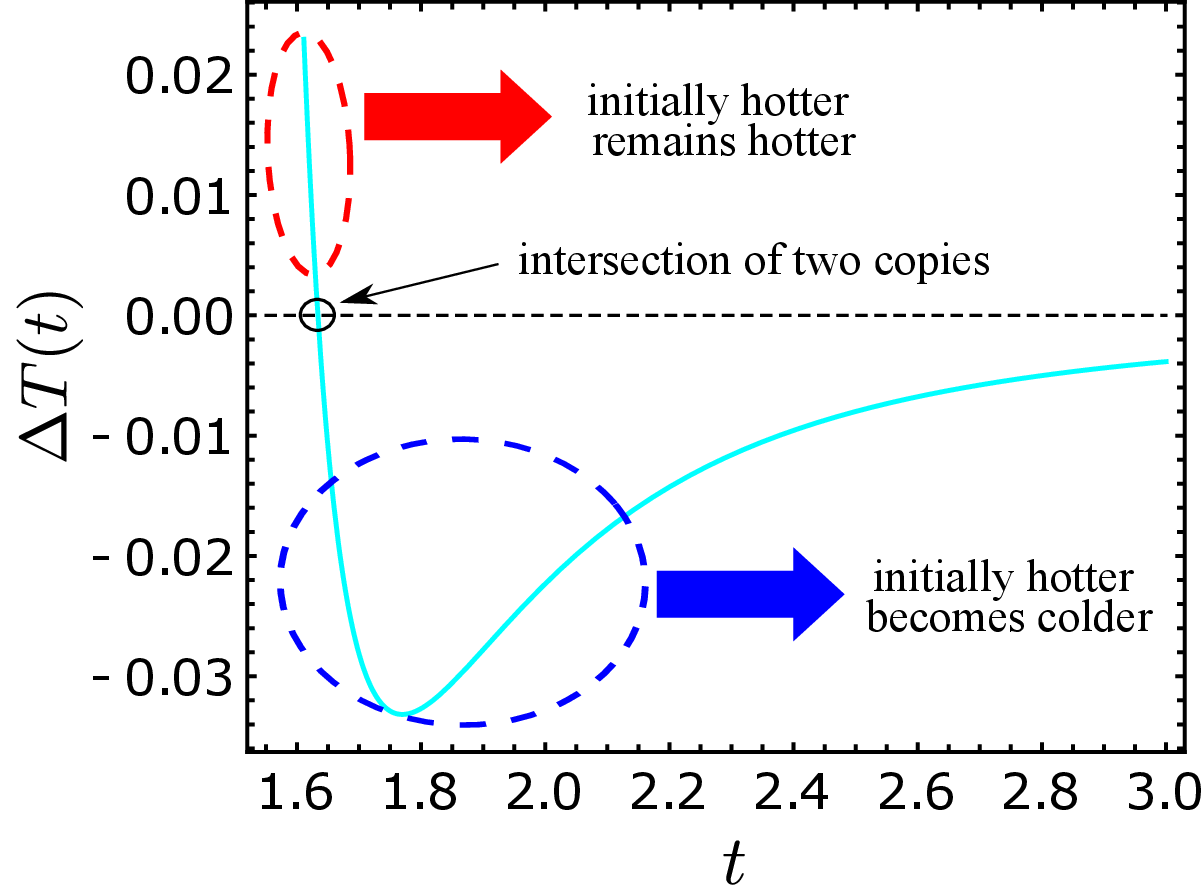}
\caption{The figure shows thermal QMPE. We have plotted the difference in temperatures between two copies I and II. After the two copies intersect each other, the initially hotter copy becomes colder leading to QMPE in temperature. Parameters used are $\tilde{d}=4.0, \tilde{\Gamma}=\sqrt{(568+64\sqrt{2})/2},  \tilde{\Gamma}_{\mathrm{I}}=15.0, \tilde{\Gamma}_{\mathrm{II}}=2.1, \tilde{d}_{\mathrm{I}}=3.0,\tilde{d}_{\mathrm{II}}=22.9$.}
\label{fig:d5}
\end{figure}
We use the definition of temperature $T(t)$ following Ref.~\cite{Chatterjee_2023} to examine the possibility of thermal QMPE in Hatano's model, as given below
\begin{eqnarray}
T(t):=\frac{\partial E(t)}{\partial S_{\mathrm{vN}}(t)}=\left.\frac{\partial  E(t)}{\partial t}\right/\frac{\partial S_{\mathrm{vN}}(t)}{\partial t},
\label{eq:d16} 
\end{eqnarray}
where $E(t)$ and $S_{\mathrm{vN}}(t)$ are defined in Eqs.~(\ref{eq:d11}) and (\ref{eq:d15}), respectively. The QMPE in system temperature has been observed recently in Ref.~\cite{Chatterjee_2023} using a semi-classical approach. Naturally, it would be interesting to explore thermal QMPE in Hatano's model that involves the effect of off-diagonal density matrix elements. However, we find that there is no thermal QMPE for the same set of parameters that have produced QMPE for ground state probability [Fig.~\ref{fig:d3}(a)], energy [Fig.~\ref{fig:d3}(b)] and entropy [Fig.~\ref{fig:d4}]. To search for thermal QMPE, we consider a variable dissipation protocol that allows us to use $\tilde{\Gamma}_{\mathrm{I}}\neq\tilde{\Gamma}_{\mathrm{II}}\neq\tilde{\Gamma}$ (wider parameter space compared to the fixed dissipation protocol $\tilde{\Gamma}_{\mathrm{I}}=\tilde{\Gamma}_{\mathrm{II}}=\tilde{\Gamma}$). Indeed, using such protocol, we have obtained the thermal QMPE shown in Fig.~\ref{fig:d5} where we present the temperature difference $\Delta T(t)$ between two copies I and II. We observe that $\Delta T(t)$ becomes zero at some intermediate relaxation time that denotes the intersection of two copies, after which the initially hotter copy becomes colder, leading to thermal QMPE.  
\subsection{Kullback-Leibler divergence}
The Kullback-Leibler (KL) divergence is a measure of difference from steady state which is a monotonically decreasing function of time, and it is given by
\begin{eqnarray}
D_{\mathrm{KL}}(t):=\mathrm{Tr}[\widehat{\rho}(t)(\mathrm{ln}(\widehat{\rho}(t))-\mathrm{ln}(\widehat{\rho}_{\mathrm{ss}}))],
\label{eq:d17} 
\end{eqnarray}
where $\widehat{\rho}_{\mathrm{ss}}$ is the steady state density matrix. 
\begin{figure}[t]
\centering
  \subfigure[]{\includegraphics[width=\linewidth]{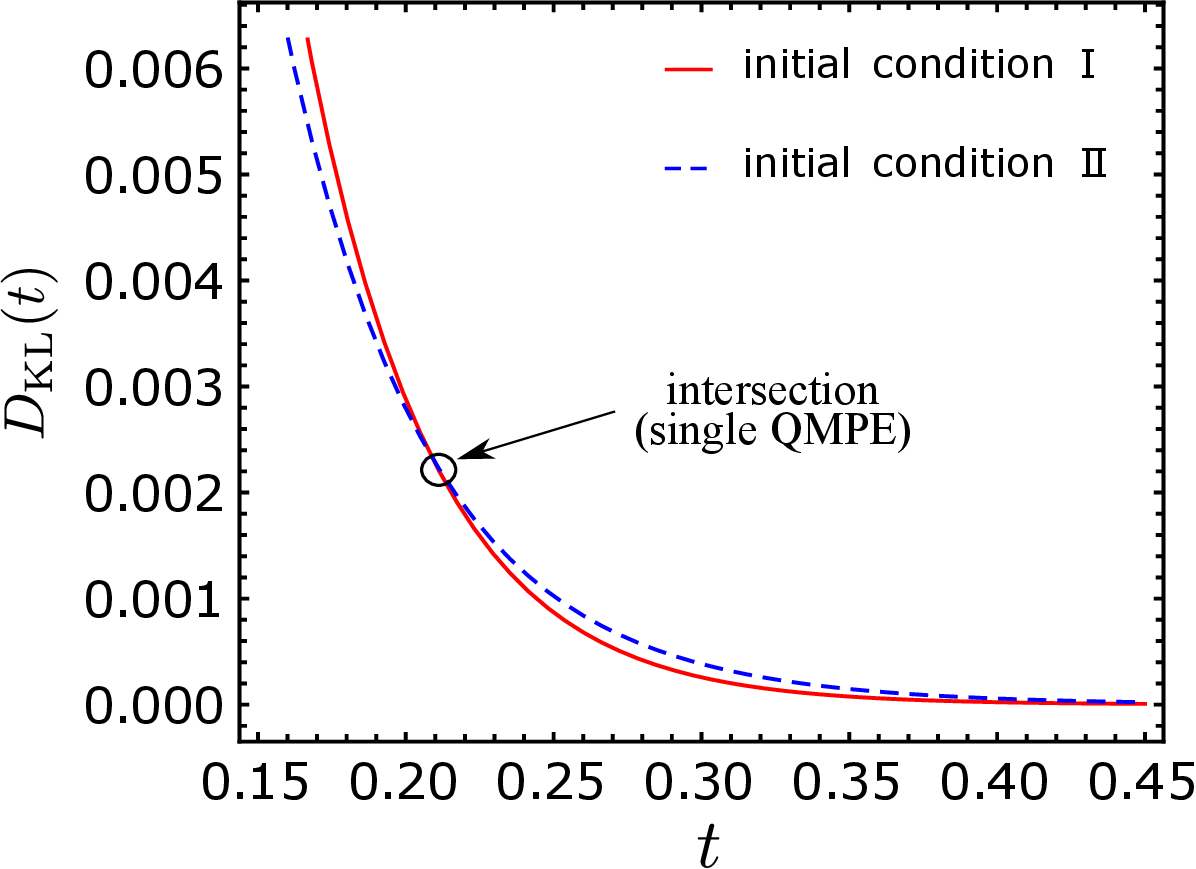}}\hfill
  \subfigure[]{\includegraphics[width=\linewidth]{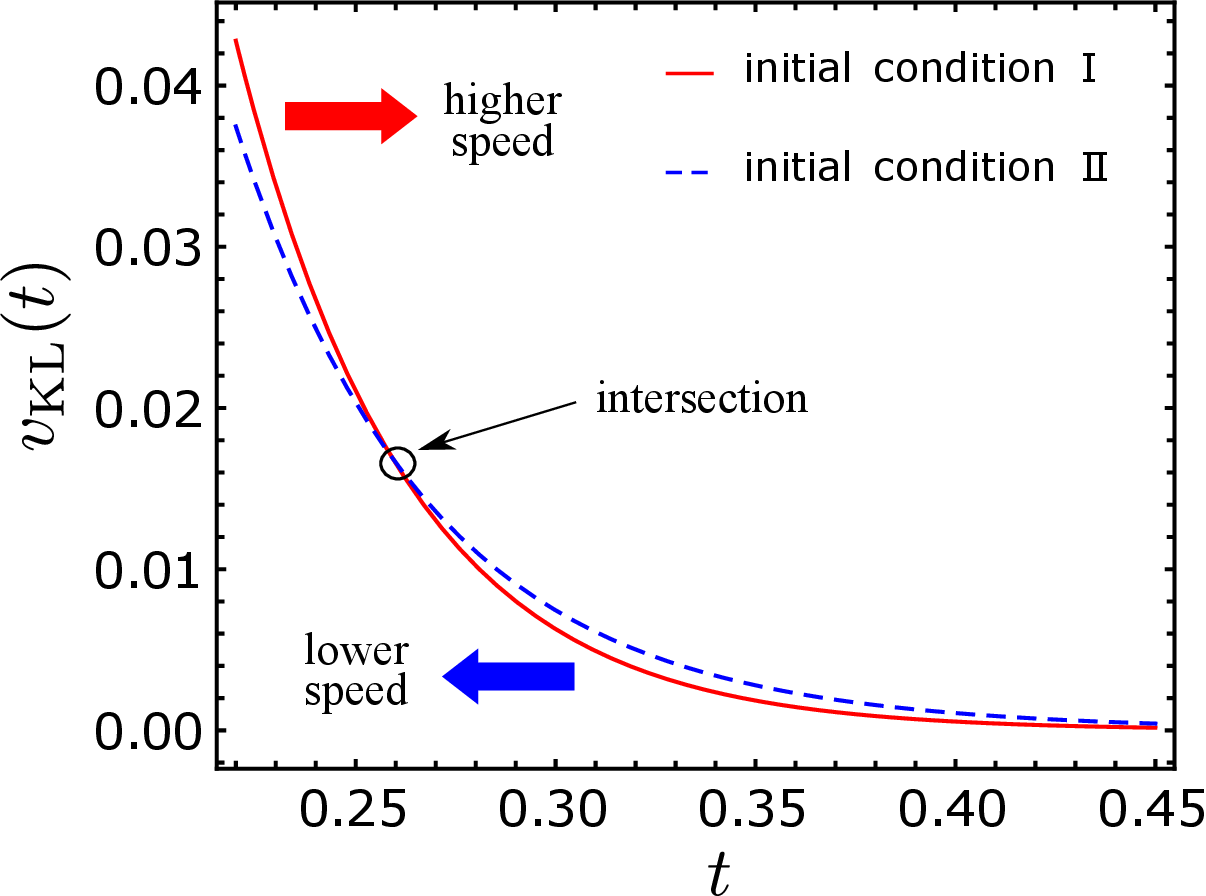}}
  \caption{The figure (a) shows QMPE in KL divergence [Eq.~(\ref{eq:d17})] where the initially higher distant copy (here I) reaches to steady state faster than the initially lower distance copy (here II). The figure (b) demonstrates the relaxation speed [Eq.~(\ref{eq:d18})] of these two copies. The speed also exhibit QMPE, but the intersection time is different compared to $D_{\mathrm{KL}}(t)$. Parameters used for both figures are $\tilde{d}=4.0, \tilde{\Gamma}=\sqrt{(568+64\sqrt{2})/2},  \tilde{\Gamma}_{\mathrm{I}}=20.0, \tilde{\Gamma}_{\mathrm{II}}=4.5, \tilde{d}_{\mathrm{I}}=10.0,\tilde{d}_{\mathrm{II}}=2.0$.}
\label{fig:d6}
\end{figure}
Although the KL divergence is not a distance in strict metric sense, however it is well-accepted as a statistical distance. Such distance function has been widely used to investigate MPE in classical systems \cite{Lu17,Kumar20,Klich19} since it is suitable to characterize the approach to steady state because of its monotonic nature. The KL divergence \cite{Kullback51} plays fundamental roles in non-equilibrium thermodynamics \cite{Sagawa22,Shiraishi23,Strasberg22,Yoshii23} and has diverse applications in various fields including information geometry \cite{Amari20}, machine learning \cite{Burnham02}, game theory \cite{Soklakov20} etc.

To obtain QMPE in KL divergence, we use the variable dissipation protocol. In Fig.~\ref{fig:d6}(a) we present an example where the copy starting at a higher distance from steady state relaxes faster compared to the other copy with lower initial distance. The copies I and II intersect each other leading to QMPE in $D_{\mathrm{KL}}(t)$ and the copy starting at higher initial distance reaches the steady state faster. We have not observed multiple QMPE in KL divergence. It would be an interesting future question to understand if the absence of multiple QMPE in $D_{\mathrm{KL}}(t)$ is its generic feature and if the reason is its monotonicity. 

An important variable of interest in any relaxation process, be it classical or quantum, is the speed of the relaxation. Since the KL divergence gives the distance measure from steady state, a relevant relaxation speed can be defined as
\begin{eqnarray}
v_{\mathrm{KL}}(t)=-\frac{\partial D_{\mathrm{KL}}(t)}{\partial t},
\label{eq:d18}
\end{eqnarray}
where the negative sign is used to complement the fact that the KL divergence decreases with time. It would be interesting to see if the QMPE in $D_{\mathrm{KL}}(t)$ is also manifested in the speed $v_{\mathrm{KL}}(t)$. For the same set of parameters that produced QMPE in KL divergence [Fig.~\ref{fig:d6}(a)], we investigate QMPE in the speed in Fig.~\ref{fig:d6}(b). Indeed, we observe that the speeds of the two copies I and II intersect each other, leading to QMPE. The higher distant copy goes towards steady state with a higher speed and after the intersection its speed becomes lower. Notably, the intersection times for $D_{\mathrm{KL}}(t)$ and $v_{\mathrm{KL}}(t)$ are different. In fact, there is a time window  where the initially higher distant copy becomes lower distant copy but still maintains its higher speed for some time. It would be intriguing to investigate whether there is some general upper bound on the relaxation speed $v_{\mathrm{KL}}(t)$. 

We end this section by mentioning that the analysis of the QMPE in region (c) would follow similar manner as in region (d) discussed elaborately in the present section. This is because region (c) is also a line of second-order EP just like region (d), the only difference is that the arrangement of eigenvalues for region (c) obeys $0<\lambda_2<\lambda_3=\lambda_4$.

\section{QMPE in region ($\mathrm{a_{1}}$): oscillations}
\label{sec5}
In this section, we explore QMPE in region ($\mathrm{a_{1}}$) of Hatano's model which contains complex eigenvalues with both non-zero real and imaginary parts. The eigenvalues are non-degenerate and they are denoted by $0,\lambda_2,\lambda_3=\lambda_{\mathrm{re}}+\mathrm{i}\lambda_{\mathrm{im}},\lambda_4=\lambda_{\mathrm{re}}-\mathrm{i}\lambda_{\mathrm{im}}$, where the subscripts $\mathrm{re}$ and $\mathrm{im}$ correspond to real and imaginary parts, respectively. In addition, the eigenvalues satisfy the relation $0<\lambda_2<\lambda_{\mathrm{re}}$ meaning $\lambda_2$ is the slowest relaxation mode. 
\begin{figure}[t]
  \centering \includegraphics[width=8.6 cm]{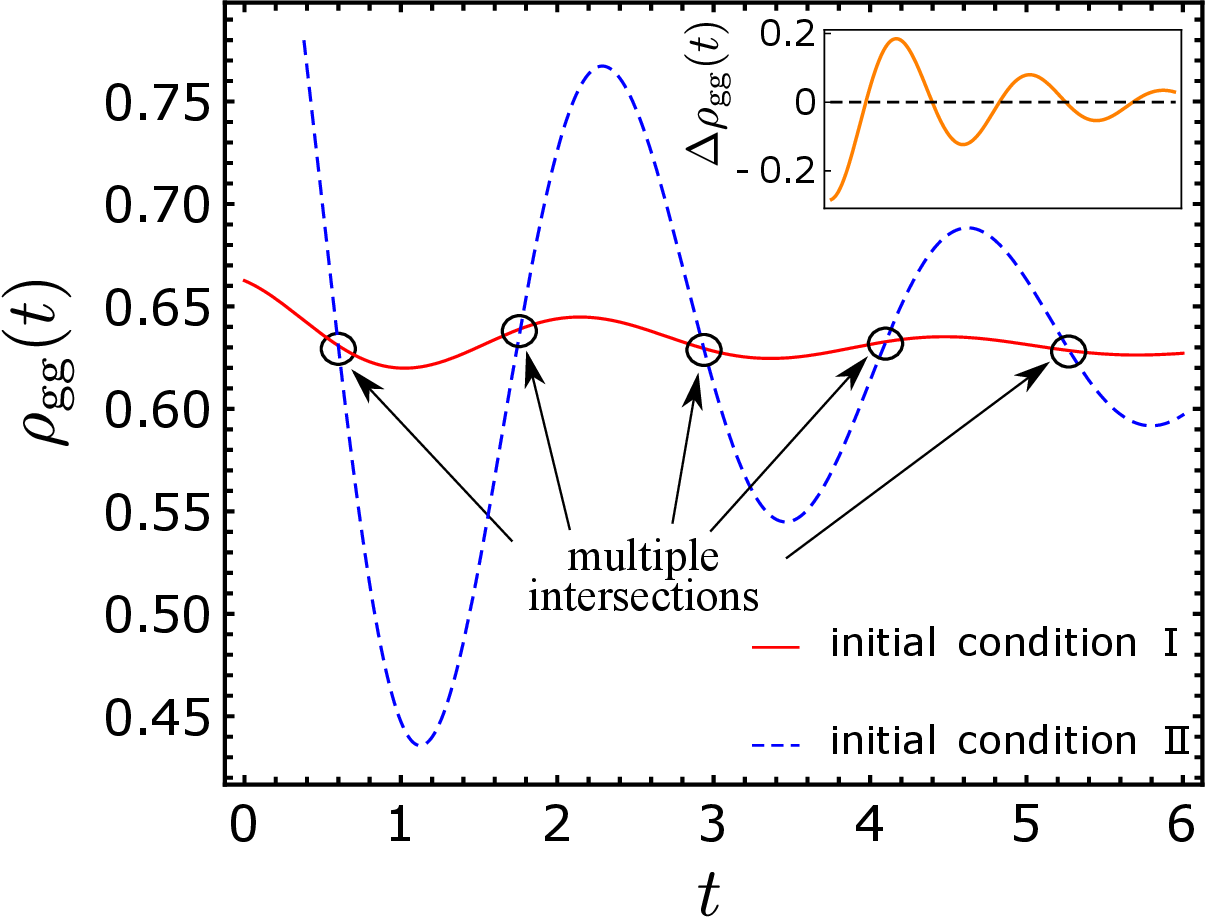}
\caption{The figure shows multiple QMPE in the ground state density matrix element, where both copies I and II exhibit oscillatory relaxation and intersect each other multiple times. The inset shows the difference between the two copies intersect zero multiple times, confirming the presence of multiple QMPE. Parameters used are $\tilde{d}=2.5, \tilde{\Gamma}_{\mathrm{I}}=\tilde{\Gamma}_{\mathrm{II}}=\tilde{\Gamma}=0.5, \tilde{d}_{\mathrm{I}}=2.1, \tilde{d}_{\mathrm{II}}=0.51$.}
\label{fig:a11}
\end{figure}
In contrast to the nontrivial approximate diagonalization [Eqs.~(\ref{eq:d3})-(\ref{eq:d4})] for second-order EP in Sec.~\ref{sec4}, here the complete diagonalization can be obtained through the usual method
\begin{eqnarray}
\widehat{\mathcal{L}}_d=\widehat{L}\widehat{\mathcal{L}}\widehat{R},
\label{eq:a11} 
\end{eqnarray}
where $\widehat{\mathcal{L}}_d$ is the diagonal form of the Lindbladian $\widehat{\mathcal{L}}$ [Eq.~(\ref{eq:super-operator})] in region ($\mathrm{a_1}$). The explicit forms of the eigenvalues, right eigenvectors $|r_k\rangle$ and left eigenvectors $\langle \ell_k|$  [($k=2,3,4$), which appear in Eq.~(\ref{eq:d2})], are given in Appendix~\ref{appen2}.

\subsection{Density matrix elements}
The the individual density matrix elements $\rho_j(t)$ ($j=1,2,3,4$) are given by
\begin{eqnarray}
\rho_j(t)=\sum_{k=1}^{4} e^{-\lambda_k t} r_{k,j} a_k, \hspace*{0.2 cm} a_k=\sum_{n=1}^{4} \ell_{k,n}\rho_n(0).
\label{eq:a12} 
\end{eqnarray}
To discuss QMPE in the density matrix, let us focus on the ground state probability i.e. $\rho_{\mathrm{gg}}(t)$. Of course, there is no algebraic time dependence in the relaxation observed in Sec.~\ref{sec4}. Rather, the interesting part of the relaxation in region $(\mathrm{a_1})$ originates from the complex nature of the eigenvalues and the corresponding coefficients $a_k$ in Eq.~(\ref{eq:a12}). Specifically, we have found that $a_3$ and $a_4$ are complex conjugates of each other, they take the forms $a_3=a_{\mathrm{re}}+\mathrm{i}a_{\mathrm{im}}$ and $a_4=a_{\mathrm{re}}-\mathrm{i}a_{\mathrm{im}}$. Using such expressions, we obtain the following formula for $\rho_{\mathrm{gg}}(t)$:
\begin{eqnarray}
\rho_{\mathrm{gg}}(t)=&&\frac{4+\tilde{d}^2+\tilde{\Gamma}^2}{4+2\tilde{d}^2+\tilde{\Gamma}^2}+ e^{-\lambda_2 t}a_2\cr &&+2e^{-\lambda_{\mathrm{re}}t}\sqrt{a^2_{\mathrm{re}}+a^2_{\mathrm{im}}}\,\mathrm{cos}(\lambda_{\mathrm{im}}t-\phi),
\label{eq:a13} 
\end{eqnarray}
where $\phi=\mathrm{tan}^{-1}(a_{\mathrm{im}}/a_{\mathrm{re}})$.  
Interestingly, the Eq.~(\ref{eq:a13}) contains an oscillatory term $\mathrm{cos}(\lambda_{\mathrm{im}}t-\phi)$ and such an  oscillatory relaxation differs from the usual exponential relaxation studied in classical MPE \cite{Kumar20,Klich19}. To investigate the effect of such oscillations in the QMPE, we consider relaxations from two different initial conditions I and II, and search for one (single QMPE) or more (multiple QMPE) intersection between the time evolved  trajectories. The detection of such intersections is made from the solutions of the equation $\Delta\rho_{\mathrm{gg}}(t)=\rho^{\mathrm{I}}_{\mathrm{gg}}(t)-\rho^{\mathrm{II}}_{\mathrm{gg}}(t)=0$. In region ($\mathrm{a_1}$), the difference $\Delta\rho_{\mathrm{gg}}(t)$ has the form
\begin{eqnarray}
\Delta&&\rho_{\mathrm{gg}}(t)=2 e^{-\lambda_2 t}\sqrt{(\Delta a_{\mathrm{re}})^2+(\Delta a_{\mathrm{im}})^2}\times\cr &&\left[\frac{\Delta a_2}{2\sqrt{(\Delta a_{\mathrm{re}})^2+(\Delta a_{\mathrm{im}})^2}}+e^{-(\lambda_{\mathrm{re}}-\lambda_2)t}\mathrm{cos}(\lambda_{\mathrm{im}}t-\theta)\right],\nonumber \\
\label{eq:a15} 
\end{eqnarray}
where $\theta=\mathrm{tan}^{-1}(\Delta a_{\mathrm{im}}/\Delta a_{\mathrm{re}})$ with $\Delta a_{\mathrm{re}}=a^{\mathrm{I}}_{\mathrm{re}}-a^{\mathrm{II}}_{\mathrm{re}}$ and $\Delta a_{\mathrm{im}}=a^{\mathrm{I}}_{\mathrm{im}}-a^{\mathrm{II}}_{\mathrm{im}}$. Due to the presence of the oscillatory cos(.) function in Eq.~(\ref{eq:a15}), it is possible that $\Delta\rho_{\mathrm{gg}}(t)$ can become zero at several points of time leading to multiple QMPE. A necessary and sufficient condition for QMPE in region ($\mathrm{a_{1}}$) is
\begin{eqnarray}
\mathrm{min}[e^{-(\lambda_{\mathrm{re}}-\lambda_2)t}&&\mathrm{cos}(\lambda_{\mathrm{im}}t-\theta)]<\frac{\Delta a_2}{2\sqrt{(\Delta a_{\mathrm{re}})^2+(\Delta a_{\mathrm{im}})^2}}\cr &&<\mathrm{max}[e^{-(\lambda_{\mathrm{re}}-\lambda_2)t}\mathrm{cos}(\lambda_{\mathrm{im}}t-\theta)],
\label{eq:a16} 
\end{eqnarray}
where $\Delta a_2$ and $\mathrm{cos}(\lambda_{\mathrm{im}}t-\theta)$ must have opposite signs. Here $\mathrm{min}[f]$ and $\mathrm{max}[f]$ correspond to minimum and maximum values of the function $f$. Note that $-1<\mathrm{min}[e^{-(\lambda_{\mathrm{re}}-\lambda_2)t}\mathrm{cos}(\lambda_{\mathrm{im}}t-\theta)]$ and $\mathrm{max}[e^{-(\lambda_{\mathrm{re}}-\lambda_2)t}\mathrm{cos}(\lambda_{\mathrm{im}}t-\theta)]<1$. 

In Fig.~\ref{fig:a11} we present an example of multiple QMPE in the ground state density matrix element $\rho_{\mathrm{gg}}(t)$, where we have chosen parameters that satisfy the QMPE criteria Eq.~(\ref{eq:a16}) for region ($\mathrm{a_1}$). We observe that the two copies I and II relax in oscillatory manner and intersect each other multiple times. The inset explicitly shows that the difference between the two copies $\Delta\rho_{\mathrm{gg}}(t)$ intersects zero multiple times, confirming the multiple QMPE. Comparing Fig.~\ref{fig:d3}(a) [region (d)] with the inset of Fig.~\ref{fig:a11} [region ($\mathrm{a_1}$)], we understand that  the number of intersections in region ($\mathrm{a_1}$) is not restricted to {\it two} as in region (d). Therefore, in region ($\mathrm{a_1}$) one can have many more intersections and higher order QMPE compared to the maximum double QMPE in the region (d). Since the amplitudes of the oscillations can be controlled by tuning the initial driving strengths $\tilde{d}_{\mathrm{I}}$ and $\tilde{d}_{\mathrm{II}}$, complex eigenvalues seem to be useful to obtain multiple intersections and thereby multiple QMPE. It would be important in the future to find the exact number of solutions to $\Delta\rho_{\mathrm{gg}}(t)=0$ either analytically or numerically. 

\subsection{Energy}
\begin{figure}[t]
  \centering \includegraphics[width=8.6 cm]{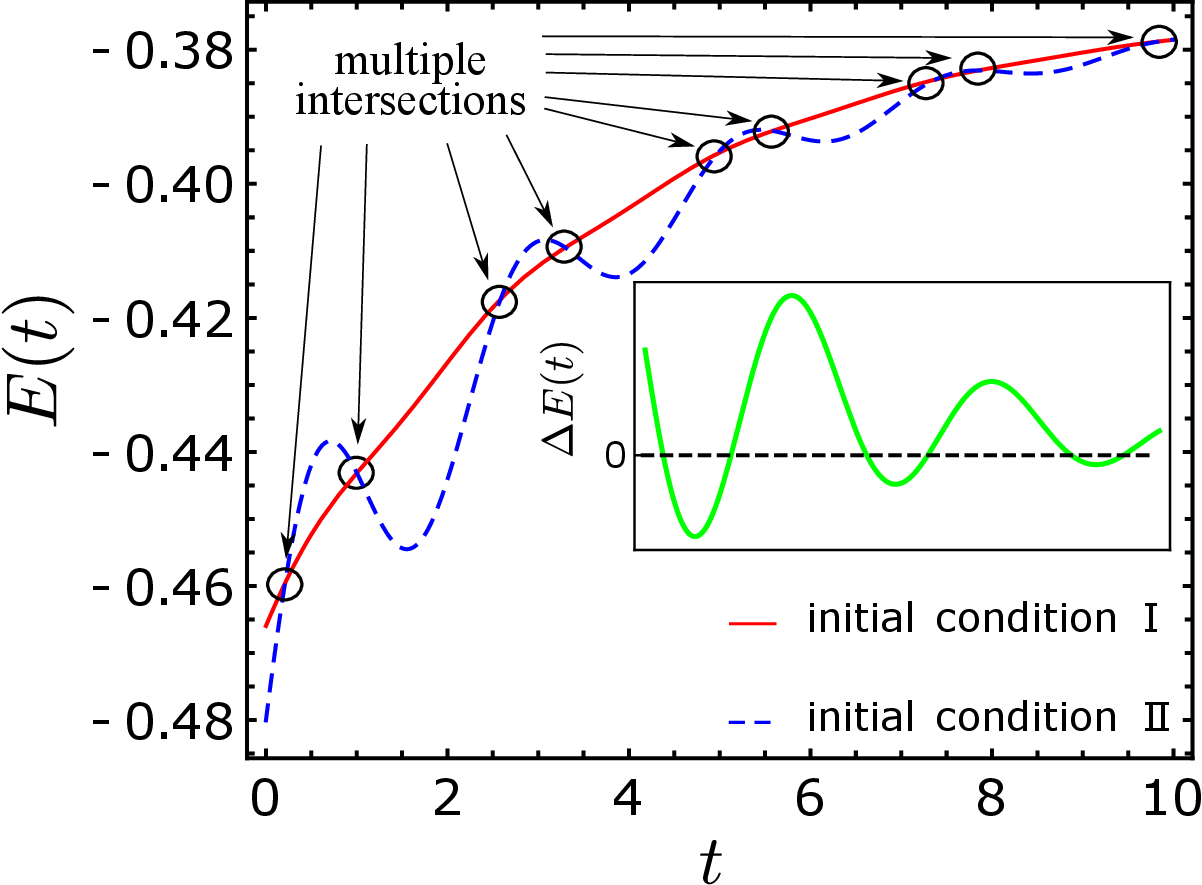}
\caption{The figure illustrates multiple QMPE in average energy where two copies I and II relaxes with oscillations and intersect each other multiple times showing multiple QMPE. The inset shows the energy difference between the two copies intersect zero multiple times, leading to multiple QMPE in energy. Parameters used are same as in Fig.~\ref{fig:a11}.}
\label{fig:a12}
\end{figure}

The formal expression for average energy $E(t)$ is given by Eq.~(\ref{eq:d11}). In the previous Sec.~\ref{sec4}, we have already shown that QMPE in ground state probability does not necessarily imply QMPE in energy, in general. However, it is possible to find control parameters that can give rise to QMPE in both observables, although the intersection time(s) are generally different. Following the same path, we would like to see if $E(t)$ shows multiple QMPE for the same set of parameters used to achieve multiple QMPE for $\rho_{\mathrm{gg}}(t)$ in Fig.~\ref{fig:a11}. Indeed, Fig.~\ref{fig:a12} exhibits multiple intersections between time evolved copies of energies starting from I and II, both relaxing towards the same final energy. The inset of Fig.~\ref{fig:a12} alternatively demonstrates the multiple QMPE through multiple intersections of the energy difference $\Delta E(t)=E_{\mathrm{I}}(t)-E_{\mathrm{II}}(t)$ with zero, remarkably different from Fig.~\ref{fig:d3}(b) with a  maximum of two such intersections in region (d).  

\subsection{von Neumann entropy}
\begin{figure}[t]
  \centering \includegraphics[width=8.6 cm]{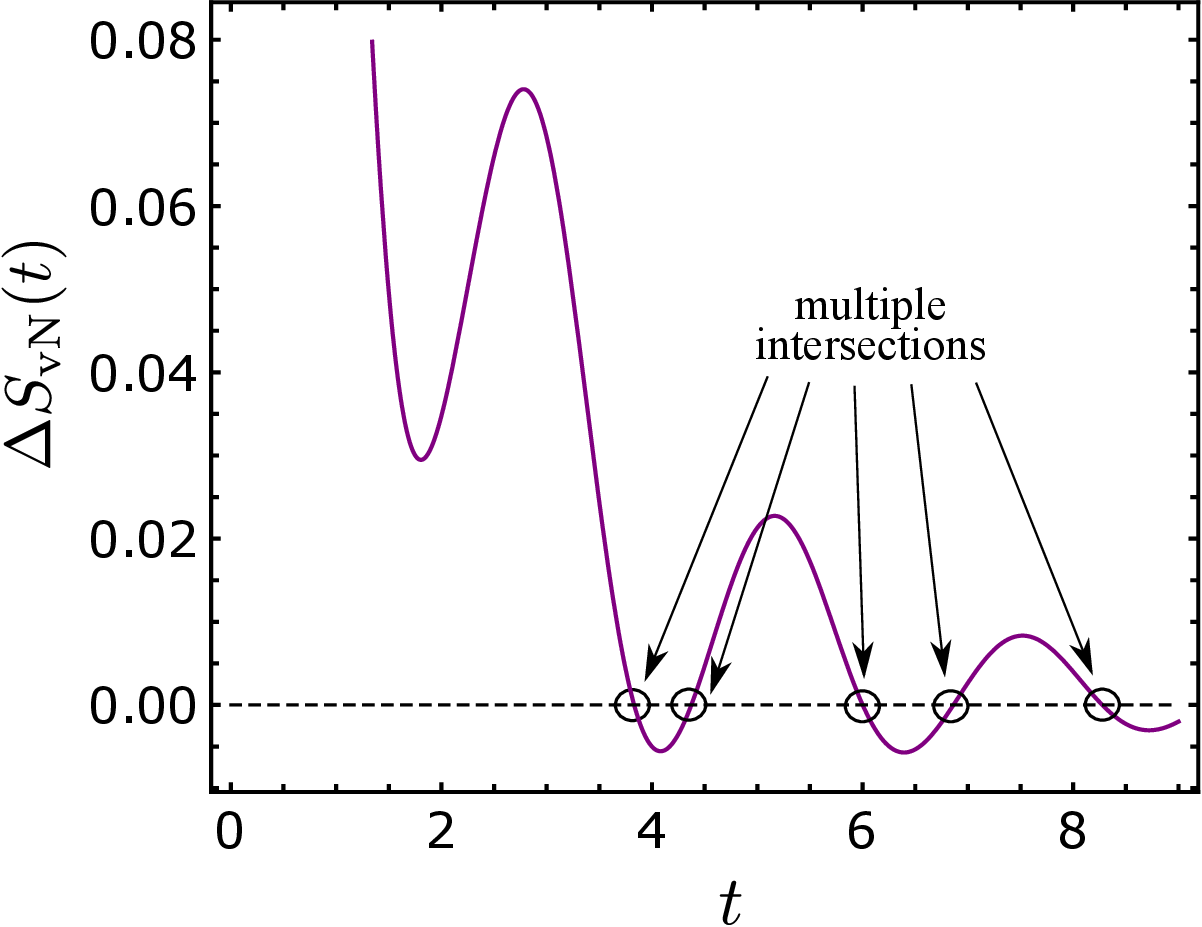}
\caption{The figure shows the temporal behavior of the difference between von Neumann entropies of two copies I and II, in region ($\mathrm{a_1}$). The entity $\Delta S_{\mathrm{vN}}(t)$ intersect zero multiple times indicating the occurrence of multiple QMPE. Parameters used are same as in Figs.~\ref{fig:a11}.}
\label{fig:a13}
\end{figure}

Next we consider the von Neumann entropy $S_{\mathrm{vN}}(t)$ [Eq.~(\ref{eq:d15})]. To observe QMPE in $S_{\mathrm{vN}}(t)$, we consider the intersection times for the entropy difference between two copies $\Delta S_{\mathrm{vN}}(t)=S^{\mathrm{I}}_{\mathrm{vN}}-S^{\mathrm{II}}_{\mathrm{vN}}$. In Fig.~\ref{fig:a13}, we present the behavior of $\Delta S_{\mathrm{vN}}(t)$ for the same set of parameter values used in Figs.~\ref{fig:a11} and \ref{fig:a12}. Indeed, we observe that $\Delta S_{\mathrm{vN}}(t)$ intersects zero multiple times showing the existence of multiple QMPE in the von Neumann entropy. 

\subsection{Temperature}
\begin{figure}[t]
  \centering \includegraphics[width=8.6 cm]{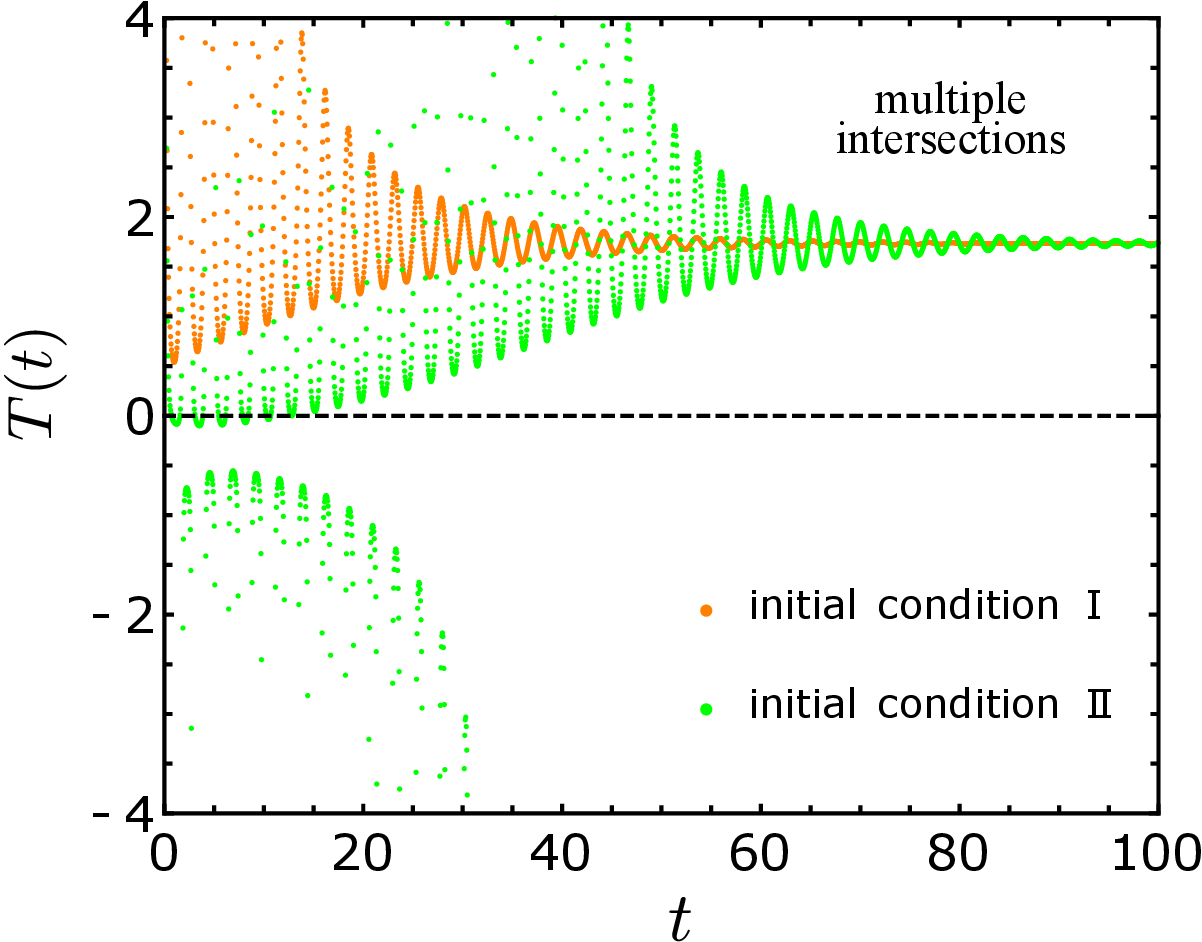}
\caption{The figure shows multiple thermal QMPE. Two copies, one initially hotter than the other, relax in oscillatory manner and intersect each other multiple times. After each intersection, the hotter and colder copies reverse their identities and these identities are preserved only until the next intersection. For initial condition II, we observe temperatures oscillating between positive and negative values in the early relaxation stage, leading to negative temperatures at certain times. Parameters used are same as in Fig.~\ref{fig:a11}.}
\label{fig:a14}
\end{figure}

Here, we study the thermal QMPE in region ($\mathrm{a_1}$) using the system temperature defined in Eq.~(\ref{eq:d16}). 
In Fig.~\ref{fig:a14}, we present the time evolution of the temperatures of two copies I and II, one initially hotter than the other. It is fascinating that the temperatures of both the copies relax in oscillatory manner towards the same steady state temperature and during the relaxations, they intersect each other multiple times, leading to multiple thermal QMPE. This, in turn, implies that the hotter and colder copies reverse their identities multiple times (after each intersection the previous hotter copy turns into colder copy and remains so until the next intersection). Such oscillatory multiple thermal QMPE is observed here for the first time in quantum systems (as far as our knowledge is concerned), although multiple thermal MPE in classical inertial suspensions has already been reported \cite{Takada21a}. The physical interpretation and utility of multiple thermal QMPE have to be explored in future in more details. Note that we have used the same  parameter set for multiple thermal QMPE as in Figs.~\ref{fig:a11}-\ref{fig:a13}. Similar to Ref.~\cite{Chatterjee_2023}, here also we observe negative temperature \cite{Landau80,Gauthier19,Johnstone19,Medley11,Ali20} in the  early stage of relaxation. To elaborate, for the parameters' values used for initial condition II in Fig.~\ref{fig:a14}, we observe temperatures oscillating between positive and negative values during initial relaxation stage. Note that these closely spaced oscillating positive and negative temperatures correspond to the same initial condition II. 
The occurrence of such negative temperature is due to non-monotonic natures of energy and entropy as functions of time. However, it is satisfying that even if we neglect such initial negative temperatures, we observe multiple QMPE at later times of the relaxation process where the temperatures of both the copies  remain positive.    

\subsection{Kullback-Leibler divergence}
In this subsection we explore the QMPE in KL divergence defined in Eq.~(\ref{eq:d17}). Since we have been able to obtain multiple QMPE in region ($\mathrm{a_1}$) in all other observables considered here, it is natural to ask about the possibility of multiple QMPE in $D_{\mathrm{KL}}(t)$ as well.  
\begin{figure}[t]
  \centering
  \subfigure[]{\includegraphics[width=\linewidth]{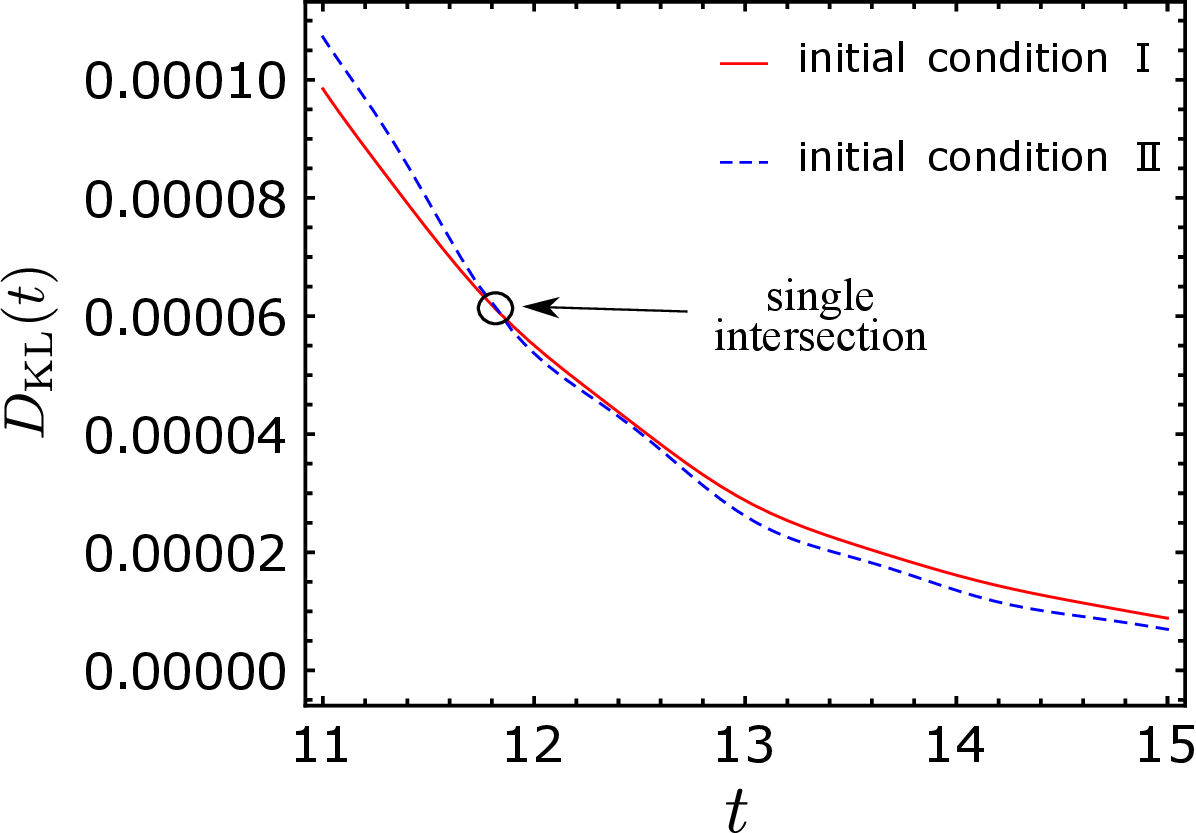}}\hfill
  \subfigure[]{\includegraphics[width=\linewidth]{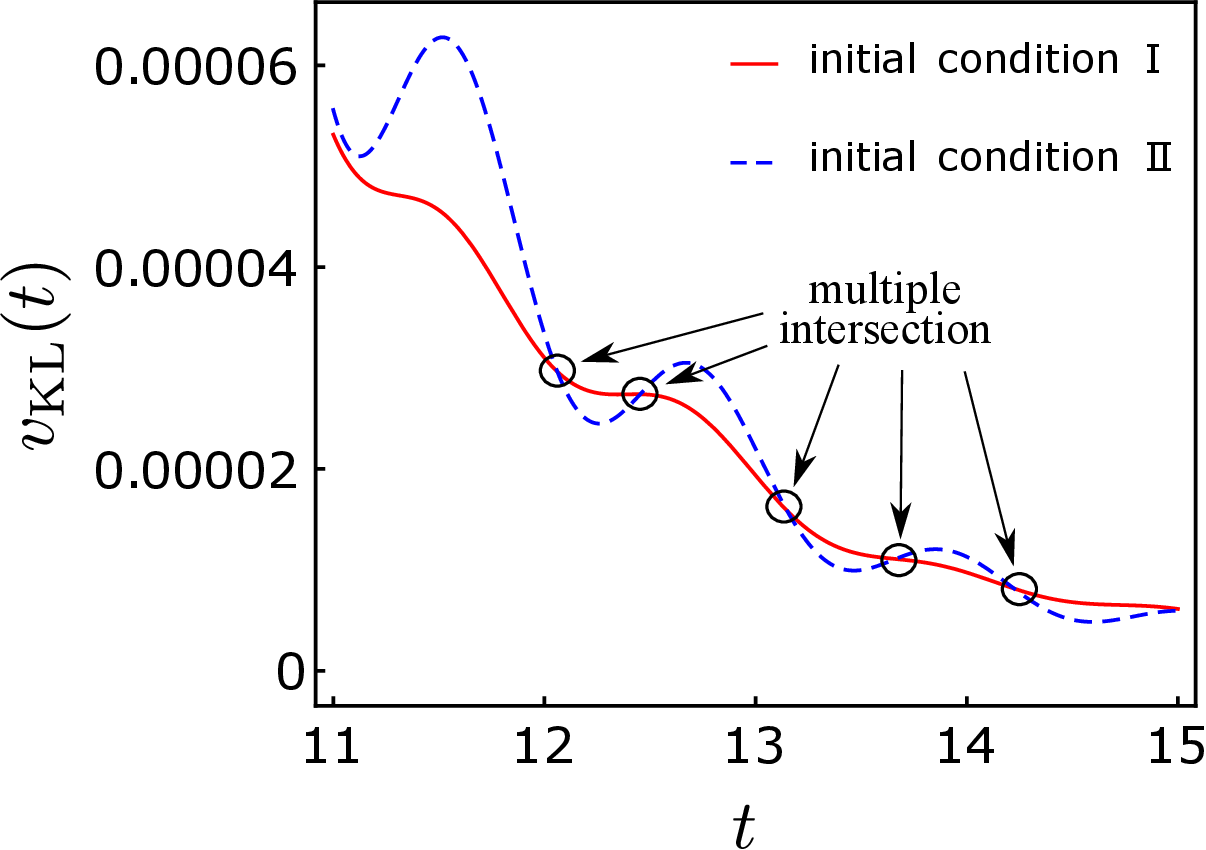}}
  \caption{The figure (a) shows single intersection between the KL divergence of the two copies I and II and thereby exhibit single QMPE instead of multiple QMPE. Parameters used are same as in Fig.~\ref{fig:a11} only except $\tilde{d}_\mathrm{I}=1.0$. For these same parameters, the figure (b) compares the time evolution of the speeds of the two copies. Unlike the distance, the speeds intersect each other multiple times leading to multiple QMPE.}
\label{fig:a15}
\end{figure}
However, we have found that the KL divergence does not show any intersection between two copies I and II for the same parameter values used in Figs.~\ref{fig:a11}-\ref{fig:a14}. Therefore, we have to use different values of the initial drives (still within the fixed dissipation protocol) to achieve QMPE in $D_{\mathrm{KL}}(t)$, presented in Fig.~\ref{fig:a15}(a). Intriguingly, we have only observed single intersection between the two time evolved copies, resulting in single QMPE in KL divergence. This means that once the initially higher distant copy intersects the initially lower distant copy producing QMPE, thereafter the new lower distant copy (that was initially higher distant) remains closer to the steady state for the rest of the relaxation process. In fact, for all other parameter sets that we have checked, we could not find multiple QMPE in $D_{\mathrm{KL}}(t)$. The proof of the possibility or impossibility of multiple QMPE in KL divergence in presence of complex eigenvalues remains unsolved and could be an important question to investigate later. Thus KL divergence intersects only once (single QMPE) whereas other observables can have multiple intersections (multiple QMPE). It would be interesting in future to find connections between QMPE in KL divergence and QMPEs in other observables e.g. thermal QMPE.

In this scenario, it would be interesting to examine the corresponding behaviors of the speeds $v_{\mathrm{KL}}(t)$ [Eq.~(\ref{eq:d18})] of the two copies I and II. In Fig.~\ref{fig:a15}(b), we present the time evolution of speeds for the same set of parameters that gave rise to single QMPE in $D_{\mathrm{KL}}(t)$. It is fascinating to observe that the speeds, unlike the distances, intersect each other multiple times producing multiple QMPE. The first intersection between the speeds of the two copies [Fig.~\ref{fig:a15}(b)] shows a time lag in comparison to the intersection between distances of the same copies [Fig.~\ref{fig:a15}(a)]. Such a time lag between the QMPE in KL divergence and the speed is in agreement with the observation in region (d) [Fig.~\ref{fig:d6}]. However, contrary to single QMPE in $v_{\mathrm{KL}}(t)$ in region (d), we have obtained multiple QMPE in the speed in region ($\mathrm{a_1}$). The contrasting behavior of single QMPE in $D_{\mathrm{KL}}(t)$ and multiple QMPE in $v_{\mathrm{KL}}(t)$ imply that the initially higher distant copy that becomes lower distant after the intersection in distances, can remain at lower distance even if its speed sometimes become smaller than the speed of the current higher distant copy. This is interesting because the system almost behaves like an intelligent and efficient runner in a race who knows how to adjust the speed (i.e. sometimes running faster and sometimes running slower than the people behind) and still remains in the first position to finish the race.  
 
\section{Summary}
\label{sec6} 
We have studied QMPE in a two-level quantum system  subjected to an oscillatory electric field and dissipative coupling with the  environment. The  Lindbladian of this two-level driven dissipative system, termed as Hatano's model, has rich variety of eigen-spectrum, especially EPs and complex eigenvalues which are generic features of non-Hermitian quantum systems. We have investigated
the roles of complex eigenvalues and EPs on the QMPE. The exact analytical formulae obtained for time evolved density matrix elements enable us to propose analytical criteria for the QMPE in observables. Interestingly, at the EPs, the density matrix elements and consequently observables (e.g. energy) attain additional algebraic time dependent relaxations apart from the usual exponential relaxations. Such additional algebraic time dependence at the EPs leads to double QMPE (twice intersections between trajectories starting at different initial conditions and relaxing towards the same steady state). We have procured the exact expressions for intersection times of  trajectories, that mark the occurrence of QMPE and we provide analytical criteria that demarcate control parameters' regimes with single or double or no QMPE. Whereas the multiple QMPE at EPs in the present model is restricted to maximum double QMPE i.e. two intersections, the multiple QMPE at the complex eigenvalues can result in even more number of intersections in a oscillatory manner. Notably, we have established the existence of thermal QMPE (single QMPE at EPs and multiple QMPE at complex eigenvalues), in presence of coherence in Hatano's model. It is fascinating that, similar to energy and entropy, temperature also exhibits multiple QMPE where the initially hotter and colder copies intersect each other multiple times. Such multiple reversals of identities (i.e. hotter becoming colder and vice versa) result in multiple thermal QMPE. In spite of achieving multiple QMPE in several observables (ground state probability, energy, entropy, temperature) using either additional algebraic time dependence at EPs or oscillations at complex eigenvalues, surprisingly the KL divergence (a distance measure from steady state) shows single QMPE only. Overall, our theoretical analysis exhibits that EPs and complex eigenvalues can be used as systematic routes to generate multiple QMPE. However, we should mention that the presence of EPs  or oscillations are not necessary for multiple QMPE. In this connection, we have shown that multiple QMPE can be achieved even with purely exponential relaxations, by putting suitable constraints on the gap statistics of the eigen-spectrum (Appendix \ref{appen3}). Although different observables such as energy, entropy, temperature, KL divergence etc. can exhibit QMPE  for the same set of control parameters (some of them may show single intersection time while others may show multiple intersection times), one should note that the intersection times are in general distinct for these observables. This is in agreement with MPE in classical systems where the MPE depends on the choice of the distance measures, as discussed in driven granular systems \cite{Biswas23}. 

An important and generic future goal in QMPE research would be to understand its utility in speeding up quantum processes. In fact, a recent work on accelerating relaxation in open quantum systems by engineering EPs \cite{Zhou23} hints towards the connection of such acceleration to QMPE. Our present work analyzes in detail how EPs lead to QMPE and provide criteria on control parameters to observe the QMPE. It is important to find out the deeper connections between the QMPE and faster relaxations at EPs and for generic open quantum systems. In this connection, an intriguing question is to find how the quantum speed limit may affect the degree of faster relaxation that can be achieved through the QMPE. Our work on QMPE by controlling the amplitudes of oscillatory electric fields hints towards possible experimental realizations of the QMPE in qubit, keeping in mind the recent advancement in qubit related experiment \cite{Gyenis21}. The analysis of QMPE in the present study paves path for future investigations of QMPE in quantum many-body systems. 

{\it Acknowledgements.-} We thank Naomichi Hatano, Sosuke Ito, Ryo Hanai, Arnab Pal, Naruo Ohga and Kohei Yoshimura for useful discussions. This work is partially
supported by the JSPS KAKENHI (Grants No. 21H01006 and
No. 20K14428). A.K.C. gratefully acknowledges a  postdoctoral fellowship from the YITP. For numerical calculations, we have used the computing facilities \cite{comp}. 



\appendix
\onecolumngrid
\setcounter{equation}{0}
\section{Eigenvectors of $\mathcal{L}$ in region (d)}
\label{appen1}
In this Appendix, we give the explicit expressions for the right eigenvectors $|r_k\rangle$ and left eigenvectors $\langle \ell_k|$ of the Lindbladian [Eq.~(\ref{eq:super-operator})] in region (d) [Sec.~\ref{sec4}], where $k=2,3,4$. The right and left eigenvectors corresponding to the zero eigenvalue i.e. the steady state. have already been provided in Eq.~(\ref{eq:m1}). Below are the expressions for the right eigenvectors
\begin{eqnarray}
|r_2\rangle&=&\left(\begin{array}{cccc}
-\displaystyle\frac{\tilde{d}}{1-\mathrm{i}(\tilde{\Gamma}/2-\lambda_2)}, &  -\displaystyle\frac{\tilde{d}}{1+\mathrm{i}(\tilde{\Gamma}/2-\lambda_2)}, & -1, & 1
                  \end{array}
\right)^T,\cr
|r_3\rangle&=&\left(\begin{array}{cccc}
\displaystyle\frac{1-\mathrm{i}(\tilde{\Gamma}/2-\lambda_4)}{\tilde{d}}, & \displaystyle\frac{-1-\mathrm{i}(\tilde{\Gamma}/2-\lambda_4)}{\tilde{d}}, & 0, & 0                 
                     \end{array}
\right)^T,\cr
|r_4\rangle&=&\left(\begin{array}{cccc}
-\displaystyle\frac{\tilde{d}}{1-\mathrm{i}(\tilde{\Gamma}/2-\lambda_4)}, & -\displaystyle\frac{\tilde{d}}{1+\mathrm{i}(\tilde{\Gamma}/2-\lambda_4)}, & -1, & 1
                    \end{array}
\right),
\label{eq:app11} 
\end{eqnarray}
where the superscript $T$ denotes transpose. The left eigenvectors are obtained to be
\begin{eqnarray}
&&\langle \ell_2|=\left(\begin{array}{cccc}
-\ell^{\mathrm{re}}_{21}-\mathrm{i} \ell^{\mathrm{im}}_{21}, & -\ell^{\mathrm{re}}_{21}+\mathrm{i} \ell^{\mathrm{im}}_{21}, & \ell_{23}, & \ell_{24}
                   \end{array}
\right),\cr
&&\langle \ell_3|=\left(\begin{array}{cccc}
\ell^{\mathrm{re}}_{31}+\mathrm{i} \ell^{\mathrm{im}}_{31}, & -\ell^{\mathrm{re}}_{31}+\mathrm{i} \ell^{\mathrm{im}}_{31}, & \ell_{33}, & \ell_{34}
                   \end{array}
\right),\cr
&&\langle \ell_4|=\left(\begin{array}{cccc}
\ell^{\mathrm{re}}_{21}+\mathrm{i} \ell^{\mathrm{im}}_{21}, & \ell^{\mathrm{re}}_{21}-\mathrm{i} \ell^{\mathrm{im}}_{21}, & \ell_{43}, & \ell_{44}
                   \end{array}
\right),\hspace*{0.3 cm}
\label{eq:app12} 
\end{eqnarray}
where the elements $l_{kn}$ are explicitly given by
\begin{eqnarray}
\ell^{\mathrm{re}}_{21}&=&-\frac{(4+(\tilde{\Gamma}-2\lambda_2)^2)(4+(\tilde{\Gamma}-2\lambda_4)^2)}{4\tilde{d}(\lambda_2-\lambda_4)\eta},\cr\nonumber\\
\ell^{\mathrm{im}}_{21}&=&-\frac{(4+(\tilde{\Gamma}-2\lambda_2)^2)(\tilde{\Gamma}-2\lambda_4)(4+(\tilde{\Gamma}-2\lambda_4)^2)}{8\tilde{d}(\lambda_2-\lambda_4)\eta},\cr\nonumber\\
\ell_{23}&=&\frac{(4+(\tilde{\Gamma}-2\lambda_2)^2)\left(\delta_1-2\tilde{\Gamma}\lambda_4(20+\tilde{\Gamma}^2-4\lambda^2_4)-4\lambda^2_4(\tilde{\Gamma}^2-12)\right)}{4(4+2\tilde{d}^2+\tilde{\Gamma}^2)(\lambda_2-\lambda_4)\eta},\cr\nonumber\\
\ell_{24}&=&-\frac{(4+(\tilde{\Gamma}-2\lambda_2)^2)\left(\delta_1+2\tilde{\Gamma}\lambda_4(4-3\tilde{\Gamma}^2-4\lambda^2_4)-4\lambda^2_4(4-3\tilde{\Gamma}^2)\right)}{4(4+2\tilde{d}^2+\tilde{\Gamma}^2)(\lambda_2-\lambda_4)\eta},\cr\nonumber\\
\ell^{\mathrm{re}}_{31}&=&4\tilde{d}(\lambda_2+\lambda_4-\tilde{\Gamma})\eta^{-1},\cr\nonumber\\
\ell^{\mathrm{re}}_{32}&=&\tilde{d}((\tilde{\Gamma}-2\lambda_2)(\tilde{\Gamma}-2\lambda_4)-4)\eta^{-1},\cr\nonumber\\
\ell_{33}&=&\frac{4\mathrm{i}\tilde{d}^2(4+2\tilde{d}^2+\tilde{\Gamma}^2+4\lambda_2\lambda_4)}{(4+2\tilde{d}^2+\tilde{\Gamma}^2)\eta},\cr\nonumber\\
\ell_{34}&=&-\frac{4\mathrm{i}\tilde{d}^2(4+2\tilde{d}^2+\tilde{\Gamma}^2-4\lambda_2\lambda_4)}{(4+2\tilde{d}^2+\tilde{\Gamma}^2)\eta},\cr\nonumber\\
\ell_{43}&=&-\frac{(4+(\tilde{\Gamma}-2\lambda_4)^2)\left(\delta_2-\delta_3-32\lambda_2(\tilde{\Gamma}-\lambda_4)\right)}{4(\lambda_2-\lambda_4)(4+2\tilde{d}^2+\tilde{\Gamma}^2)\eta},\cr\nonumber\\
\ell_{44}&=&\frac{(4+(\tilde{\Gamma}-2\lambda_4)^2)\left(\delta_2+\delta_3+4\tilde{\Gamma}\lambda_2(4-\tilde{\Gamma}^2+2\tilde{\Gamma}\lambda_4)\right)}{4(\lambda_2-\lambda_4)(4+2\tilde{d}^2+\tilde{\Gamma}^2)\eta},
\label{eq:app13} 
\end{eqnarray}
where the eigenvalues $\lambda_2$ and $\lambda_4$ are given by Eq.~(\ref{eq:d5}) and the parameters $\eta,\delta_1,\delta_2,\delta_3$ used in Eq.~(\ref{eq:app13})  have the following forms
\begin{eqnarray}
\eta&=&\left(\tilde{\Gamma}^3-2\tilde{\Gamma}^2(\lambda_2+2\lambda_4)+4\tilde{\Gamma}(-3+2\lambda_2\lambda_4+\lambda^2_4)+8(\lambda_2+2\lambda_4-\lambda_2\lambda^2_4)\right),\cr
\delta_1&=&-16+\tilde{\Gamma}^4+2\tilde{d}^2((\tilde{\Gamma}-2\lambda_4)^2-4),\cr
\delta_2&=&-16+\tilde{\Gamma}^4-8\tilde{\Gamma}\lambda_4-2\tilde{\Gamma}^3\lambda_4+2\tilde{d}^2\left((\tilde{\Gamma}-2\lambda_2)(\tilde{\Gamma}-2\lambda_4)-4\right),\cr
\delta_3&=&4\lambda^2_2\left(\tilde{\Gamma}^2-2\tilde{\Gamma}\lambda_4-4\right).
\end{eqnarray}
\section{Eigenvectors of $\mathcal{L}$ in region ($\mathrm{a_1}$)}
\label{appen2}
In this Appendix, we give the explicit expressions for the right eigenvectors $|r_k\rangle$ and left eigenvectors $\langle \ell_k|$ of the Lindbladian [Eq.~(\ref{eq:super-operator})] in region ($\mathrm{a_1}$) [Sec.~\ref{sec5}], where $k=2,3,4$. The right and left eigenvectors corresponding to the zero eigenvalue i.e. steady state. have already been provided in Eq.~(\ref{eq:m1}). Below are the expressions for the right eigenvectors
\begin{eqnarray}
|r_2\rangle&=&\left(\begin{array}{cccc}
-\displaystyle\frac{\tilde{d}}{1-\mathrm{i}\left(\frac{\tilde{\Gamma}}{2}-\lambda_2\right)}, & -\displaystyle\frac{\tilde{d}}{1+\mathrm{i}\left(\frac{\tilde{\Gamma}}{2}-\lambda_2\right)}, & -1, & 1                   
                  \end{array}
\right)^{T},\cr
|r_3\rangle&=&\left(\begin{array}{cccc}
-\displaystyle\frac{\tilde{d}}{1-\lambda_{\mathrm{im}}-\mathrm{i}\left(\frac{\tilde{\Gamma}}{2}-\lambda_{\mathrm{re}}\right)}, & -\displaystyle\frac{\tilde{d}}{1+\lambda_{\mathrm{im}}+\mathrm{i}\left(\frac{\tilde{\Gamma}}{2}-\lambda_{\mathrm{re}}\right)}, & -1, & 1                   
                  \end{array}
\right)^{T},\cr
|r_4\rangle&=&\left(\begin{array}{cccc}
-\displaystyle\frac{\tilde{d}}{1+\lambda_{\mathrm{im}}-\mathrm{i}\left(\frac{\tilde{\Gamma}}{2}-\lambda_{\mathrm{re}}\right)}, & -\displaystyle\frac{\tilde{d}}{1-\lambda_{\mathrm{im}}+\mathrm{i}\left(\frac{\tilde{\Gamma}}{2}-\lambda_{\mathrm{re}}\right)}, & -1, & 1                   
                  \end{array}
\right)^{T}.
\label{eq:app21} 
\end{eqnarray}
The left eigenvectors are given by
\begin{eqnarray}
&&\langle \ell_2|=\left(\begin{array}{cccc}
\ell^{\mathrm{re}}_{21}+\mathrm{i} \ell^{\mathrm{im}}_{21}, & \ell^{\mathrm{re}}_{21}-\mathrm{i} \ell^{\mathrm{im}}_{21}, & \ell_{23}, & \ell_{24}
                   \end{array}
\right),\cr
&&\langle \ell_3|=\left(\begin{array}{cccc}
\ell^{\mathrm{re}}_{31}+\mathrm{i} \ell^{\mathrm{im}}_{31}, & \ell^{\mathrm{re}}_{32}+\mathrm{i} \ell^{\mathrm{im}}_{32}, & \ell^{\mathrm{re}}_{33}+\mathrm{i} \ell^{\mathrm{im}}_{33}, & \ell^{\mathrm{re}}_{34}+\mathrm{i} \ell^{\mathrm{im}}_{34}
                   \end{array}
\right),\cr
&&\langle \ell_4|=\left(\begin{array}{cccc}
\ell^{\mathrm{re}}_{32}-\mathrm{i} \ell^{\mathrm{im}}_{32}, & \ell^{\mathrm{re}}_{31}-\mathrm{i} \ell^{\mathrm{im}}_{31}, & \ell^{\mathrm{re}}_{33}-\mathrm{i} \ell^{\mathrm{im}}_{33}, & \ell^{\mathrm{re}}_{34}-\mathrm{i} \ell^{\mathrm{im}}_{34}
                   \end{array}
\right),\hspace*{0.3 cm}
\label{eq:app22} 
\end{eqnarray}
where
\begin{eqnarray}
\ell^{\mathrm{re}}_{21}&=&-\frac{(4+(\tilde{\Gamma}-2\lambda_2)^2)(4(\lambda^2_{\mathrm{im}}-1)+(\tilde{\Gamma}-2\lambda_{\mathrm{re}})^2)}{32\tilde{d}\left(\lambda^2_{\mathrm{im}}+(\lambda_2-\lambda_{\mathrm{re}})^2\right)},\cr
\ell^{\mathrm{im}}_{21}&=&-\frac{(4+(\tilde{\Gamma}-2\lambda_2)^2)(\tilde{\Gamma}-2\lambda_{\mathrm{re}})}{8\tilde{d}\left(\lambda^2_{\mathrm{im}}+(\lambda_2-\lambda_{\mathrm{re}})^2\right)},\cr
\ell_{23}&=&-\frac{(4+(\tilde{\Gamma}-2\lambda_2)^2)(2\tilde{d}^2+\tilde{\Gamma}^2+4(1+\lambda^2_{\mathrm{im}}+\lambda^2_{\mathrm{re}}))}{8(4+2\tilde{d}^2+\tilde{\Gamma}^2)\left(\lambda^2_{\mathrm{im}}+(\lambda_2-\lambda_{\mathrm{re}})^2\right)},\cr
\ell_{24}&=&\frac{(4+(\tilde{\Gamma}-2\lambda_2)^2)(2\tilde{d}^2+\tilde{\Gamma}^2-4(1+\lambda^2_{\mathrm{im}}+\lambda^2_{\mathrm{re}}))}{8(4+2\tilde{d}^2+\tilde{\Gamma}^2)\left(\lambda^2_{\mathrm{im}}+(\lambda_2-\lambda_{\mathrm{re}})^2\right)},\cr
\ell^{\mathrm{re}}_{31}&=&\frac{(\xi_1+\xi_2)(4(1+\lambda_{\mathrm{im}})^2+(\tilde{\Gamma}-2\lambda_{\mathrm{re}})^2)}{64\tilde{d}\lambda_{\mathrm{im}}\left(\lambda^2_{\mathrm{im}}+(\lambda_2-\lambda_{\mathrm{re}})^2\right)},\cr
\ell^{\mathrm{re}}_{32}&=&\frac{(\xi_1-\xi_2)(4(\lambda_{\mathrm{im}}-1)^2+(\tilde{\Gamma}-2\lambda_{\mathrm{re}})^2)}{64\tilde{d}\lambda_{\mathrm{im}}\left(\lambda^2_{\mathrm{im}}+(\lambda_2-\lambda_{\mathrm{re}})^2\right)},\cr
\ell^{\mathrm{im}}_{31}&=&\frac{\left(\xi_3-2\tilde{\Gamma}\lambda_{\mathrm{im}}(\lambda_{\mathrm{im}}-2)+4\lambda_2(\lambda_{\mathrm{im}}-1)^2\right)(4(1+\lambda_{\mathrm{im}})^2+(\tilde{\Gamma}-2\lambda_{\mathrm{re}})^2)}
{64\tilde{d}\lambda_{\mathrm{im}}\left(\lambda^2_{\mathrm{im}}+(\lambda_2-\lambda_{\mathrm{re}})^2\right)},\cr
\ell^{\mathrm{im}}_{32}&=&\frac{\left(\xi_3-2\tilde{\Gamma}\lambda_{\mathrm{im}}(\lambda_{\mathrm{im}}+2)+4\lambda_2(\lambda_{\mathrm{im}}+1)^2\right)(4(-1+\lambda_{\mathrm{im}})^2+(\tilde{\Gamma}-2\lambda_{\mathrm{re}})^2)}
{64\tilde{d}\lambda_{\mathrm{im}}\left(\lambda^2_{\mathrm{im}}+(\lambda_2-\lambda_{\mathrm{re}})^2\right)},
\label{eq:app23} 
\end{eqnarray}
where the parameters $\xi_1,\xi_2,\xi_3$ used in Eq.~(\ref{eq:app23}) have the following forms
\begin{eqnarray}
\xi_1&=&\tilde{\Gamma}^2\lambda_{\mathrm{im}}-4\tilde{\Gamma}\lambda_2\lambda_{\mathrm{im}}+4\lambda_{\mathrm{im}}(\lambda^2_2-1),\cr
\xi_2&=&4\tilde{\Gamma}(\lambda_2-\lambda_{\mathrm{re}})+4(\lambda^2_{\mathrm{re}}-\lambda^2_2+\lambda^2_{\mathrm{im}}),\cr
\xi_3&=&-\tilde{\Gamma}^2\lambda_{2} +2\lambda^2_2(\tilde{\Gamma}-2\lambda_{\mathrm{re}})-4\lambda_{\mathrm{re}}+\tilde{\Gamma}\lambda_{\mathrm{re}}(\tilde{\Gamma}-2\lambda_{\mathrm{re}})+4\lambda_2\lambda^2_{\mathrm{re}}.
\end{eqnarray}
The four other elements $\ell^{\mathrm{re}}_{33},\ell^{\mathrm{im}}_{33},\ell^{\mathrm{re}}_{34}$ and $\ell^{\mathrm{im}}_{34}$ in Eq.~(\ref{eq:app22}) have too lengthy explicit expressions. Therefore, we choose to provide their forms in terms of the elements whose explicit expressions are already given in Eq.~(\ref{eq:app23}), as
\begin{eqnarray}
\ell^{\mathrm{re}}_{33}&=& \frac{\tilde{d}\left((\ell^{\mathrm{re}}_{31}+\ell^{\mathrm{re}}_{32})(8\lambda^2_2-8\tilde{\Gamma}\lambda_2-2(4+2\tilde{\Gamma}^2+\tilde{d}^2))+(\ell^{\mathrm{im}}_{32}-\ell^{\mathrm{im}}_{31})(4\tilde{\Gamma}\lambda^2_2+4\lambda_2(4+\tilde{d}^2)-\tilde{\Gamma}(4+2\tilde{\Gamma}^2+\tilde{d}^2))\right)}{(4+2\tilde{\Gamma}^2+\tilde{d}^2)(4+(\tilde{\Gamma}-2\lambda_2)^2)},\cr
\ell^{\mathrm{im}}_{33}&=&\frac{\tilde{d}\left((\ell^{\mathrm{im}}_{31}+\ell^{\mathrm{im}}_{32})(8\lambda^2_2-8\tilde{\Gamma}\lambda_2-2(4+2\tilde{\Gamma}^2+\tilde{d}^2))+(\ell^{\mathrm{re}}_{31}-\ell^{\mathrm{re}}_{32})(4\tilde{\Gamma}\lambda^2_2+4\lambda_2(4+\tilde{d}^2)-\tilde{\Gamma}(4+2\tilde{\Gamma}^2+\tilde{d}^2))\right)}{(4+2\tilde{\Gamma}^2+\tilde{d}^2)(4+(\tilde{\Gamma}-2\lambda_2)^2)},\cr
\ell^{\mathrm{re}}_{34}&=& \frac{\tilde{d}\left((\ell^{\mathrm{re}}_{31}+\ell^{\mathrm{re}}_{32})(8\lambda^2_2-8\tilde{\Gamma}\lambda_2+2(4+2\tilde{\Gamma}^2+\tilde{d}^2))+(\ell^{\mathrm{im}}_{32}-\ell^{\mathrm{im}}_{31})(4\tilde{\Gamma}\lambda^2_2-4\lambda_2(\tilde{\Gamma}^2+\tilde{d}^2)+\tilde{\Gamma}(4+2\tilde{\Gamma}^2+\tilde{d}^2))\right)}{(4+2\tilde{\Gamma}^2+\tilde{d}^2)(4+(\tilde{\Gamma}-2\lambda_2)^2)},\cr
\ell^{\mathrm{im}}_{34}&=&\frac{\tilde{d}\left((\ell^{\mathrm{im}}_{31}+\ell^{\mathrm{im}}_{32})(8\lambda^2_2-8\tilde{\Gamma}\lambda_2+2(4+2\tilde{\Gamma}^2+\tilde{d}^2))+(\ell^{\mathrm{re}}_{31}-\ell^{\mathrm{re}}_{32})(4\tilde{\Gamma}\lambda^2_2-4\lambda_2(\tilde{\Gamma}^2+\tilde{d}^2)+\tilde{\Gamma}(4+2\tilde{\Gamma}^2+\tilde{d}^2))\right)}{(4+2\tilde{\Gamma}^2+\tilde{d}^2)(4+(\tilde{\Gamma}-2\lambda_2)^2)}.
\end{eqnarray}
The explicit expressions for $\lambda_2,\lambda_{\mathrm{re}}$ and $\lambda_{\mathrm{im}}$ in terms of control parameters $(\tilde{d},\tilde{\Gamma})$ are obtained as 
\begin{eqnarray}
 \lambda_2&=&\frac{1}{12}\left((-\mathrm{i}+\sqrt
 3)\chi^{1/3}+8\tilde{\Gamma}-\frac{(\mathrm{i}+\sqrt{3})(12+12\tilde{d}^2-\tilde{\Gamma}^2)}{\chi^{1/3}}\right)\cr
 \lambda_{\mathrm{re}}&=&\frac{12\tilde{d}^2(\mathrm{i}+\sqrt{3})+(\mathrm{i}-\sqrt{3})\chi^{2/3}+16\tilde{\Gamma}\chi^{1/3}-(\mathrm{i}+\sqrt{3})(\tilde{\Gamma}^2-12)}{24\chi^{1/3}},\cr
 \lambda_{\mathrm{im}}&=&\frac{12\sqrt{3}\tilde{d}^2(\mathrm{i}+\sqrt{3})-\sqrt{3}(\mathrm{i}-\sqrt{3})\chi^{2/3}-\sqrt{3}(\mathrm{i}+\sqrt{3})(\tilde{\Gamma}^2-12)}{24\chi^{1/3}},
\end{eqnarray}
where $\chi$ has the following form
\begin{equation}
\chi=\left(\mathrm{i}\tilde{\Gamma}(36-18\tilde{d}^2+\tilde{\Gamma}^2)+6\sqrt{3\tilde{\Gamma}^2(\tilde{d}^4+20\tilde{d}^2-8)-48(1+\tilde{d}^2)^3-3\tilde{\Gamma}^4}\right). 
\end{equation}
\section{QMPE in region ($\mathrm{b}$): purely exponential relaxation}
\label{appen3}
In this Appendix we discuss QMPE in region (b) of Hatano's model that contain eigenvalues $0,-\mathrm{i}\lambda_2,-\mathrm{i}\lambda_3,-\mathrm{i}\lambda_4$ obeying $0<\lambda_2<\lambda_3<\lambda_4$. 
\subsection{Density matrix elements}
The individual density matrix elements $\rho_j(t)$ ($j=1,2,3,4$) are given by
\begin{eqnarray}
\rho_j(t)=\sum_{k=1}^{4} e^{-\lambda_k t} r_{k,j} a_k, \hspace*{0.2 cm} a_k=\sum_{n=1}^{4} \ell_{k,n}\rho_n(0).
\label{eq:l4} 
\end{eqnarray}
The right eigenvectors $|r_k\rangle$ and left eigenvectors $\langle \ell_k|$, are of the form below
\begin{eqnarray}
|r_k\rangle=\left(\begin{array}{cccc}
-\frac{\tilde{d}}{1-\mathrm{i}\left(\frac{\tilde{\Gamma}}{2}-\lambda_k\right)}, & -\frac{\tilde{d}}{1+\mathrm{i}\left(\frac{\tilde{\Gamma}}{2}-\lambda_k\right)}, & -1, & 1                   
                  \end{array}
\right)^{T}, \hspace{0.3 cm}
\langle \ell_k|=\left(\begin{array}{cccc}
\ell^{\mathrm{re}}_{k1}+\mathrm{i} \ell^{\mathrm{im}}_{k1}, & \ell^{\mathrm{re}}_{k1}-\mathrm{i} \ell^{\mathrm{im}}_{k1}, & \ell_{k3}, & \ell_{k4}
                   \end{array}
\right), k=2,3,4.
\label{eq:l6} 
\end{eqnarray}
The elements $\ell_{kn}$ in Eq.~(\ref{eq:l6}) are obtained as
\begin{eqnarray}
\ell^{\mathrm{re}}_{k1}&=&-\frac{(4+(\tilde{\Gamma}-2\lambda_k)^2)(-4+\prod_{n=2, n\neq k}^{4} (\tilde{\Gamma}-2\lambda_n))}{32\tilde{d}\prod_{n=2, n\neq k}^{4}(\lambda_k-\lambda_n)},\hspace*{0.3 cm}
\ell^{\mathrm{im}}_{k1}=-\frac{(4+(\tilde{\Gamma}-2\lambda_k)^2)(\tilde{\Gamma}-\sum_{n=2, n\neq k}^{4} \lambda_n)}{8\tilde{d}\prod_{n=2, n\neq k}^{4}(\lambda_k-\lambda_n)},\cr
\ell_{k3}&=&-\frac{(4+(\tilde{\Gamma}-2\lambda_k)^2)(4+2\tilde{d}^2+\tilde{\Gamma}^2+4\prod_{n\neq k}\lambda_n)}{8(4+2\tilde{d}^2+\tilde{\Gamma}^2)\prod_{n\neq k}(\lambda_k-\lambda_n)},\hspace*{0.3 cm}
\ell_{k4}=\frac{(4+(\tilde{\Gamma}-2\lambda_k)^2)(4+2\tilde{d}^2+\tilde{\Gamma}^2-4\prod_{n\neq k}\lambda_n)}{8(4+2\tilde{d}^2+\tilde{\Gamma}^2)\prod_{n\neq k}(\lambda_k-\lambda_n)}.
\label{eq:l7} 
\end{eqnarray} 
\begin{figure}[t]
  \centering 
  \subfigure[]{\includegraphics[width=0.5\linewidth]{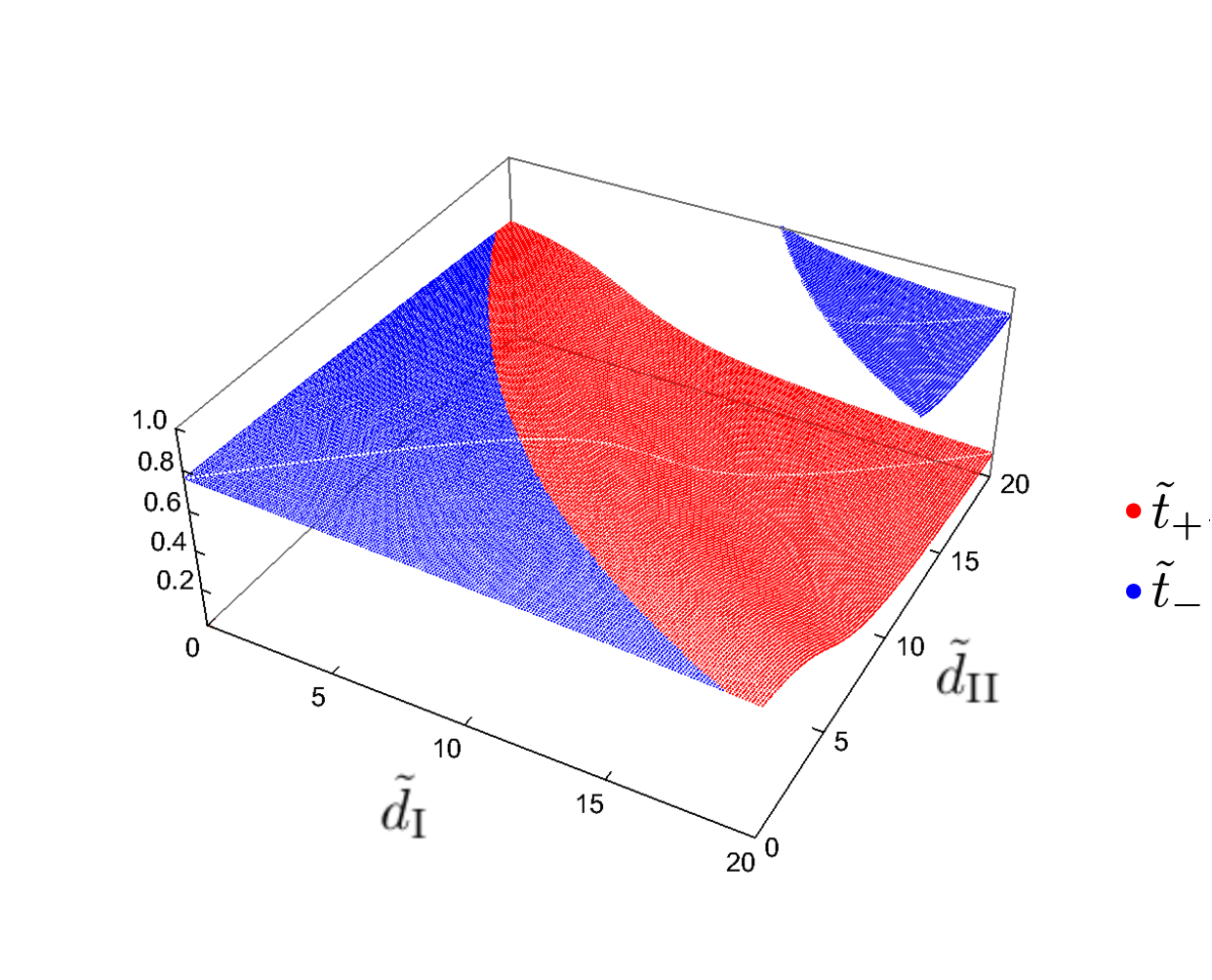}}\hfill
  \subfigure[]{\includegraphics[width=0.5\linewidth]{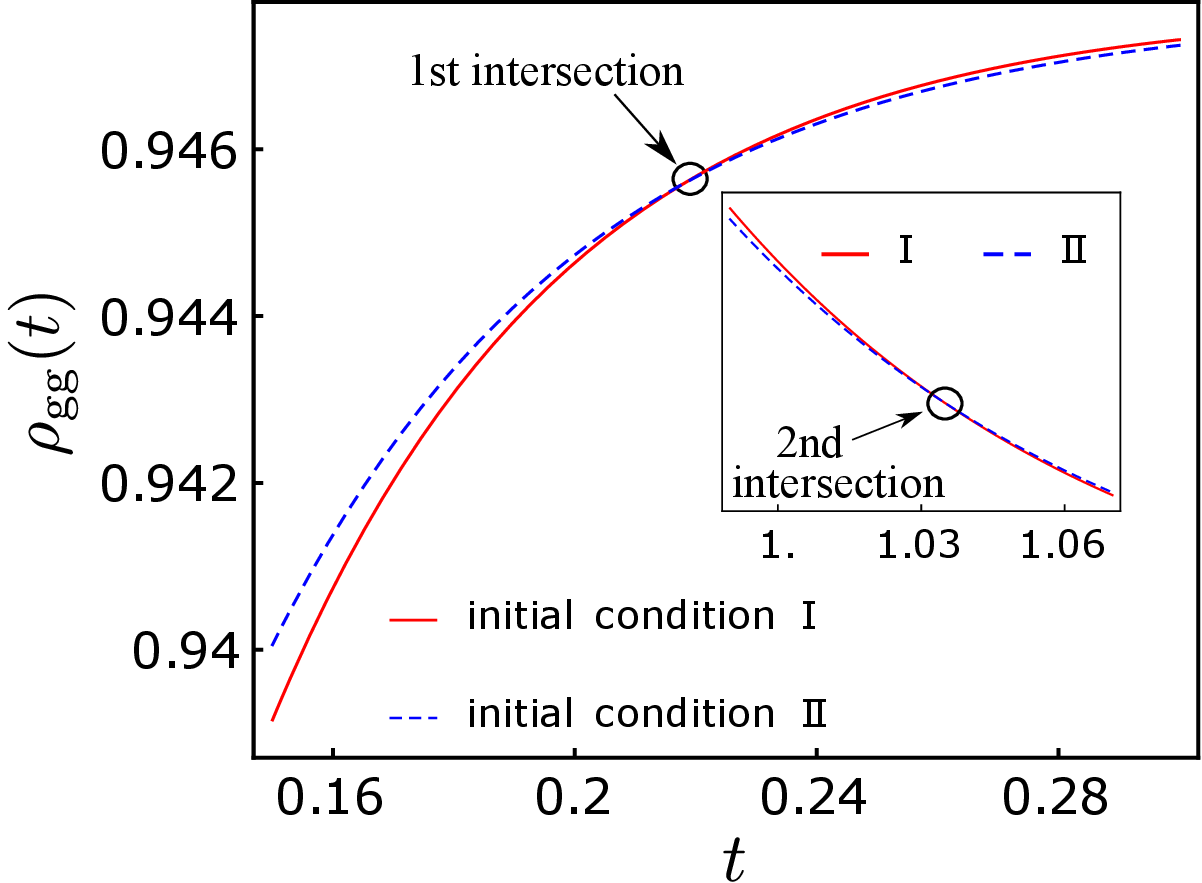}}
\caption{The figure (a) shows the intersection time solutions [Eq.~(\ref{eq:l14})] in region (b) as functions of the control parameters $\tilde{d}_{\mathrm{I}}$ and $\tilde{d}_{\mathrm{II}}$. Some of the regimes show finite and positive values of both $\tilde{t}_+$ and $\tilde{t}_-$, implying double QMPE. Parameters used are $\tilde{d}=6.0, \tilde{\Gamma}_{\mathrm{I}}=\tilde{\Gamma}_{\mathrm{II}}=\tilde{\Gamma}=6\sqrt{17}$ (from Eq.~(\ref{eq:l12})). The figure (b) shows a specific example of double intersection (double QMPE) in $\rho_{\mathrm{gg}}(t)$. The main figure shows the first intersection. The inset shows much weaker second intersection. Parameters additionally used for (b) are $\tilde{d}_{\mathrm{I}}=16.3, \tilde{d}_{\mathrm{II}}=13.4$.}
\label{fig:lb1}
\end{figure}
To explore the possibility of QMPE, we consider the ground state probability $\rho_4(t)\equiv\rho_{\mathrm{gg}}(t)$. The expression for $\rho_{\mathrm{gg}}(t)$ becomes
\begin{eqnarray}
\rho_{\mathrm{gg}}(t)=\frac{4+\tilde{d}^2+\tilde{\Gamma}^2}{4+2\tilde{d}^2+\tilde{\Gamma}^2}+ \sum_{k=2}^{4}e^{-\lambda_k t}\left[2(\ell^{\mathrm{re}}_{k1}\rho_{\mathrm{re}}(0)-\ell^{\mathrm{im}}_{k1}\rho_{\mathrm{im}}(0))+\ell_{k3}+(\ell_{k4}-\ell_{k3})\rho_{\mathrm{gg}}(0)\right].
\label{eq:l8} 
\end{eqnarray}
The difference $\Delta\rho_{\mathrm{gg}}(t)=$ has the form
\begin{eqnarray}
\Delta\rho_{\mathrm{gg}}(t)&=&e^{-\lambda_2 t}\left[\alpha_4 x^{\frac{\lambda_4-\lambda_2}{\lambda_3-\lambda_2}}+\alpha_3 x+\alpha_2\right],
\label{eq:l9} 
\end{eqnarray}
where 
\begin{eqnarray}
x=e^{-(\lambda_3-\lambda_2)t},\hspace*{0.3 cm} 
\alpha_k=2(\ell^{\mathrm{re}}_{k1}\Delta\rho_{\mathrm{re}}(0)-\ell^{\mathrm{im}}_{k1}\Delta\rho_{\mathrm{im}}(0))+(\ell_{k4}-\ell_{k3})\Delta\rho_{\mathrm{gg}}(0), \cr
\label{eq:l10}
\end{eqnarray}
where $k=2,3,4,$ and $\Delta y=y^{\mathrm{I}}-y^{\mathrm{II}}$ in Eq.~(\ref{eq:l10}). 
We want to create the possibility of multiple solutions to $\Delta \rho_{\mathrm{gg}}(t)(t)=0$. To get $m$ number of intersections, we have to produce at least an $m$-th order polynomial in Eq.~(\ref{eq:l9}). To do so, clearly we need
\begin{equation}
\lambda_4-\lambda_3=(m-1)(\lambda_3-\lambda_2).
\label{eq:l11} 
\end{equation}
Interestingly, Eq.~(\ref{eq:l11}) puts a constraint on the gap statistics of the eigenvalues of the Lindbladian. Thus the gap statistics of eigenvalues connect directly to the possibility of multiple QMPE. One can numerically check the maximum possible value of $m$ in region (b) by estimating $(\lambda_4-\lambda_3)/(\lambda_3-\lambda_2)$ with the variation of $(\tilde{d},\tilde{\Gamma})$ in this region. 

To consider an example, we study $m=2$ i.e. the double QMPE. The line in $\tilde{d}-\tilde{\Gamma}$ plane that satisfy this condition is
\begin{equation}
\tilde{\Gamma}=3\sqrt{2}\sqrt{\tilde{d}^2-2}. 
\label{eq:l12}
\end{equation}
Any $(\tilde{d},\tilde{\Gamma})$ on the line Eq.~(\ref{eq:l12}) simplifies Eq.~(\ref{eq:l9}) such that the QMPE condition $\Delta \rho_{\mathrm{gg}}(t)=0$ reads as $\alpha_4 x^2+\alpha_3 x+\alpha_2=0$. The solutions to this equation are 
\begin{equation}
x_{\pm}=\frac{(-\alpha_3\pm\sqrt{\alpha^2_3-4\alpha_4 \alpha_2})}{2\alpha_4}.
\label{eq:l13}
\end{equation}
\begin{figure}[t]
  \centering
  \subfigure[]{\includegraphics[width=0.45\linewidth]{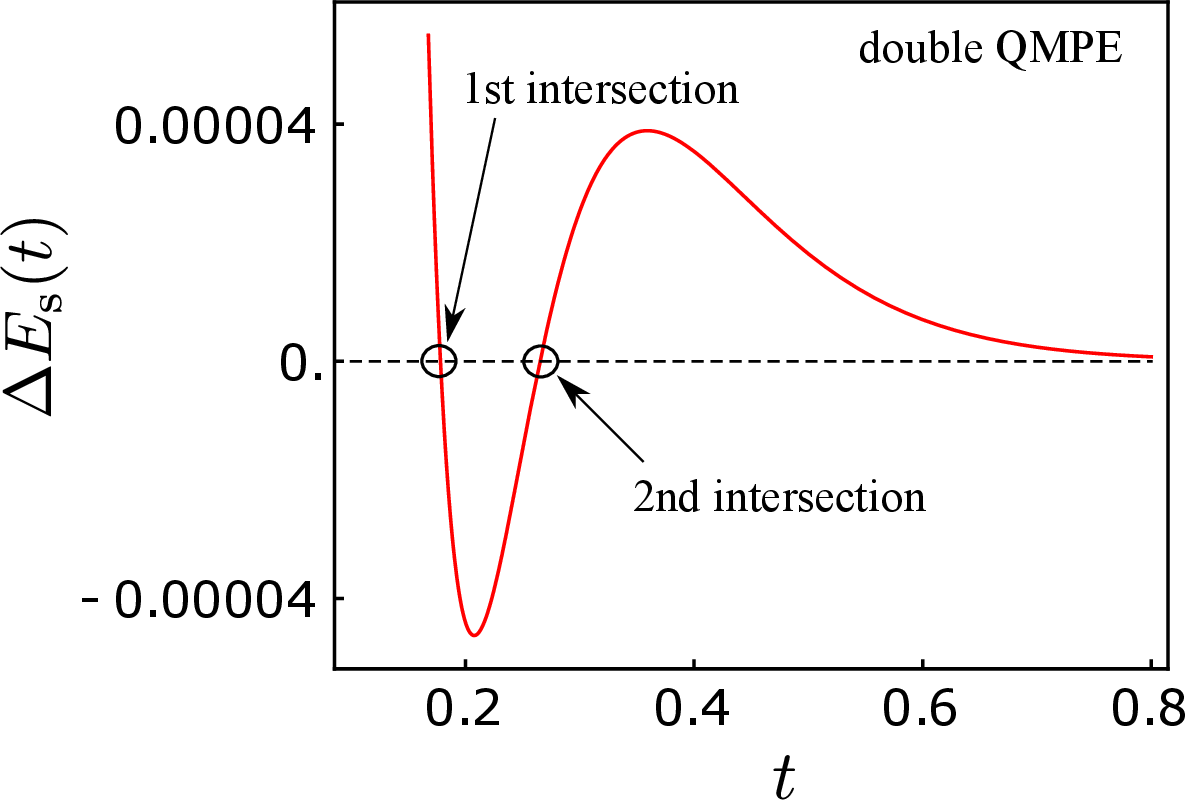}}\hfill
  \subfigure[]{\includegraphics[width=0.45\linewidth]{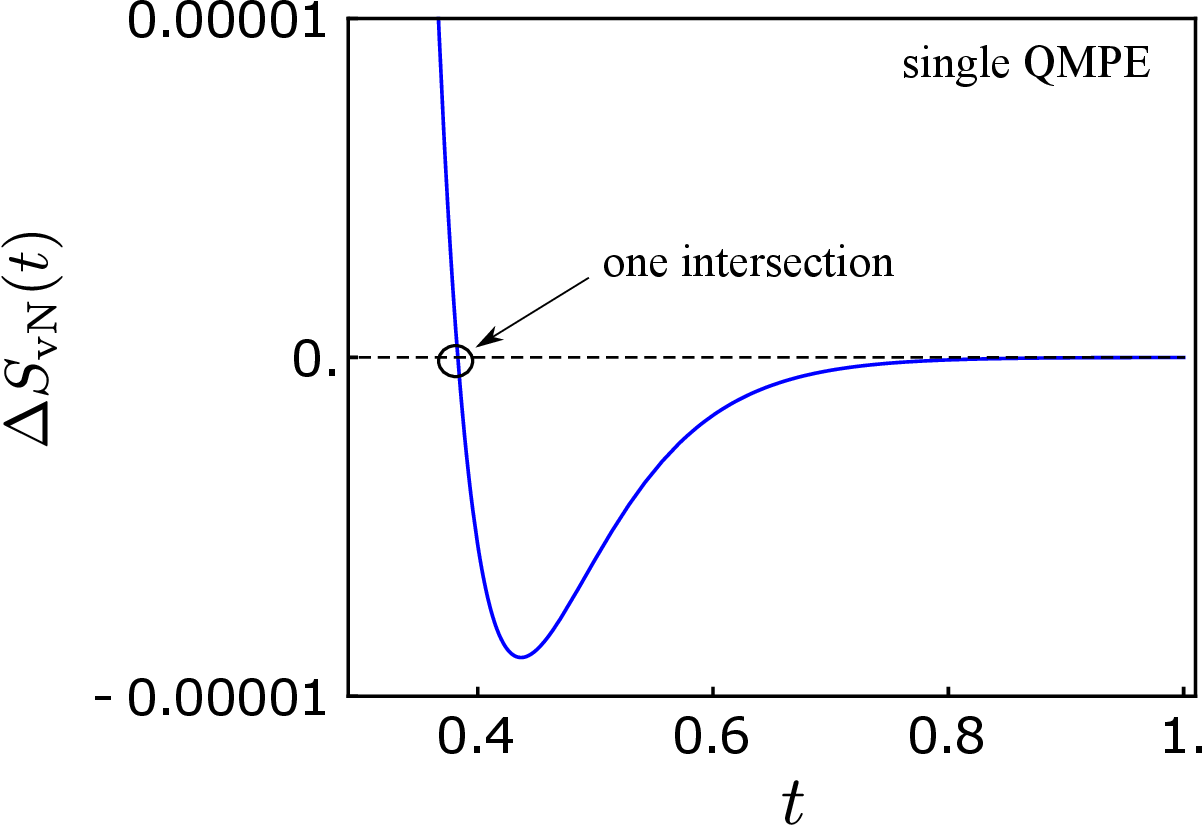}}
  \caption{The figures (a) and (b) show nature of QMPE in energy and von Neumann entropy in region (b), respectively, for the same parameters exhibiting double QMPE in $\rho_{\mathrm{gg}}(t)$ in Fig.~\ref{fig:lb1}(b). We observe that although energy shows double QMPE (figure (a)), the von Neumann entropy exhibits single QMPE only (figure (b)).}
\label{fig:lb3}
\end{figure}
The criteria for different types of QMPE in ground state probability $\rho_{\mathrm{gg}}(t)$ in the region (b), are given by:
\begin{eqnarray}
&&0<x_+<1\,\, \&\,\, 0<x_{-}<1: \,\,\mathrm{double}\,\, \mathrm{QMPE},\cr
&&0<x_+<1 \,\,\mathrm{or}\,\, 0<x_{-}<1: \,\,\mathrm{single}\,\, \mathrm{QMPE},\cr
&& x_+\not\in (0,1)\,\, \&\,\, x_{-}\not\in (0,1): \,\,\mathrm{no}\,\, \mathrm{QMPE}.
\label{eq:en10}
\end{eqnarray}
Then, the intersection times $\tilde{t}_{\pm}$ are obtained as (using Eq.~(\ref{eq:l10})) 
\begin{eqnarray}
\tilde{t}_{\pm}=-\frac{1}{\lambda_3-\lambda_2}\mathrm{ln}(x_{\pm}).
\label{eq:l14} 
\end{eqnarray}
In Fig.~\ref{fig:lb1}(a), we present the intersection time solutions denoting QMPE in $\rho_{\mathrm{gg}}(t)$ as functions of the control parameters $\tilde{d}_{\mathrm{I}}$ and $\tilde{d}_{\mathrm{II}}$. Indeed, we observe that there are parameter regions where both $\tilde{t}_+$ and $\tilde{t}_-$ are positive and finite, indicating double QMPE. Other regions have only one intersection time, either $\tilde{t}_+$ and $\tilde{t}_-$, implying single QMPE. To demonstrate double QMPE, the Fig.~\ref{fig:lb1}(b) exhibits a specific example of double intersection in $\rho_{\mathrm{gg}}(t)$. However, the two copies become close enough so that the second intersection (shown in inset of Fig.~\ref{fig:lb1}(b)) is weak in magnitude and hard to detect.

\subsection{Energy}
We look for solutions to $\Delta E_{\mathrm{s}}(t)=0$, which can give to single or multiple QMPE depending on the number of positive finite intersection time  solutions. The Fig.~\ref{fig:lb3}(a) shows that the average energy exhibits double QMPE indicated by the twice zero intersections of $\Delta E_{\mathrm{s}}(t)$, for the same set of parameter values producing double QMPE in $\rho_{\mathrm{gg}}(t)$ in Fig.~\ref{fig:lb1}(b).
\begin{figure}[t]
  \centering
  \subfigure[]{\includegraphics[width=0.45\linewidth]{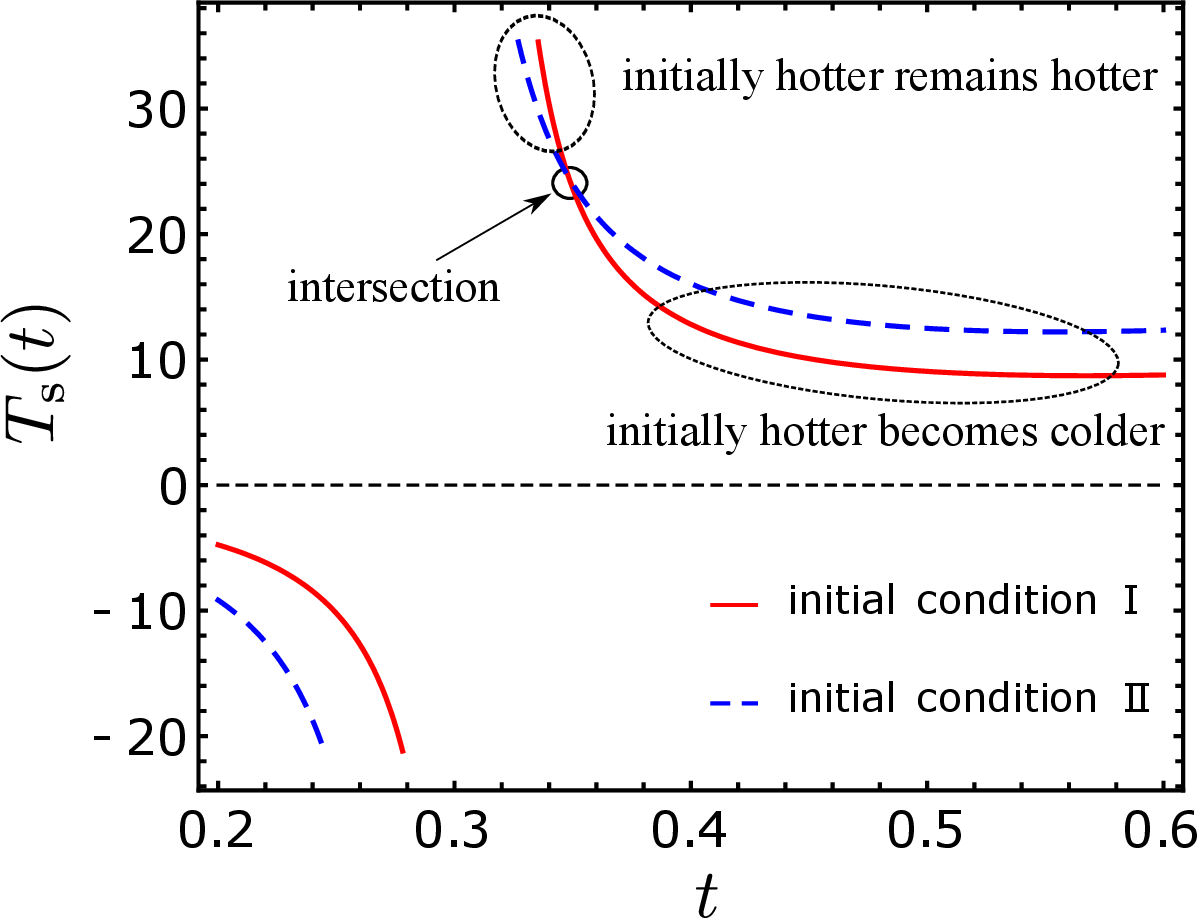}}\hfill
  \subfigure[]{\includegraphics[width=0.45\linewidth]{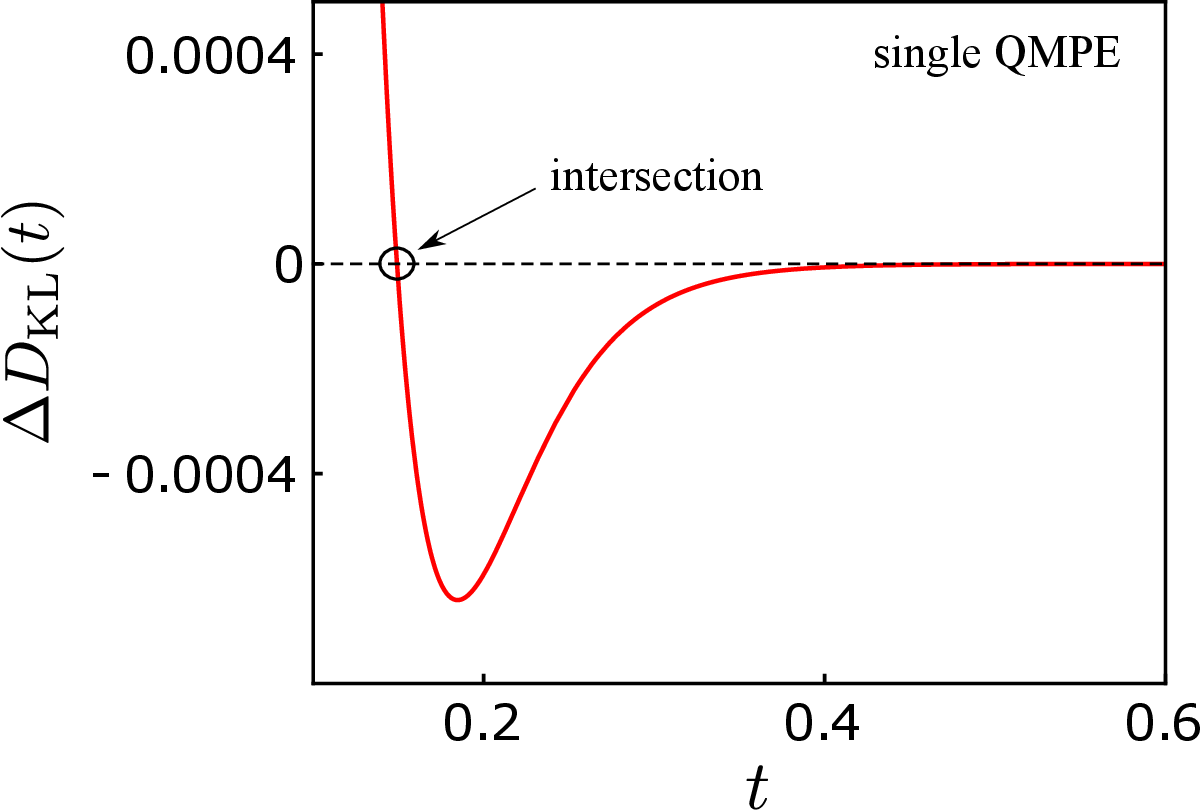}}
  \caption{The figure (a) shows thermal QMPE in region (b) where the temperatures of the two copies intersect each other once leading to single thermal QMPE. Parameters used are $\tilde{d}=6.0, \tilde{\Gamma}_{\mathrm{I}}=0.5,\tilde{\Gamma}_{\mathrm{II}}=0.4,\tilde{\Gamma}=6\sqrt{17}, \tilde{d}_{\mathrm{I}}=3.0, \tilde{d}_{\mathrm{II}}=2.0$. The figure (b) exhibits single QMPE in KL divergence in region (b). Parameters used for (b) are $\tilde{d}=6.0, \tilde{\Gamma}_{\mathrm{I}}=0.5,\tilde{\Gamma}_{\mathrm{II}}=0.4,\tilde{\Gamma}=6\sqrt{17}, \tilde{d}_{\mathrm{I}}=3.0, \tilde{d}_{\mathrm{II}}=2.0$.}
\label{fig:lb4}
\end{figure}

\subsection{von-Neumann entropy}
Here we ask if the entropy, like density matrix and energy, also exhibit double QMPE for the same set of parameters used in Figs.~\ref{fig:lb1}(b) and \ref{fig:lb3}(a). However, in Fig.~\ref{fig:lb3}(b), we observe that the entropy difference $\Delta S_{\mathrm{vN}}(t)$ between two copies intersect zero only once, giving rise to single QMPE instead of double QMPE. 

\subsection{Temperature}
In Fig.~\ref{fig:lb4}(a), we present the time evolution of temperatures for the two copies I and II, one initially starting as hotter with respect to the other. We observe that the temperatures of the two copies intersect each other i.e. the hotter becomes colder and vice versa. Notably, unlike ground state probability [Fig.~\ref{fig:lb1}(b)], energy [Fig.~\ref{fig:lb3}(a)] that can exhibit double QMPE, we have observed only single QMPE in temperature. If observation of multiple QMPE is possible or not with purely exponential relaxations, remains an open question. 

\subsection{Kullback-Leibler divergence} 
Here we study the possibility of QMPE in KL divergence for the same parameters used to explore thermal QMPE in Figs.~\ref{fig:lb4}(a). The Fig.~\ref{fig:lb4}(b) shows QMPE in KL divergence, however the intersection time corresponds to negative temperatures in Fig.~\ref{fig:lb4}(a).

\section{QMPE in region $\mathrm{(e)}$: 3rd order EP}
\label{appen4}
The region (e) in the $(\tilde{d},\tilde{\Gamma})$ plane of Hatano's model is composed of a single point 
\begin{eqnarray}
\tilde{d}=2\sqrt{2}, \,\,\tilde{\Gamma}=6\sqrt{3}.
\label{eq:1rhoe} 
\end{eqnarray}
It is a third order EP where all the non-zero eigenvalues become equal and their eigenvectors coalesce. Since this is a single point with fixed values of 
$\tilde{d}$ and $\tilde{\Gamma}$, the right eigenvectors are explicitly given by
\begin{eqnarray}
|r_1\rangle&=&\left(\begin{array}{cccc}
                   \frac{-1-3\mathrm{i}\sqrt{3}}{15\sqrt{2}}, & \frac{-1+3\mathrm{i}\sqrt{3}}{15\sqrt{2}}, & \frac{1}{15}, & 1 \\
                  \end{array}
\right)^{T},\hspace*{0.4 cm}
|r_2\rangle=\left(\begin{array}{cccc}
                   \frac{-1+\mathrm{i}\sqrt{3}}{\sqrt{2}}, & \frac{-1-\mathrm{i}\sqrt{3}}{\sqrt{2}}, & -1, & 1 \\
                  \end{array}
\right)^{T},\cr
|r_3\rangle&=&\left(\begin{array}{cccc}
                   \frac{1+\mathrm{i}\sqrt{3}}{2\sqrt{2}}, & \frac{-1+\mathrm{i}\sqrt{3}}{2\sqrt{2}}, & 0, & 0 \\
                  \end{array}
\right)^{T},\hspace*{0.4 cm}
|r_4\rangle=\left(\begin{array}{cccc}
                   \frac{1}{2\sqrt{2}}, & \frac{1}{2\sqrt{2}}, & 0, & 0 \\
                  \end{array}
\right)^{T}.
\label{eq:e1} 
\end{eqnarray}
Of course $|r_1\rangle$ and $\langle \ell_1|$ have same expressions both in Eqs.~(\ref{eq:m1}) and (\ref{eq:e1}). The left eigenvectors are
\begin{figure}[t]
  \centering 
  \subfigure[]{\includegraphics[width=0.5\linewidth]{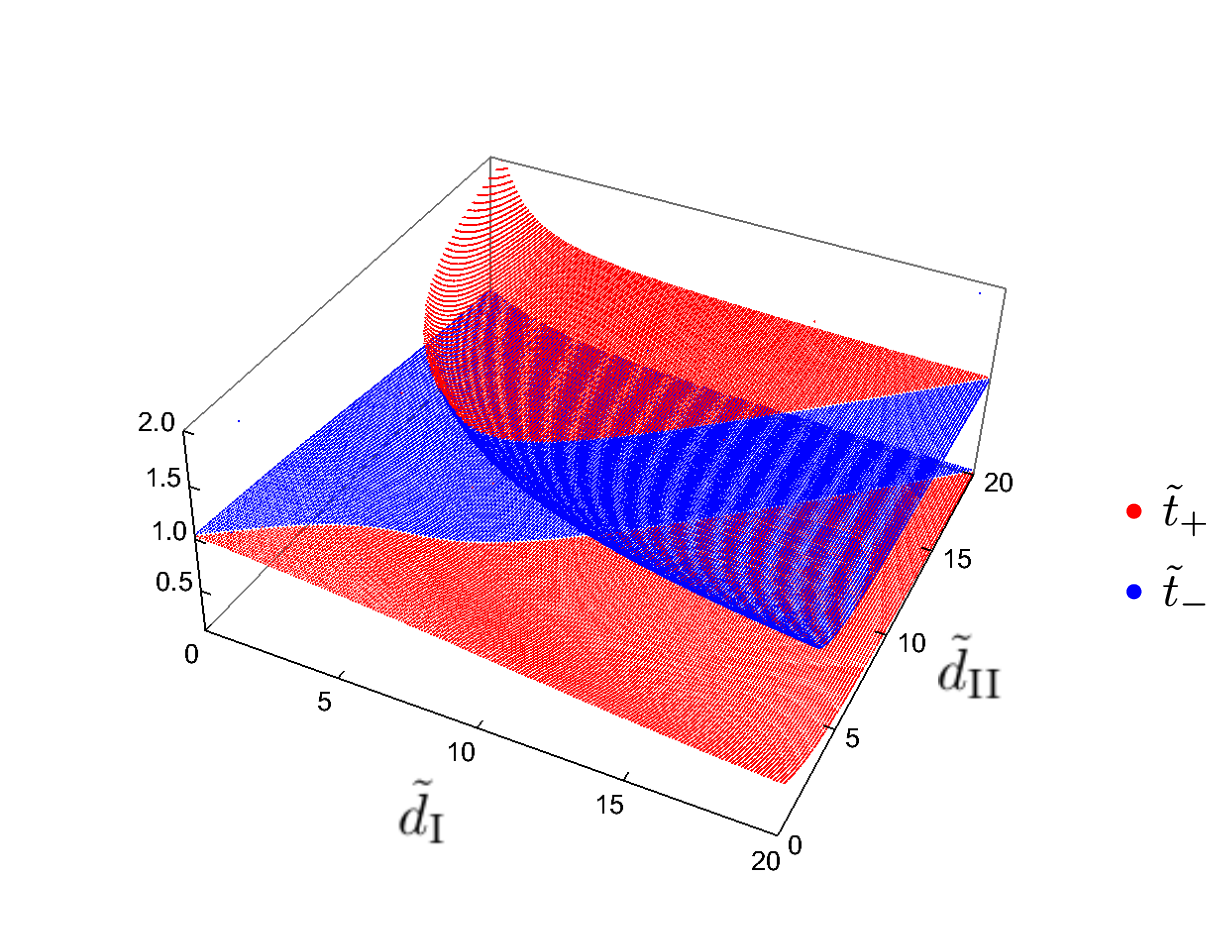}}\hfill
  \subfigure[]{\includegraphics[width=0.5\linewidth]{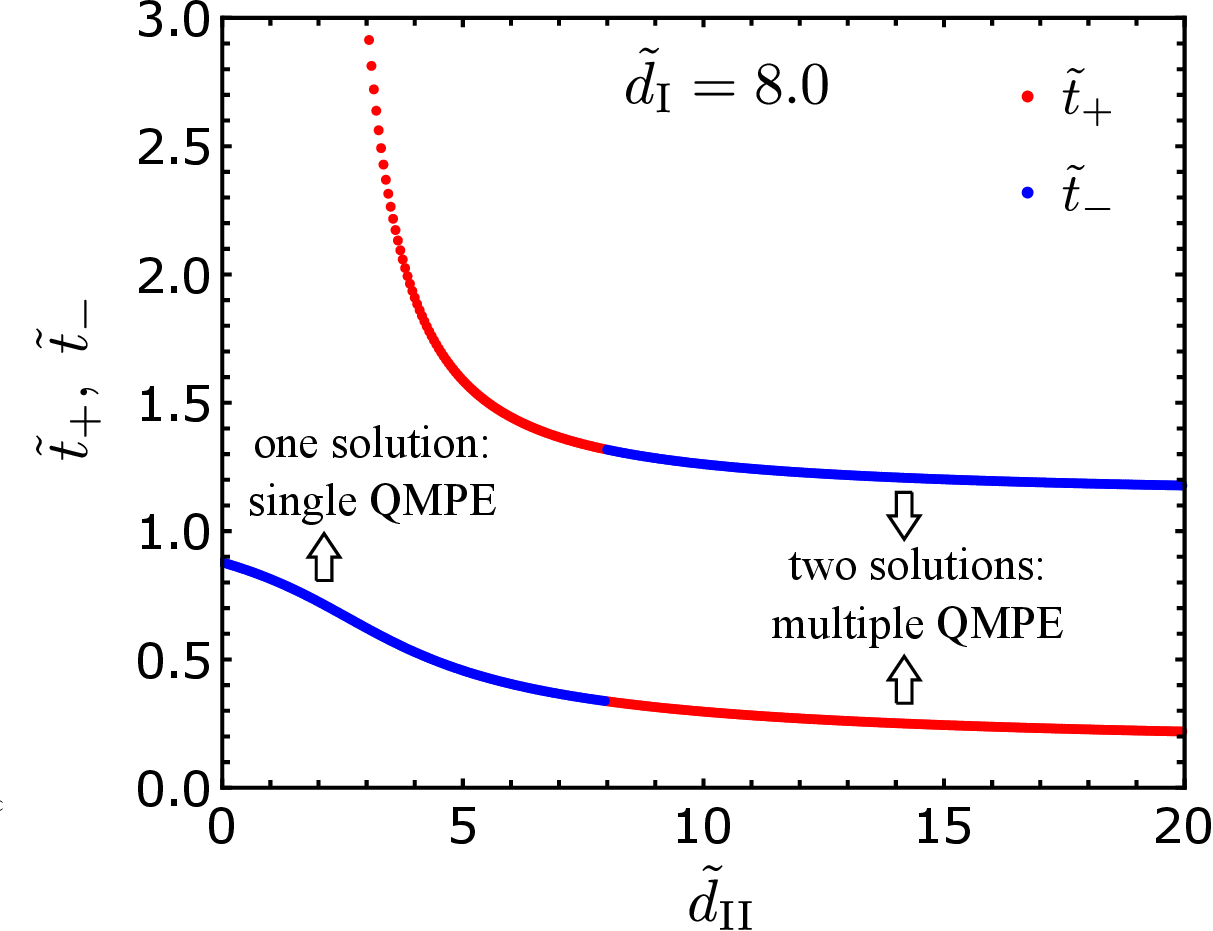}}
\caption{The figure (a) illustrates the two intersection time solutions $\tilde{t}_{\pm}$ [Eq.~(\ref{eq:e7})] in region (e) as functions of $\tilde{d}_{\mathrm{I}}$ and $\tilde{d}_{\mathrm{II}}$. For comparatively high values of these control parameters, both solutions exist, leading to double QMPE. Parameters used are $\tilde{d}=2\sqrt{2}, \tilde{\Gamma}=\tilde{\Gamma}_{\mathrm{I}}=\tilde{\Gamma}_{\mathrm{II}}=6\sqrt{3}$. The figure (b) shows a surface from the three dimensional plot in figure (a). It demonstrates the intersection times [Eq.~(\ref{eq:e7})] by fixing $\tilde{d}_{\mathrm{I}}$ and varying $\tilde{d}_{\mathrm{II}}$. We observe single QMPE as well as double QMPE depending on the parameter choices. Depending on the sign of $(\tilde{d}_{\mathrm{I}}-\tilde{d}_{\mathrm{II}})$, the two solutions interchange their branches. Parameters used are $\tilde{d}=2\sqrt{2}, \tilde{\Gamma}=\tilde{\Gamma}_{\mathrm{I}}=\tilde{\Gamma}_{\mathrm{II}}=6\sqrt{3}, \tilde{d}_{\mathrm{I}}=8.0$.}
\label{fig:e1}
\end{figure}
\begin{eqnarray}
\langle \ell_1|&=&\left(\begin{array}{cccc}
                       0, & 0, & \frac{15}{16}, & \frac{15}{16} \\
                      \end{array}
\right),\hspace*{0.3 cm}
\langle \ell_2|=\left(\begin{array}{cccc}
                       0, & 0, & -\frac{15}{16}, & \frac{1}{16} \\
                      \end{array}
\right),\cr \langle \ell_3|&=&\left(\begin{array}{cccc}
                       \sqrt{2}, & -\sqrt{2}, & \frac{\mathrm{i}9\sqrt{3}}{4}, & \frac{\mathrm{i}\sqrt{3}}{4} \\
                      \end{array}
\right),\hspace*{0.3 cm} \langle \ell_4|=\left(\begin{array}{cccc}
                       \sqrt{2}+\mathrm{i}\sqrt{6}, & \sqrt{2}-\mathrm{i}\sqrt{6}, & 5, & 1 \\
                      \end{array}
\right).
\label{eq:e2} 
\end{eqnarray}
The non-zero eigenvalue of the Lindbladian [Eq.~(\ref{eq:super-operator})] in region (e) is $-\mathrm{i}\lambda=-\mathrm{i}4\sqrt{3}$. Importantly, as a consequence of the 3rd order EP, we can only achieve an approximately diagonal form i.e. Jordan normal form $\widehat{\mathcal{L}}_{\mathrm{Jn}}$ leading to the equation
\begin{eqnarray}
\widehat{\mathcal{L}}_{\mathrm{Jn}}=\widehat{L}\widehat{\mathcal{L}}\widehat{R},
\label{eq:e3} 
\end{eqnarray}
where $\widehat{\mathcal{L}}_{\mathrm{Jn}}$ has the explicit form as below
\begin{eqnarray}
\widehat{\mathcal{L}}_{\mathrm{Jn}}=\left(\begin{array}{cccc}
                                            0 & 0 & 0 & 0 \\
                                            0 & -\mathrm{i}4\sqrt{3} & 1 & 0 \\
                                            0 & 0 & -\mathrm{i}4\sqrt{3} & 1 \\
                                            0 & 0 & 0 & -\mathrm{i}4\sqrt{3} \\
                                          \end{array}
\right).
\label{eq:e4} 
\end{eqnarray} 
Due to the off-diagonal nonzero defects in Eq.~(\ref{eq:e4}), the time evolution of density matrix elements and other observables pick up additional algebraic terms in time.
\subsection{Density matrix elements}
In region (e), the density matrix elements have the expressions below
\begin{eqnarray}
\rho_j(t)&=&\sum_{k=1}^{4} e^{-\lambda_k t} r_{k,j} a_k-\mathrm{i} t e^{-\lambda t}(r_{2,j}a_3+r_{3,j}a_4)-\frac{t^2}{2}e^{-\lambda t}r_{2,j}a_4, \cr a_k&=&\sum_{n=1}^{4} \ell_{k,n}\rho_n(0).
\label{eq:e5} 
\end{eqnarray}
To investigate QMPE, we consider two initial conditions I and II as usual and calculate the difference in time evolution between these two copies. We have obtained the following expression for ground state probability difference $\Delta \rho_{\mathrm{gg}}(t)=\rho^{\mathrm{I}}_{\mathrm{gg}}-\rho^{\mathrm{II}}_{\mathrm{gg}}$ in region (e):
\begin{eqnarray}
&&\Delta \rho_{\mathrm{gg}}(t)=e^{-4\sqrt{3} t}\left[\gamma_2 t^2+\gamma_1 t+\gamma_0\right],\cr
&&\gamma_2=2\rho_{\mathrm{gg}}(0)-\sqrt{2}\left(\rho_{\mathrm{re}}(0)+\sqrt{3}\rho_{\mathrm{im}}(0)\right),\hspace*{0.3 cm}
\gamma_1=2\sqrt{2}\rho_{\mathrm{im}}(0)-2\sqrt{3}\rho_{\mathrm{gg}}(0),\hspace*{0.3 cm}
\gamma_0=\Delta \rho_{\mathrm{gg}}(0).
\label{eq:e6} 
\end{eqnarray}
The possibility of single or multiple QMPE in region (e) is dictated by the equation $\Delta \rho_{\mathrm{gg}}(t)=0$ and its solutions which are the intersection times, are 
\begin{eqnarray}
\tilde{t}_{\pm}=\frac{-\gamma_1\pm\sqrt{\gamma^2_1-4\gamma_2\gamma_0}}{2\gamma_2}.
\label{eq:e7} 
\end{eqnarray}

The criteria for different types of QMPE in energy in region (e), are listed below
\begin{eqnarray}
&&0<\tilde{t}_+<\infty\,\, \&\,\, 0<\tilde{t}_{-}<\infty: \,\,\mathrm{double}\,\, \mathrm{QMPE},\cr
&&0<\tilde{t}_+<\infty \,\,\mathrm{or}\,\, 0<\tilde{t}_{-}<\infty: \,\,\mathrm{single}\,\, \mathrm{QMPE},\cr
&& \tilde{t}_+\not\in (0,\infty)\,\, \&\,\, \tilde{t}_{-}\not\in (0,\infty): \,\,\mathrm{no}\,\, \mathrm{QMPE}.
\label{eq:e8}
\end{eqnarray}
\begin{figure}[t]
  \centering
    \subfigure[]{\includegraphics[width=0.45\linewidth]{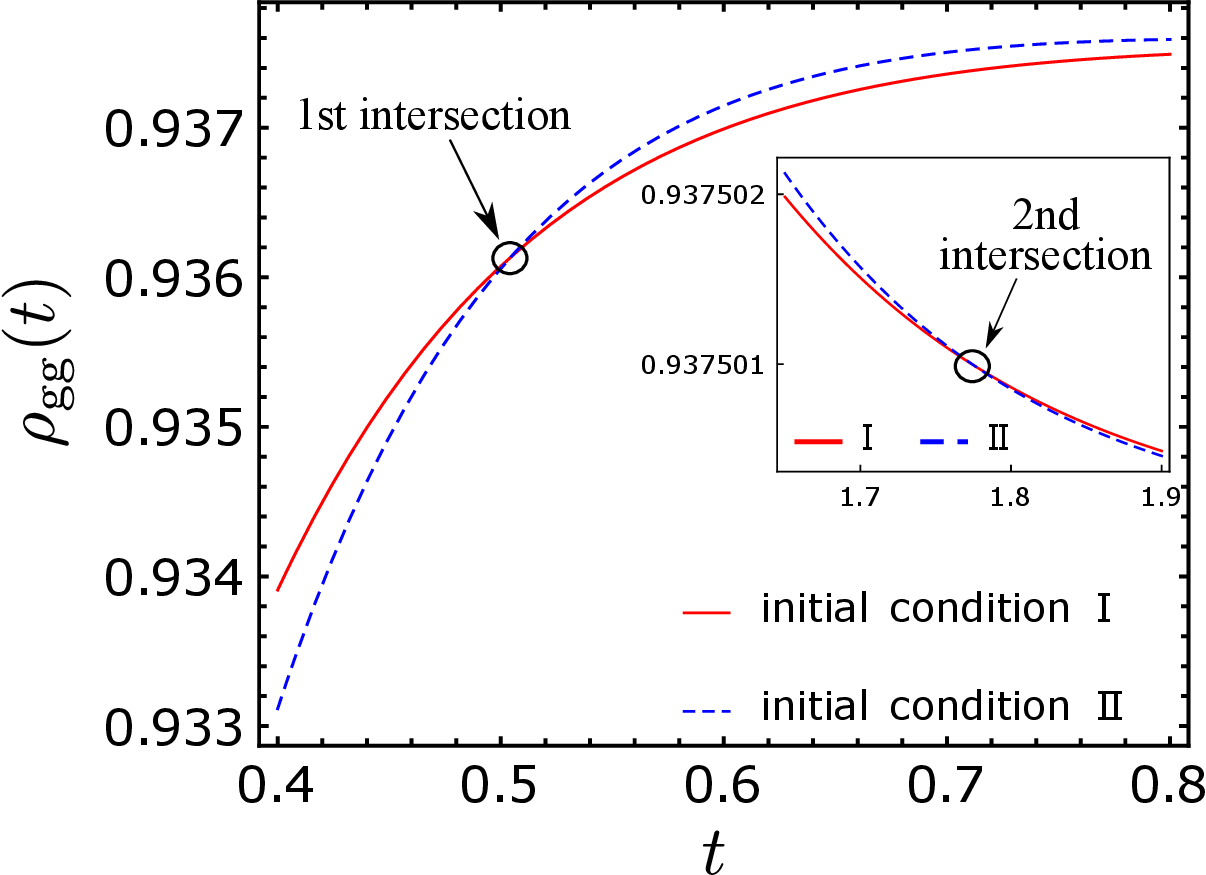}}\hfill
  \subfigure[]{\includegraphics[width=0.45\linewidth]{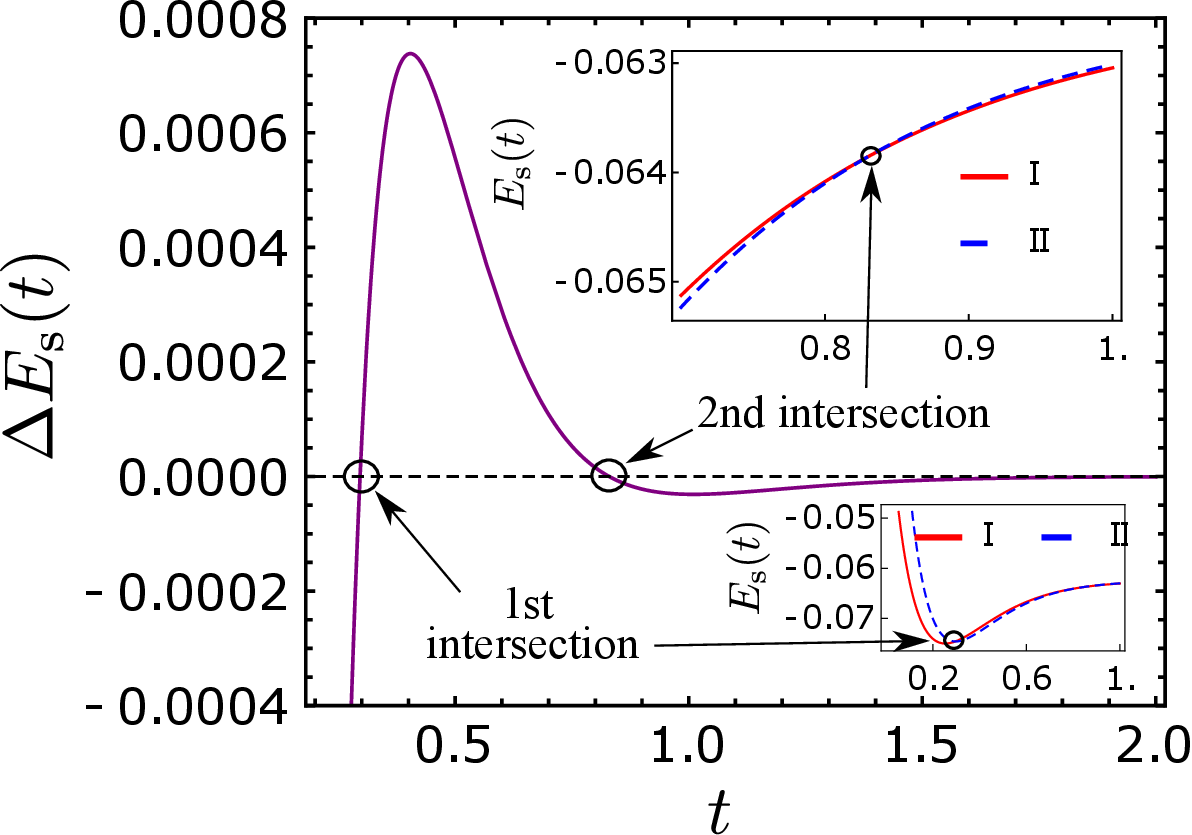}}
\caption{The figure (a) shows an example of double QMPE in the ground state density matrix element in region (e). The main figure shows the first intersection between two copies I and II whereas the inset shows their second intersection that is weaker in magnitude. The figure (b) shows QMPE in average energy in region (e). The main figure shows that the energy difference between two copies I and II intersect zero two times, resulting in double QMPE. The insets show the intersections in the time evolution of the individual energies of the two copies. Parameters used are $\tilde{d}=2\sqrt{2}, \tilde{\Gamma}=\tilde{\Gamma}_{\mathrm{I}}=\tilde{\Gamma}_{\mathrm{II}}=6\sqrt{3}, \tilde{d}_{\mathrm{I}}=5.0, \tilde{d}_{\mathrm{II}}=7.0$.}
\label{fig:e3}
\end{figure}
In Fig.~\ref{fig:e1}(a), we present existence of $\tilde{t}_+$ and $\tilde{t}_-$ as function of the control parameters $\tilde{d}_{\mathrm{I}}$ and $\tilde{d}_{\mathrm{II}}$ using the fixed dissipation protocol. The solutions' surfaces in Fig.~\ref{fig:e1}(a) reveal that there are regions where only one of the solutions $\tilde{t}_+$ or $\tilde{t}_-$ exist leading to single QMPE, and in other regions both the solutions exist resulting in double QMPE. Particularly, lower values of $\tilde{d}_{\mathrm{I}}$ and $\tilde{d}_{\mathrm{II}}$ correspond to single intersection time solution and therefore single QMPE. To understand the behaviors of $\tilde{t}_{+,-}$ more clearly, we fix $\tilde{d}_{\mathrm{I}}$ and display $\tilde{t}_{+,-}$ as functions of $\tilde{d}_{\mathrm{II}}$ in Fig.~\ref{fig:e1}(b). The Fig.~\ref{fig:e1}(b) shows that the two branches $\tilde{t}_+$ and $\tilde{t}_-$ interchange as $\tilde{d}_{\mathrm{I}}<\tilde{d}_{\mathrm{II}}$. More specifically, for $\tilde{d}_{\mathrm{I}}>\tilde{d}_{\mathrm{II}}$ we have $\tilde{t}_+>\tilde{t}_-$ and for $\tilde{d}_{\mathrm{I}}<\tilde{d}_{\mathrm{II}}$ we have $\tilde{t}_+<\tilde{t}_-$. The underlying reason is that the discriminant $\sqrt{\gamma^2_1-4\gamma_2\gamma_0}$ in Eq.~(\ref{eq:e7}) is proportional to sign of $(\tilde{d}_{\mathrm{I}}-\tilde{d}_{\mathrm{II}})$. Thus the two intersection times interchange their roles by changing the sign accompanying the discriminant as we go from $\tilde{d}_{\mathrm{I}}>\tilde{d}_{\mathrm{II}}$ to $\tilde{d}_{\mathrm{I}}<\tilde{d}_{\mathrm{II}}$. The Fig.~\ref{fig:e3}(a) demonstrates an explicit example of double QMPE in ground state probability in region (e). Like other regions, here also the second intersection (shown in the inset of Fig.~\ref{fig:e3}(a)) is much feeble in comparison to the first intersection.  

\subsection{Average energy}
For the same parameters used to obtain QMPE for $\rho_{\mathrm{gg}}(t)$ in Fig.~\ref{fig:e3}(a), we would like to examine the possibility of QMPE in average energy $E(t)$. Indeed, we observe in Fig.~\ref{fig:e3}(b) that energy also exhibits double QMPE, marked by the twice intersection of zero by the energy difference $\Delta E(t)$ between the two copies. The insets of Fig.~\ref{fig:e3}(b) shows separately the two intersections in the time evolution of the energies for the two copies. The second intersection is much weaker in magnitude compared to the first intersection just like the case of ground state probability. Note that the first intersection between the energies of I and II occur after they overshoot their steady state value which they reach during some early time of the relaxation process.

\subsection{von Neumann entropy}
\begin{figure}[t]
  \centering 
  \subfigure[]{\includegraphics[width=0.45\linewidth]{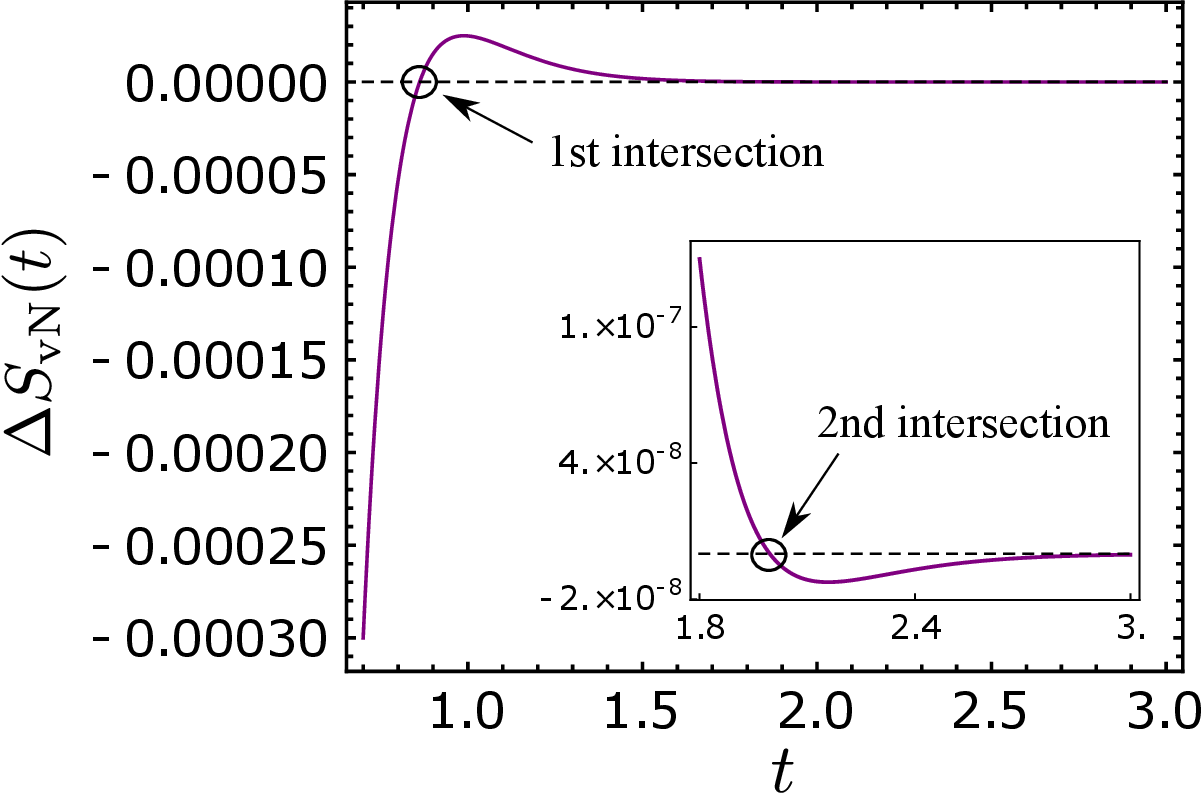}}\hfill
  \subfigure[]{\includegraphics[width=0.45\linewidth]{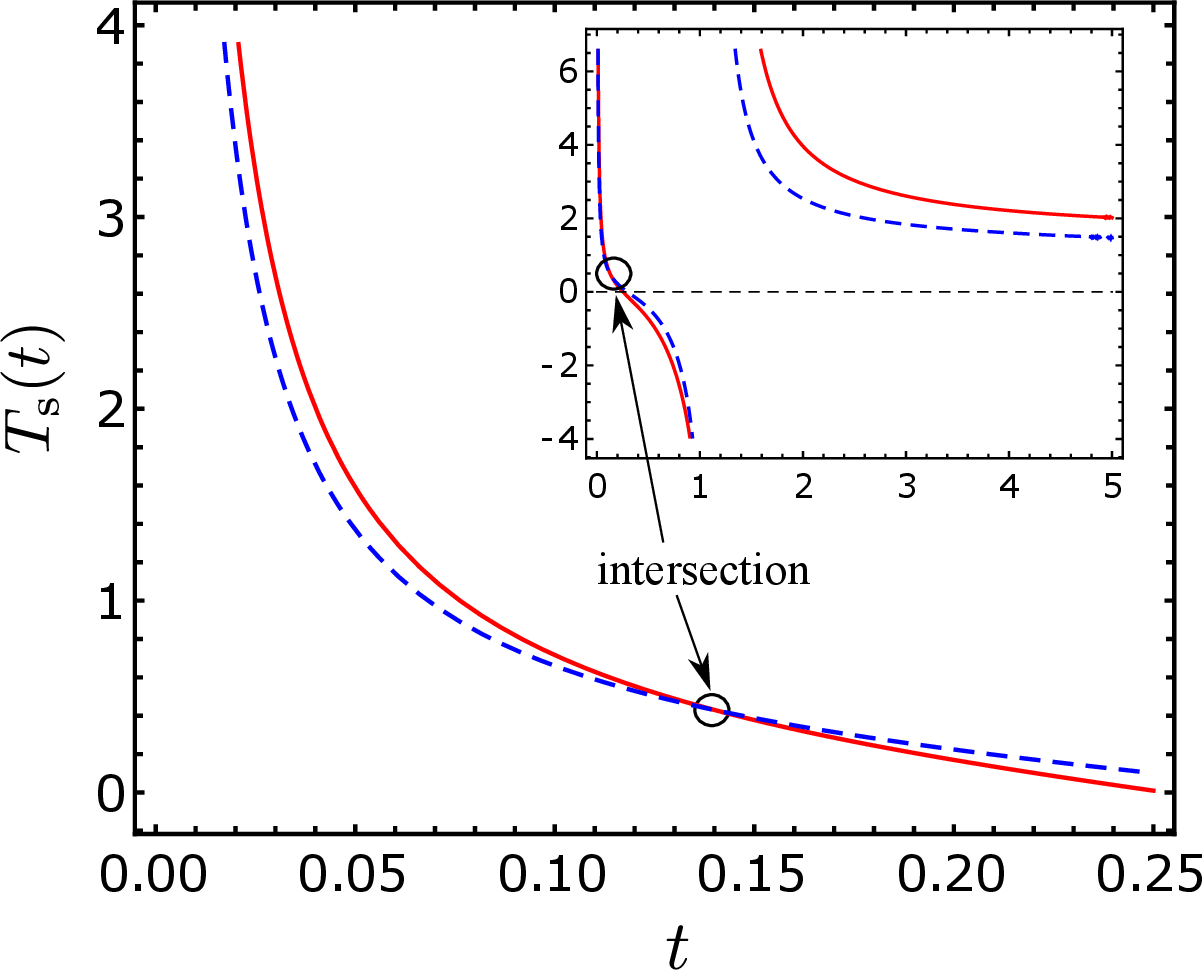}}
\caption{The figure (a) illustrates QMPE in von Neumann entropy in region (e). The difference of entropies between two copies I and II intersect zero twice, leading to double QMPE. The inset shows the second intersection that is weaker in magnitude with respect to the first intersection shown in the main figure. Parameters used in figure (a) are same as in Fig.~\ref{fig:e3}. The figure (b) illustrates an example where the intersection between initially hotter and colder copies occurs {\it before} the initial stage of negative temperature. The inset of figure (b) shows the full time evolution including the initial negative temperatures. Parameters used are $\tilde{d}=2\sqrt{2}, \tilde{\Gamma}=6\sqrt{3}, \tilde{\Gamma}_{\mathrm{I}}=14.0, \tilde{\Gamma}_{\mathrm{II}}=15.0, \tilde{d}_{\mathrm{I}}=6.0, \tilde{d}_{\mathrm{II}}=3.0$.}
\label{fig:e5}
\end{figure}
Here we investigate the von Neumann entropy $S_{\mathrm{vN}}(t)$, starting from two copies I and II. We use same parameters that have produced double QMPE both in ground state probability [Fig.~\ref{fig:e3}(a)] and average energy [Fig.~\ref{fig:e3}(b)]. In fact, the Fig.~\ref{fig:e5}(a) shows that the entropy also exhibits double QMPE indicated by twice zero intersections (the second one shown in inset of Fig.~\ref{fig:e5}(a)) of the entropy difference $\Delta S_{\mathrm{vN}}(t)$ between the two copies.

\subsection{Temperature}
In this section we study the thermal QMPE in region (e). We  investigate thermal QMPE in region (e) by using variable dissipation protocol in Fig.~\ref{fig:e5}(b). There we observe single  intersection and reversal of identities (i.e. hotter becoming colder and vice versa) occur before the negative temperatures have happened. Note that this is contrary to the thermal QMPEs obtained in Figs.~\ref{fig:d5}, \ref{fig:a11} and \ref{fig:lb4}(a) where we obtain single or
multiple intersections after the initial stage of negative temperatures have passed. So, such temperatures may not be considered as well-defined and therefore this intersection in Fig.~\ref{fig:e5}(b) may not correspond  to single QMPE in temperature in region (e). We have to further look for other parameters’ values which can lead us to intersections {\it after} the initial stage of negative temperatures so that thermal QMPE can be
obtained unambiguously like Figs.~\ref{fig:d5}, \ref{fig:a11} and \ref{fig:lb4}(a).

\section{Different types of eigenvalue regions in $\tilde{d}-\tilde{\Gamma}$ parameters' plane}
\label{appen5}
\begin{figure}[h]
  \centering \includegraphics[width=8.6 cm]{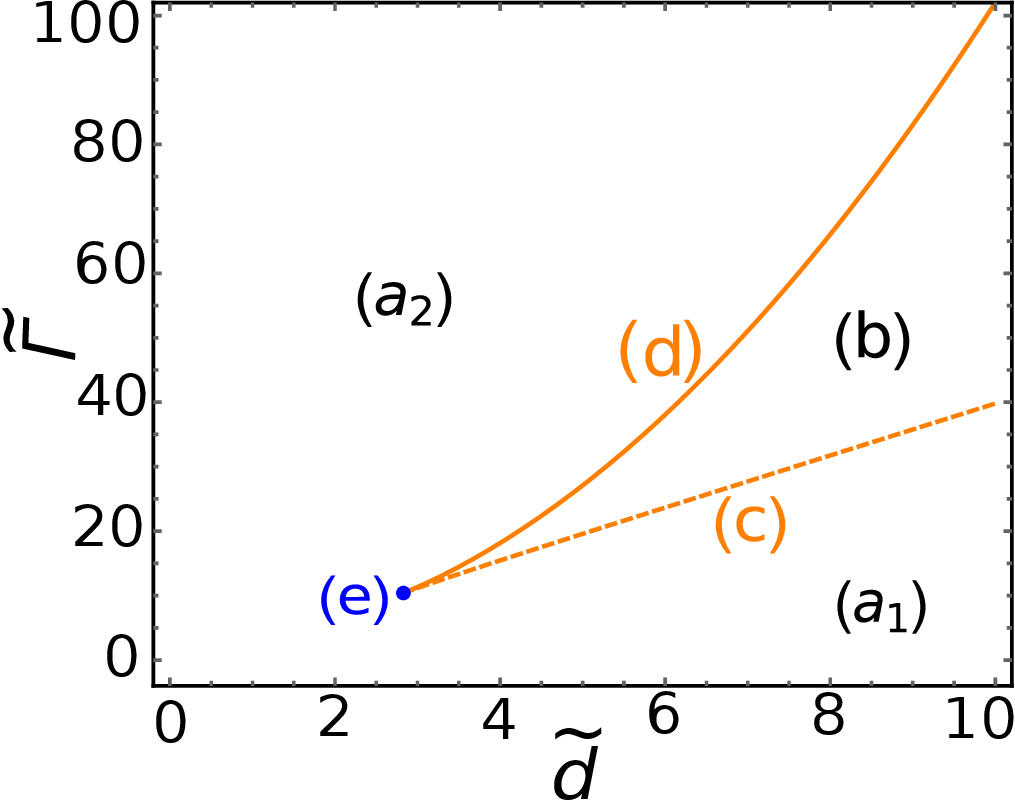}
\caption{The figure shows different eigenvalue regions ($\mathrm{a_1}$), ($\mathrm{a_2}$), (b), (c), (d) and (e) of Hatano's model in the parameters' plane of driving amplitude $\tilde{d}$ and dissipation strength $\tilde{\Gamma}$. This figure is equivalent to Fig.~4 in \cite{Hatano_2019}.}
\label{fig:appen5}
\end{figure}

In this appendix we explicitly show the different regions ($\mathrm{a_1}$), ($\mathrm{a_2}$), (b), (c), (d) and (e); appearing in the parameters' plane $\tilde{d}-\tilde{\Gamma}$, characterized by the nature of the eigenvalues of the Lindbladian [Eq.~(\ref{eq:lindblad_eq})]. For this purpose, we present Fig.~\ref{fig:appen5} that is reproduced from Ref.~\cite{Hatano_2019}.
In Fig.~\ref{fig:appen5}, we observe that the exceptional points [regions (c), (d) and (e)] occur at relatively higher values of $\tilde{\Gamma}$ compared to $\tilde{d}$ and $\delta$ (=1). Same is true for region ($\mathrm{a_2}$) with complex eigenvalues and region (b) with all real (and unequal) eigenvalues. On the contrary, the region ($\mathrm{a_1}$) with complex eigenvalues (discussed in Sec.~\ref{sec5}) correspond to lower values of the dissipation strength  ($<1$) compared to driving amplitude and the detuning (=1).

\twocolumngrid

\end{document}